\documentclass[%
 reprint,
superscriptaddress,
 amsmath,
 amssymb,
prl
]{revtex4-1}

\usepackage{graphicx}
\usepackage{dcolumn}
\usepackage{bm}
\usepackage[usenames,dvipsnames]{color}
\usepackage{bbold}
\usepackage[hidelinks]{hyperref}
\usepackage{mathtools}

\newcommand{\nyuphysics}{Center for Soft Matter Research, Department of Physics, New York University, New York 10003, USA}

\newcommand{\nyusimons}{Simons Center for Computational Physical Chemistry, Department of Chemistry, New York University, New York 10003, USA}
\newcommand{\nyucourant}{Courant Institute of Mathematical Sciences, New York University, New York 10003, USA}

\begin{document}
\preprint{APS/123-QED}

\title{Fast Generation of Spectrally-Shaped Disorder}

\author{Aaron Shih}
\thanks{Equal contribution.}
\affiliation{\nyucourant}
\affiliation{\nyuphysics}
\author{Mathias Casiulis}
\thanks{Equal contribution.}
\affiliation{\nyuphysics}
\affiliation{\nyusimons}
\author{Stefano Martiniani}
\email{sm7683@nyu.edu}
\affiliation{\nyucourant}
\affiliation{\nyuphysics}
\affiliation{\nyusimons}

\date{\today}

\begin{abstract}
Media with correlated disorder display unexpected transport properties, but it is still a challenge to design structures with desired spectral features at scale.
In this work, we introduce an optimal formulation of this inverse problem by means of the non-uniform fast Fourier transform, thus arriving at an algorithm capable of generating systems with arbitrary spectral properties, with a computational cost that scales $\mathcal{O}(N \log N)$ with system size.
The method is extended to accommodate arbitrary real-space interactions, such as short-range repulsion, to simultaneously control short- and long-range correlations. 
We thus generate the largest-ever stealthy hyperuniform configurations in $2d$ ($N = 10^9$) and $3d$ ($N > 10^7$).
By an Ewald sphere construction we link the spectral and optical properties at the single-scattering level, and show that these structures in $2d$ and $3d$ generically display transmission gaps, providing a concrete example of fine-tuning of a physical property at will.
We also show that large $3d$ power-law hyperuniformity in particle packings leads to single-scattering properties near-identical to those of simple hard spheres.
Finally, we show that enforcing large spectral power at a small number of peaks with the right symmetry leads to the non-deterministic generation of quasicrystalline structures in both $2d$ and $3d$.
\end{abstract}


\maketitle

The study of condensed matter is often facilitated by the periodicity of atomic structures: for instance, photonic bandgaps in crystals can be predicted by Block's theorem~\cite{Kittel}.
Contrarily, analytic models are still being developed to understand the emergent optical properties of \textit{disordered media}~\cite{Carminati2021,Vynck2023}, \textit{i.e.}, materials that do not exhibit conventional forms of long-range order.
Among disordered materials, systems with \textit{correlated disorder}, whose structures are non-Poissonian random point patterns, have garnered attention following experimental and computational reports of unconventional scattering properties: structural coloration~\cite{Djeghdi2022,Vynck2022}, isotropic bandgaps~\cite{Florescu2009,Man2013}, or Anderson localization~\cite{Froufe-Perez2017,Monsarrat2022}.

Beyond condensed matter, correlated point patterns are crucial to computer graphics~\cite{Yan2015, Pharr2018}, and various protocols have been introduced to impose prescribed spatial correlations between points~\cite{Yan2015, Pilleboue2015, Pharr2018}.
These strategies amount to an optimal sampling problem: given some natural image, where should a finite number of sample points be placed in order to minimize aliasing errors?
A common answer is to use \textit{blue-noise sampling}~\cite{Yan2015}, \textit{i.e.} point patterns with highly suppressed long-ranged pair correlations but no clear periodicity.
In practice, the best such point patterns have strictly zero low-frequency content~\cite{Pilleboue2015}.

\begin{figure*}
    \centering
    \includegraphics[width=0.96\textwidth]{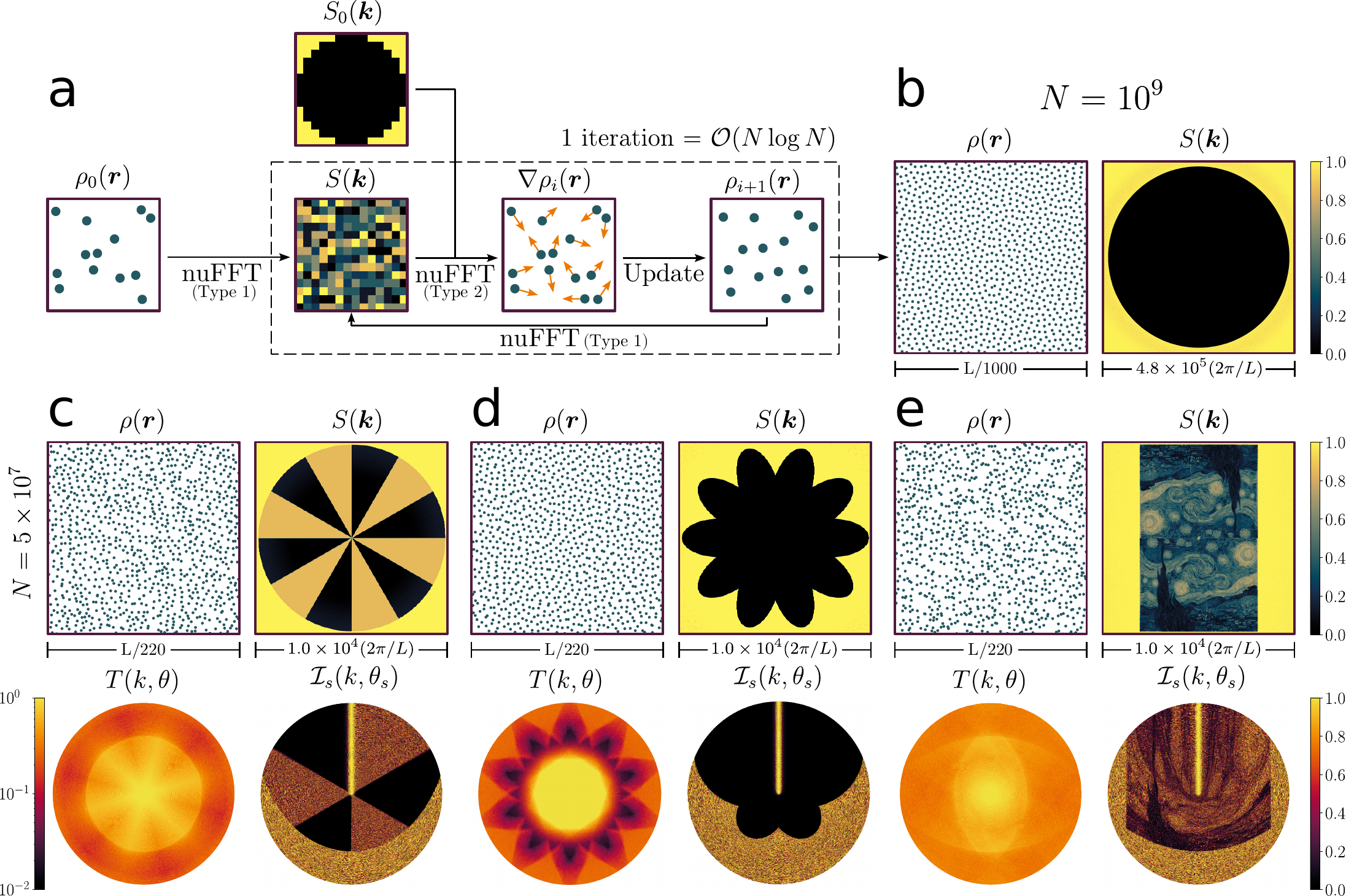}
    \caption{\textbf{Fast Reciprocal-Space Correlator (FReSCo):}
    $(a)$ Sketch of the implementation of our algorithm.
    A point pattern $\rho_0(\bm{r})$ is subjected to a nuFFT transformation, so that a loss can be computed from the difference between the observed $k$-space structure, $S(\bm{k})$, and a target function, $S_0(\bm{k})$.
    The gradient of this loss is obtained as another nuFFT, so that each iteration of the optimization can be performed in $\mathcal{O}(N \log N)$ operations.
    $(b)$ Example output of the algorithm: we show a small portion of an $N=10^9$ point pattern, as well as the final structure factor.
    $(c)-(e)$ A few example outputs, imposing a variety of target structures to smaller systems ($N = 5 \times 10^7$ points): from left to right, a pinwheel, a 10-petaled flower structure factor, and the lightness scale of Van Gogh's Starry Night~\cite{StarryNight}.
    In each panel, we show a portion of the final point pattern (top left), structure factor (top right), forward scattering transmission, $T(k, \theta)$, as a function of the magnitude, $k$, and orientation, $\theta$, of an incoming wave (bottom left), and scattered intensity, $\mathcal{I}_s(k, \theta_s)$ (bottom right) for an upward incident wave-vector, as a function of the incident frequency, $k$, and of the scattered direction, $\theta_s$. }
    \label{fig:Sketch}
\end{figure*}
Physicists refer to such point patterns as \textit{disordered hyperuniform} structures~\cite{Torquato2018}, for which photonic bandgaps and localization have been attributed to their suppressed long-range density fluctuations~\cite{Florescu2009,Man2013,Klatt2022, Froufe-Perez2017,Haberko2020,Monsarrat2022}.
To produce hyperuniformity, the best known algorithm relies on the collective coordinate approach~\cite{Uche2004, Uche2006}, which minimizes differences between an observed pair correlation function and a desired one using gradient optimization.
These algorithms suffer from a major drawback: their algorithmic complexity is $\mathcal{O}(N^2)$~\cite{Morse2023}, or even $\mathcal{O}(N^3)$~\cite{Uche2004, Uche2006} in the number, $N$, of points.
Consequently, the vast majority of hyperuniform systems studied in the literature contain modest numbers of points ($10^2$ to $10^4$ points~\cite{Uche2004,Uche2006,Froufe-Perez2016}, more recently up to $\sim 10^6$ points using a massively parallel GPU implementation~\cite{Morse2023}), and were overwhelmingly limited to one specific kind of hyperuniformity to make calculations tractable~\cite{Leseur2016, Monsarrat2022, Morse2023}.
These limitations have also critically affected the scale of additively manufactured hyperuniform materials, typically a few hundreds of particles only~\cite{Florescu2009,Man2013,Gorsky2019}, which is particularly problematic in $3d$~\cite{Haberko2020,Torquato2021} as the linear size of the system reaches only tens of particles across.
This raises the question of whether the structures used in past studies truly encoded hyperuniformity, an inherently long-range property (see SI).

In this paper, we introduce a powerful optimization algorithm, sketched in Fig.~\ref{fig:Sketch}$(a)$, that can generate spectrally-shaped disordered point structures with arbitrary spectral features (see Fig.~\ref{fig:Sketch}$(b)-(e)$).
In short, the algorithm resorts to non-uniform Fast Fourier Transforms (nuFFTs) to efficiently compute the structure factor $S(\bm{k})$ of a point pattern $\rho(\bm{r})$.
The distance from $S(\bm{k})$ to a prescribed target $S_0(\bm{k})$ then defines a loss, the gradient of which can also be written as a nuFFT, so that the cost of one step in a minimization procedure scales quasilinearly in the system size, $\mathcal{O}(N \log N)$.
This cost function can be jointly optimized with additional physical constraints, such as short-ranged pair repulsion, with no increase in computational complexity.

We demonstrate its application for as many as $10^9$ points (Fig.~\ref{fig:Sketch}$(b)$), outperforming (by at least 3 orders of magnitude) all previously published methods possessing the same specificity in $k$-space for point patterns~\cite{Uche2004,Uche2006,Morse2023}.
The target structure factor, $S_0$, can be chosen at will as long as the number of $k$-space features being constrained does not exceed the number of degrees of freedom.
We show a few smaller ($N = 5 \times 10^7$) examples in Fig.~\ref{fig:Sketch}$(c)-(e)$, where we embed a pinwheel $(c)$, a 10-fold flower $(d)$, or Van Gogh's Starry Night~\cite{StarryNight}$(e)$ into the structure factor.
We also show (bottom row of $(c)-(e)$) that the optical properties of such structures can be characterized in the single-scattering regime on the scale of realistic devices, without assuming periodicity.
We show the forward-scattered transmission pattern, $T$, of these structures against the wave-vector of an incoming plane wave, as well as the intensity, $\mathcal{I}_s$, of the scattered field in each direction, for an upward incident wave, across frequencies (see precise definitions below).

In the following, we highlight the range of applications of this algorithm.
We show that unquestionably stealthy hyperuniform systems (i.e., with hyperuniform density fluctuations scalings spanning over 3 decades, see SI) have transmission gaps even at the single-scattering level.
However, we show that particle systems with power-law behaviour in $S$ (like particles at jamming~\cite{Jiao2011a} or critical absorbing-state models~\cite{Corte2008,Hexner2015}), have single-scattering properties indistinguishable from equilibrium hard spheres.
We finally discuss extensions of FReSCo, \textit{e.g.} one that produces quasicrystalline structures~\cite{Levine1984} from very few constraints.

\noindent\textit{Algorithm --} 
Consider a set of $N$ points at $d$-dimensional positions, $\bm{r}_1, \ldots, \bm{r}_N$ $\in \mathbb{R}^d$, each carrying a weight $c_n \in \mathbb{C}$.
One may define a density field as the sum of $N$ Dirac deltas, $\rho(\bm{r}) \equiv \sum_n  c_n \delta(\bm{r}-\bm{r}_n)$.
In Fourier space, $\widehat{\rho}\,(\bm{k}) = \sum_n c_n \exp( i\bm{k}\cdot\bm{r}_n)$, so that one may define the structure factor
\begin{align}
    S(\bm{k}) \equiv \frac{|\widehat{\rho}\,(\bm{k})|^2}{N}, 
    \label{eq:structure_factor}
\end{align}
which encodes the two-point correlations of $\rho$~\cite{Hansen2006}.
The Fast Reciprocal-Space Correlator (FReSCo) is a minimization protocol against a loss, $\mathcal{L}_S$, defined as the least square error between $S(\bm{k})$ and a prescribed target, $S_0(\bm{k})$, in a finite region $\mathcal{K}$ of reciprocal space:
\begin{align}
    \mathcal{L}_S\left[(\bm{r}_1, c_1), \ldots, (\bm{r}_N, c_N)\right] = \sum_{\bm{k} \in \mathcal{K}}w(\bm{k}) L [S(\bm{k}), S_0(\bm{k})],
    \label{eq:SF_loss}
\end{align}
where $w(\bm{k})$ is a weighting function, and
\begin{align}
    L [S(\bm{k}), S_0(\bm{k})] =
    \begin{cases}
        \left[(S(\bm{k})-S_0(\bm{k}))/S_0(\bm{k})\right]^2 & \text{if } S_0(\bm{k}) \neq 0\\
        S(\bm{k})^2 & \text{otherwise}
    \end{cases}
    \label{eq:loss}
\end{align}
penalizes the relative distance to $S_0(k)$.
We choose $w(\bm{k}) \sim |\bm{k}|^{-(d-1)}$ when $S_0(\bm{k})$ has pronounced radial symmetry around $k =0$, so that $k$-space constraints are equally strong on every spherical shell.
For instance, this applies to the structure factors of Fig.~\ref{fig:Sketch}$(b)-(d)$, while for panel $(e)$ we choose $w(\bm{k}) = 1$.

Gradients of this loss can be written both with respect to weights and positions (see Methods).
We first focus on the optimization of continuous positions with ${c_n = 1}$, a problem formally equivalent to that introduced in the collective coordinate approach~\cite{Uche2004,Uche2006,Morse2023}.
In that case, $\rho(\bm{r})$ is real-valued and $S(\bm{k}) = S(-\bm{k})$, a property known in crystallography as Friedel's law~\cite{Schwarzenbach1996}, so that only centrosymmetric $S$ are realizable.
For instance, embedding Starry Night~\cite{StarryNight} in the $k_y < 0$ half-plane leads to its inversion being constrained for $k_y > 0$ in Fig.~\ref{fig:Sketch}$(e)$.
Using Eq.~\ref{eq:structure_factor}, the gradients of the loss function can be written as Fourier transforms (see Methods),
\begin{equation}
\begin{aligned}
    \frac{\partial \mathcal{L}_S}{\partial \bm{r}_{n}} = Re\left[\sum_{\bm{k}}\bm{C}(\bm{k}) c_n \exp(-i\bm{k}\cdot\bm{r}_n)\right] = c_n Re\left[\widehat{\bm{C}}(\bm{r}_n)\right]
    \label{eq:gradient}
\end{aligned}
\end{equation}
where $\bm{C}(\bm{k})=-4i\bm{k} w(\bm{k})\left[S(\bm{k})-S_0(\bm{k})\right]\widehat{\rho}\,(\bm{k})/N$ are coefficients of a Fourier series, and $\widehat{\bm{C}}$ is the Fourier transform of $\bm{C}$.

In total, one FFT is required to compute $\mathcal{L}_S$ and $d$ additional FFTs are required for the gradient of $\mathcal{L}_S$.
As these are the most costly steps in calculating the loss, the time complexity of our algorithm is $\mathcal{O}(N \log N)$ for loss and gradient calls (see benchmarks in SI), nearly N times (viz., many orders of magnitude) faster than the $\mathcal{O}(N^2)$ or $\mathcal{O}(N^3)$ of previous algorithms \cite{Uche2004,Uche2006,Morse2023}.
Coupled with state-of-the-art optimizations built into FINUFFT, the computational speed increase is enormous, enabling generation of correlated disordered systems up to ${N=10^9}$ on CPUs (Figure~\ref{fig:Sketch}$(b)$), the main limitation being memory requirements.
The optimization is performed by feeding the configuration and gradient to L-BFGS~\cite{Liu1989}, a quasi-Newton method, with a maximal step size and a backtracking line-search~\cite{Nocedal1999}.

\begin{figure*}
    \centering
    \includegraphics[width=0.90\textwidth]{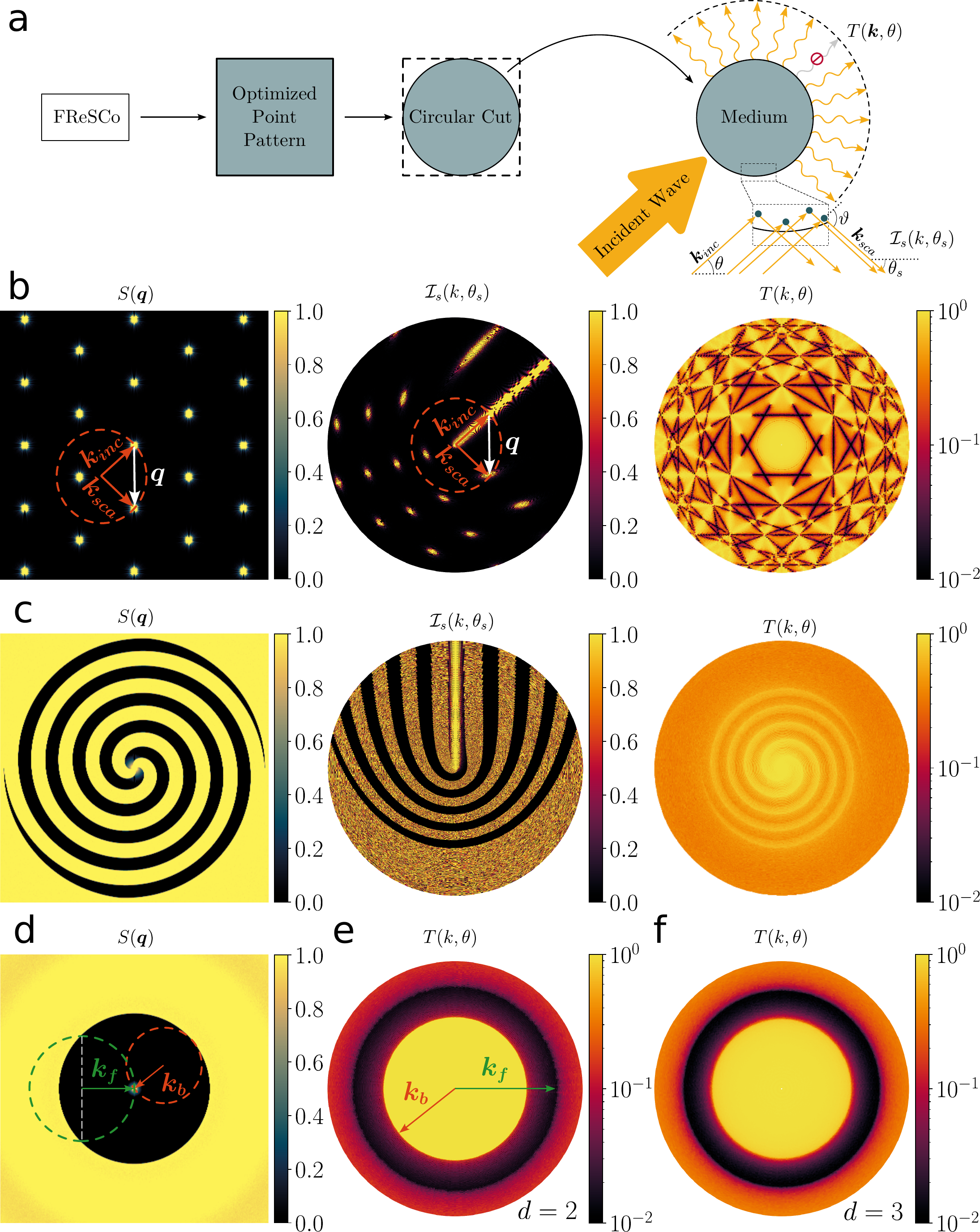}
    \caption{\textbf{Ewald sphere construction and single scattering.}
    $(a)$ Sketch.
    From an optimized structure, we cut the central disk then measure its single-scattering properties.
    To an incoming wavevector, $\bm{k}_{inc}$, we associate the far-field, normalized scattered intensity, $\mathcal{I}_s$, in direction $\theta_s$.
    $(b)$ Illustration of the Ewald circle construction on a $2d$ triangular lattice.
    From the structure factor $S(\bm{q})$ (left) evaluated at $\bm{q} = \bm{k}_{sca} - \bm{k}_{inc}$, we obtain $\mathcal{I}_s$ (middle) which, once integrated, yields the transmission, $T$, as a function of the incident wave-vector (right).
    $(c)$ Example of an optimized $2d$ point pattern ($N = 5\times 10^7$) with a spiral of zeros in its $S(\bm{q})$.
    $(d)$ Structure factor, $S(\bm{q})$, and $(e)-(f)$ transmission, $T$, results for generated stealthy hyperuniform structures, for $(e)$ $N = 5 \times 10 ^7$ in $2d$, and $(f)$ $N = 5 \times 10^6$ in $3d$ (right).
    In $(d)-(e)$, we indicate the values $k_b$ (red) and $k_f$ (green) above which back- and forward-scattering develop, respectively, based on the Ewald construction.
    In $(c)$ and $(e)$, $K =5050$ and plots are shown up to $k_{max} = 4500$.
    In $(f)$, $K = 142$, $k_{max} = 130$, and $\theta$ sweeps one arbitrary circle.
    }
    \label{fig:ewald}
\end{figure*}

\noindent\textit{Structure and scattering--}
To characterize the optical behaviour of systems at scale without introducing artificial periodicity, we proceed as follows.
First, like in experiments~\cite{Man2013}, we cut the optimized point patterns into disks to avoid anisotropy coming from the shape of the medium; then, we use the Ewald sphere construction~\cite{Kittel} on the resulting object by means of FINUFFT transformations (see sketch in Fig.~\ref{fig:ewald} $(a)$).
We remind that the Ewald sphere construction is equivalent to the far-field single-scattering response in SI, and illustrate the technique in Fig.~\ref{fig:ewald}$(b)$.
In short, at single-scattering level, for an incident wavevector $\bm{k}_{inc}$, the far-field intensity scattered with wavevector $\bm{k}_{sca}$ is proportional to $S(\bm{q})$ at $\bm{q} = \bm{k}_{sca} - \bm{k}_{inc}$~\cite{Carminati2021}.
Therefore, we read off the single-scattering response at any given frequency, and in any observation direction, by drawing a sphere centered at $-\bm{k}_{inc}$ with radius $k = |\bm{k}_{inc}| = |\bm{k}_{sca}|$.
In Fig.~\ref{fig:ewald}$(b)$, we illustrate this procedure and show the scattered intensity profile, $\mathcal{I}_s$, for a single incident illumination direction, against the wave-vector magnitude $k$ (radial direction), and the observation direction $\theta_s$ (orthoradial direction).
In order to account for a finite detection width, we also define a normalized $2d$ transmission,
\begin{align}
    T(\bm{k}_{inc}) = \frac{\int_{\mathcal{F} \setminus 0} S\left[k  \hat{\bm{e}}(\theta + \vartheta) - k \hat{\bm{e}}(\theta )\right]d\vartheta}{\oint_{\mathcal{C} \setminus 0}S\left[k  \hat{\bm{e}}(\theta + \vartheta) - k\hat{\bm{e}}(\theta )\right]d\vartheta},
    \label{eq:ewald}
\end{align}
where the angle $\vartheta$ between the incident and scattered waves is integrated over the forward half-circle $\mathcal{F} \subset \mathcal{C}$ around the incident direction $\theta$, and the normalization is the total scattered intensity on the full circle, $\mathcal{C}$, removing from both integrals the direction $\vartheta = 0$ (that reduces to the peak $S(\bm{0}) = N$).
We extend this definition to $3d$ by replacing $\vartheta$ by a solid angle and the (half-)circle by a (half-)sphere.

In Fig.~\ref{fig:ewald}$(c)$, we show one example of the possibilities offered by FReSCo: a spiral-shaped domain of zeros in the structure factor of $N = 5\times 10^7$ particles leads to fringes of low scattered intensities in a range of frequencies, and to a spiral-shaped transmission pattern.
This example and those of Fig.~\ref{fig:Sketch}$(c)$ demonstrate that achieving fine control over $S(\bm{q})$ in large point patterns enables the design of intricate scattering behaviours.

We take advantage of this approach to study stealthy hyperuniform structures, systems with $S(k) = 0$ in a disk of radius $K$, at scale: Fig.~\ref{fig:ewald}$(d)$ we show a typical structure factor and, in Figs.~\ref{fig:ewald}$(e)-(f)$, transmission plots, $T(k,\theta)$.
On $S(\bm{q})$, we highlight two special values of $k$.
The first such value, $k_b = K/2$, separates the domain $k < k_b$ in which the system is transparent (up to multiple scattering effects~\cite{Leseur2016}) from the domain $k> k_b$ in which the system backscatters, as the back of the Ewald circle overlaps high values of $S(\bm{q})$.
The second special value, $k_f = K/\sqrt{2}$, separates the regimes $k_b<k<k_f$ where only backscattering happens, and $k>k_f$, where forward-scattering sets in, down to narrower and narrower angles as the frequency increases.
Therefore, one expects a trough of lower forward-scattered transmission at intermediate $k_b < k < k_f$,  suggestive of an isotropic bandgap, in a stealthy hyperuniform configuration.
This picture is confirmed in the transmission plot of panel $(e)$, obtained for a $2d$ configuration with $N = 5 \times 10^7$ points, where we report these values, and in panel $(f)$, obtained for a $3d$ configuration with $N= 5 \times 10^6$ points.

While similar observations were made for $2d$ systems~\cite{Florescu2009,Man2013,Froufe-Perez2016,Froufe-Perez2017,Klatt2022,Monsarrat2022,Morse2023}, this result constitutes the largest-scale direct check that stealthy hyperuniform systems do feature isotropic transmission gaps in $2d$, and, to our knowledge, the first such measurement in a stealthy hyperuniform system in $3d$, as well as the largest $3d$ hyperuniform systems altogether~\cite{Liew2011,Muller2014,Haberko2020}.
Note that a smaller angular integration domain in $T$ (i.e., a smaller detector) leads to a broader transmission trough, $\left[ k_b; k_f +\epsilon\right]$ with $\epsilon \geq 0$.
By integrating over the full half-disk or half-sphere we are thus reporting the \textit{narrowest} observable transmission gap, see SI.

\begin{figure*}
    \centering
    \includegraphics[width=0.96\textwidth]{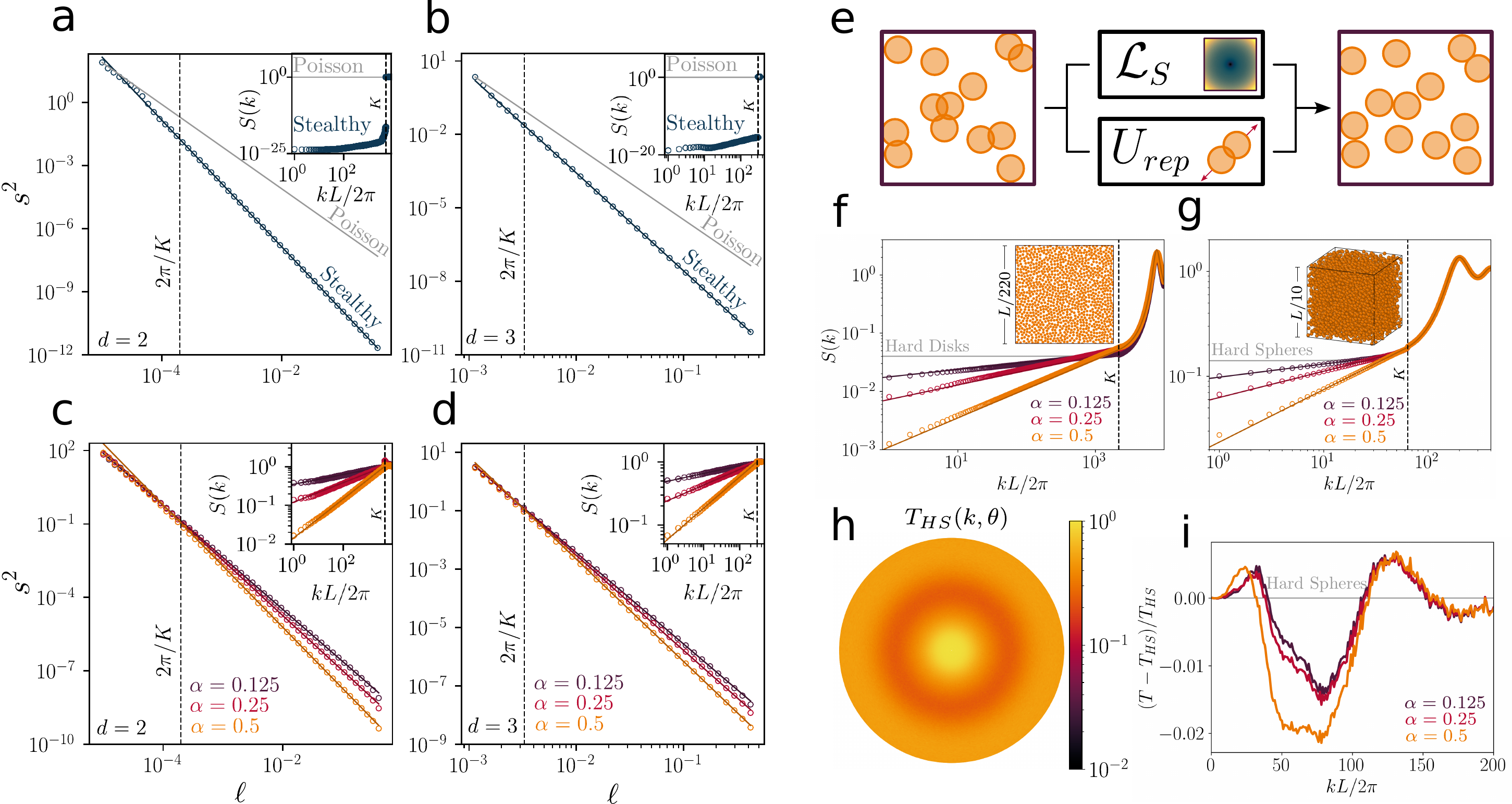} 

    \caption{\textbf{Hyperuniform structures.}
    $(a)-(b)$ Number fluctuations against measurement window size (main panel), and  structure factor (inset) for stealthy hyperuniform systems.
    We indicate the stealthy $\ell^{-(d+1)}$ (dark blue) and Poisson $\ell^{-d}$ (gray) scalings as solid lines, as well as a Poisson structure factor $S(k)=1$ in the inset.
    $(c)-(d)$ Same plots for power-law hyperuniform structures with exponents $\alpha \in \{0.125, 0.25, 0.5\}$ in $(c)$  $2d$ and $(d)$ $3d$.
    We show next to each curve the expected power law, $\ell^{-(d+\alpha)}$, in the main panels, and $S(k) \sim k^{\alpha}$ in the insets, with solid lines.
    Across $(a)-(d)$, we show the radius $K$ of the constrained disk in Fourier space, and the corresponding length scale $\ell = 2\pi /K$, as dashed lines.
    $(e)$ Sketch of the algorithm in the presence of both real- and reciprocal-space loss functions.
    From arbitrary initial conditions, we jointly optimize for prescribed features in $k$-space and short-range repulsion.
    $(f)-(g)$ Structure factors of hyperuniform monodisperse $(f)$ disk packings ($\phi = 0.6$) and $(g)$ sphere packings ($\phi = 0.25$).
    Insets depict packings of the hyperuniform $\alpha=0.5$ power law systems.
    Structure factors are averaged over 10 realizations.
    $(h)$ Forward-scattered transmission for equilibrium hard spheres at $\phi = 0.25$ up to $k_{max} = 200$ and $(i)$ relative change of forward-scattered transmission between $3d$ power-law hyperuniform structures of panel $(f)$ and the starting equilibrium hard sphere configuration.
    Across $(a)-(f)$, $N = 5\times 10^7$, and $N = 4\times 10^6$ in $(g)-(i)$.}
    \label{fig:Hyperuniformity}
\end{figure*}
\noindent\textit{Disordered Hyperuniform Structures --}
We now focus on the quality of our disordered hyperuniform structures.
From a $k$-space perspective, hyperuniformity is associated with an anomalous decay of the structure factor, $S(\bm{k})$, at long range, or $S(\bm{k})\to 0$ when $|\bm{k}|\to 0$.
How the structure factor decays depends on the class of hyperuniform system \cite{Torquato2018}.
We investigate two types of disordered hyperuniformity: stealthy and power-law.
Stealthy hyperuniformity occurs when $S(\bm{k})=0$ for $|\bm{k}|<K$, like the examples of Figs.~\ref{fig:Sketch}$(b)$ and~\ref{fig:ewald}$(d)$, while power-law hyperuniformity implies $S(\bm{k})\sim |\bm{k}|^\alpha$ for $|\bm{k}|<K$ and $\alpha > 0$.
As hyperuniform structures considered in the literature were small, while structure factors appeared consistent with a power law for a few $k$-vectors~\cite{Uche2006,Liew2011}, it is not obvious that density fluctuations were actually suppressed at long range.
Concretely, hyperuniformity is achieved only if the variance in the number of points sampled across spheres, grows \textit{slower} than their volume, $s^2 \equiv  \langle N^2 \rangle /\langle N \rangle^2 - 1\sim \ell^{-\beta}$, with $d \leq \beta \leq d +1$, while in an uncorrelated point pattern $s^2\sim \ell^{-d}$.
In the following, we show that FReSCo is able to generate structures for which these power laws are verified over several decades of $\ell$.

In Fig.~\ref{fig:Hyperuniformity}$(a)$ and $(b)$, we show the radially averaged structure factors (insets) and the associated number fluctuations, shown as the reduced variance, $s^2$, against the radius of a measurement sphere (main panels) for $N = 5\times 10^7$ in $2d$ and $3d$ respectively.
These disordered stealthy hyperuniform configurations are orders of magnitude larger than any previous realization~\cite{Uche2004, Uche2006, Leseur2016, Froufe-Perez2016, Gorsky2019,Haberko2020, Monsarrat2022, Klatt2022, Morse2023}, as well as the most solid evidence of stealthy hyperuniformity in a system being associated with a $s^2 \sim \ell^{-(d+1)}$ decay of number fluctuations.

Inspired by critical configurations of absorbing-phase models~\cite{Corte2008,Hexner2015} and jammed packings~\cite{Jiao2011a}, we also design power-law hyperuniform point patterns by constraining the structure factor $S(\bm{k})\sim|\bm{k}|^\alpha$ such that the structure factor at the largest wavevector magnitude being constrained is $S(K)=1$.
We minimize 10 configurations of $N=5 \times 10^7$ point systems for power laws $\alpha\in\{0.125,0.25,0.5\}$.
Figure \ref{fig:Hyperuniformity}$(c)$ and $(d)$ depict the final structure factors (insets) and the associated number fluctuations against $\ell$ (main panels) in $2d$ and $3d$, respectively.
The decay in the variance matches the predicted trends, $s^2 \sim \ell^{-(d + \alpha)}$~\cite{Torquato2018,Pilleboue2015}, decades beyond the length scale $2\pi / K$.
This is by far the largest, and most rigorous, test of the real-space properties of power-law hyperuniform point patterns reported to date.

\textit{Hyperuniform Particle Packings --}
Thus far, we only constrained Fourier-space properties of point patterns, so there was no notion of excluded volume: two points could come arbitrarily close together.
This generically precludes the fabrication of raw point patterns without the use of arbitrary geometric transformations~\cite{Florescu2009}.
In order to generate more physical systems, in line with previous works~\cite{Kim2020}, we introduce a hybrid loss that combines the structure factor loss, Eq.~\ref{eq:SF_loss}, with a repulsive pair potential $U_{rep}$,
\begin{align}
    \mathcal{L} &= \mathcal{L}_{S} + \sum\limits_{m < n} U_{rep}(\bm{r}_m - \bm{r}_n). \label{eq:hybrid_loss}
\end{align}
This variant of FReSCo is sketched in Fig.~\ref{fig:Hyperuniformity}$(e)$.
As long as the potential is finite-ranged, computing the loss or its gradient still takes $\mathcal{O}(N \log N)$ operations.
One may also introduce polydispersity into the system by specifying individual particle diameters in $U_{rep}$, in which case, to get the correct definition of $S$ for homogeneous polydisperse spheres, each particle should be weighed by the ratio of its $d$-dimensional volume, $V_n$, to the mean volume, $\langle V \rangle$, i.e, $c_n = V_n /\langle V \rangle$ in Eq.~\ref{eq:gradient}~\cite{Vrij1978}.
Here, we choose a monodisperse Hertzian potential, $U_{rep}(r) \propto (r-\sigma)^{2.5}$, with $\sigma$ the repulsive diameter.
We also adjust the prefactor of the power law and the extent, $K$, of the domain in which we constrain the structure factor such that the target, $S_0(K)$, smoothly interpolates the Percus-Yevick approximation for the structure factor of hard sphere liquids in $3d$~\cite{Hansen2006}, and a similar approximation in $2d$~\cite{Rosenfeld1990}.
Results thus obtained are shown in Fig.~\ref{fig:Hyperuniformity}, in both $2d$ (panel $(f)$) and $3d$ (panel $(g)$).
Our results show that arbitrary long-range features can still be achieved in the presence of short-range constraints like excluded volume, which guarantees the fabricability of structures with actual physical objects.

We also generate the single-scattering Ewald transmissions of these $3d$ configurations.
In Fig.~\ref{fig:Hyperuniformity}$(h)$ we show the resulting $T_{HS}$ for an equilibrium hard sphere configuration (obtained using event-chain Monte Carlo methods~\cite{Bernard2009}) at $\phi = 0.25$ and, in Fig.~\ref{fig:Hyperuniformity}$(i)$ the relative change between $T_{HS}$ and the transmission, $T$, of power-law hyperuniform structures (same as in Fig.~\ref{fig:Hyperuniformity}$(g)$), radially averaged over incoming angles.
Power-law hyperuniformity, even in such large systems, does not significantly affect the scattering properties of hard sphere systems in the single-scattering limit, as the largest relative change is only a few percents.
Thus hyperuniformity \textit{per se}, as realized in critical systems like jammed packings, is not a necessary condition to observe a transmission gap in the single-scattering regime.
This result, reminiscent of past work on stealthy hyperuniform structures~\cite{Froufe-Perez2016}, is the first direct measurement in large power-law hyperuniform systems.

\begin{figure*}
    \centering
    \includegraphics[width=0.96\textwidth]{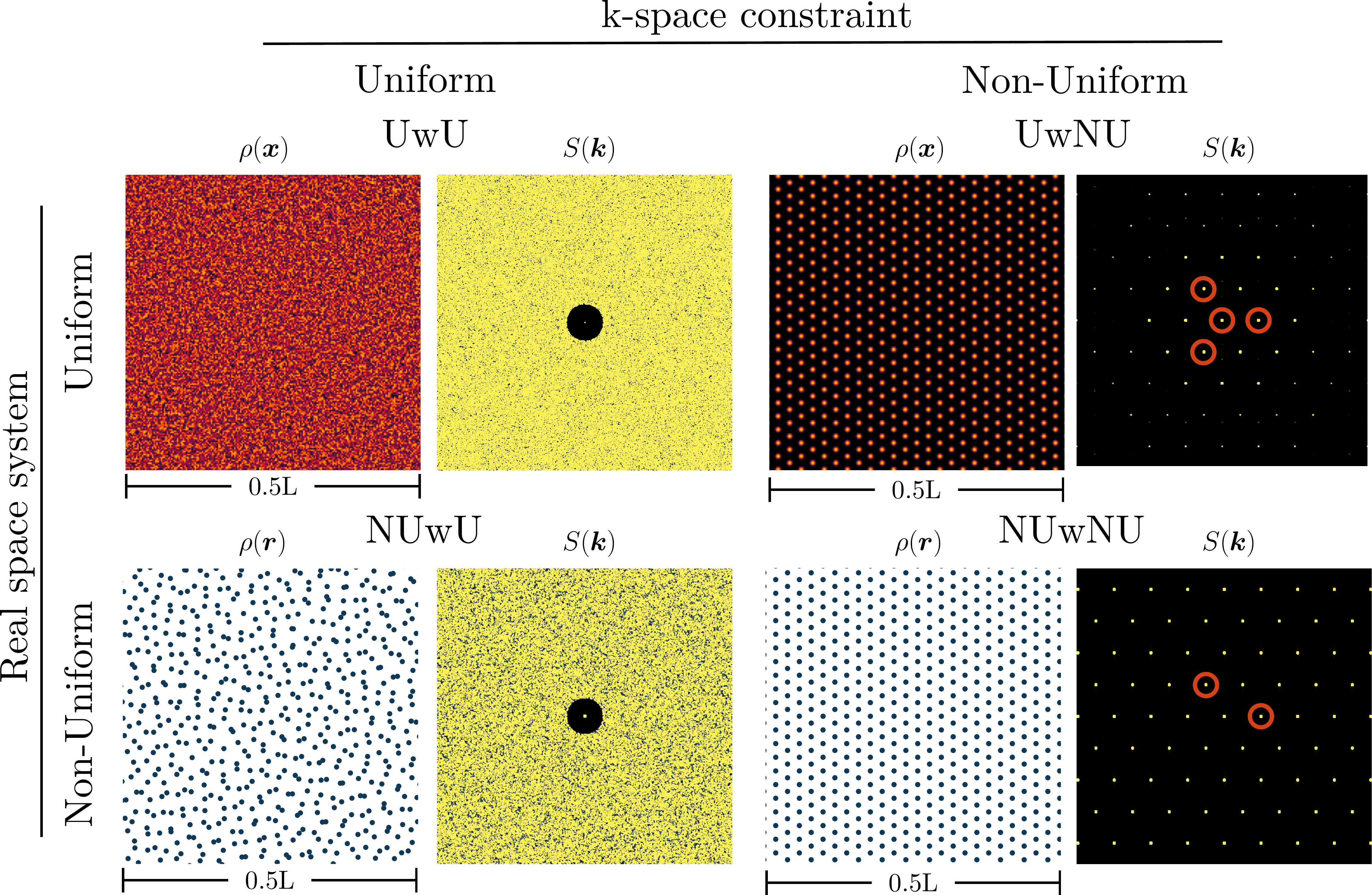}
    \caption{\textbf{Variations of FReSCo.}
    We optimize the weights carried by a Uniform (U) grid, or the positions of Non-Uniform (NU) sets of points in real space, while imposing constraints on a uniform grid or a non-uniform set of points in $k$-space.
    Thus, we obtain four variants of the FReSCo algorithm: UwU (Uniform real space with Uniform k-space constraints), UwNU (Uniform real space with Non-Uniform k-space constraints), NUwU (Non-Uniform real space with Uniform k-space constraints), and NUwNU (Non-Uniform real space with Non-Uniform k-space constraints).
    Small example systems are provided (grid size $403\times 403$ for uniform cases, $N\sim 2000$ points for non-uniform cases); see SI for further analysis.
    In UwU and UwNU, we control the range of values of pixels via an external potential, and the total mass via a constraint on $S(k = 0)$, see Methods.
    Problems in which positions and weights are optimized simultaneously will be considered in future work.}
    \label{fig:UwUandfriends}
\end{figure*}

\begin{figure*}
    \centering
    \includegraphics[width=0.96\textwidth]{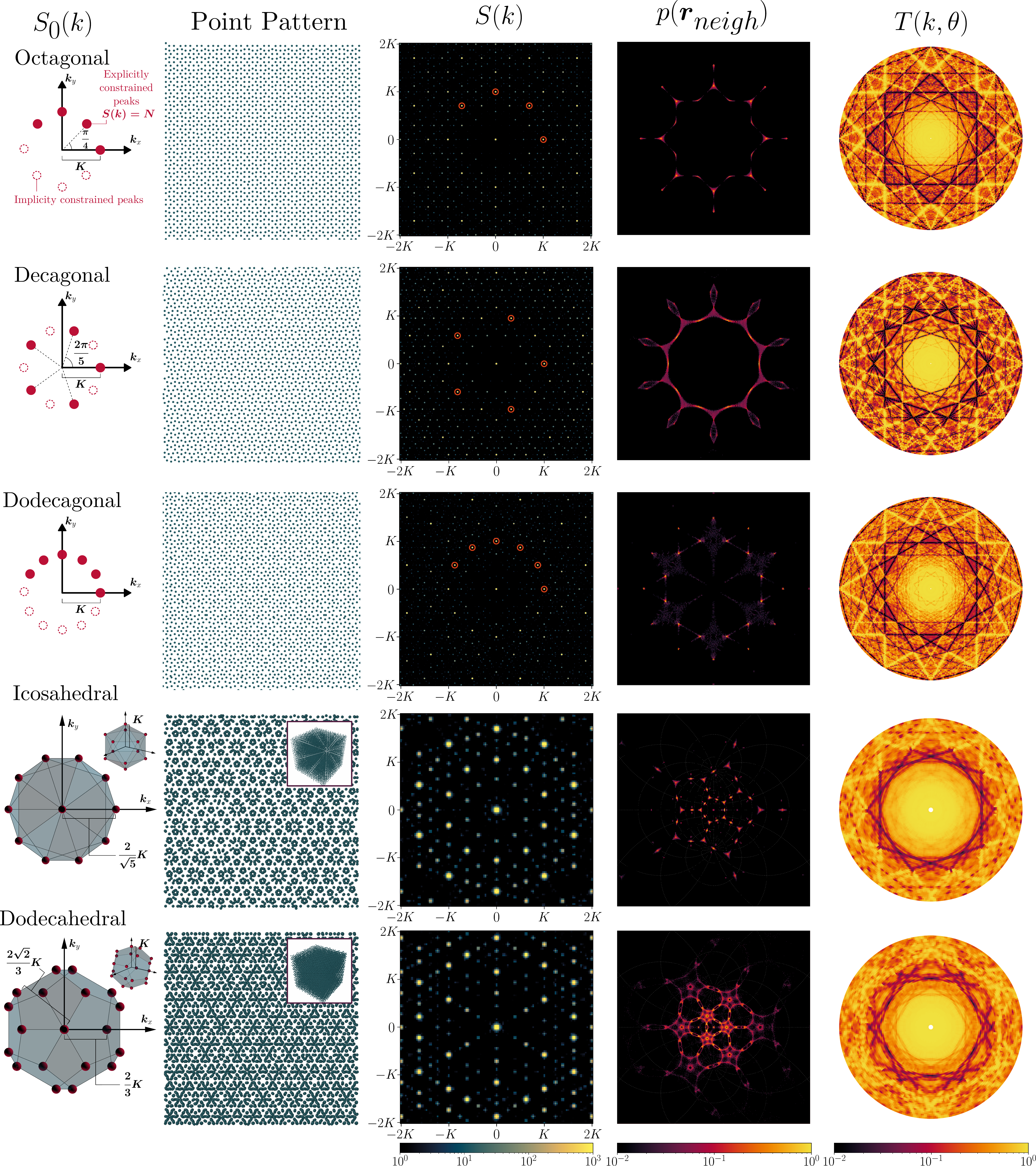}
    \caption{\textbf{Generation of special symmetries using NUwNU.}
    From left to right, sketches of the constrained peaks in $S(k)$; portion of the output point pattern; intensity map of the structure factor; density map of the distribution of Voronoi nearest neighbors; and Ewald transmission plot (all in log-intensity), for systems of $N\approx 10^4$ particles constrained with NUwNU to maximize peak height at specific locations in $k$-space.
    Each row shows one specific type of imposed $n$-fold symmetry: from top to bottom, we show $8$-fold, $10$-fold, and $12$-fold symmetry in $2d$, then  icosahedral and dodecahedral symmetry in $3d$.
    In the structure factor, we highlight constrained peaks in $2d$.
    In $3d$, we replace the $2d$ panels by close equivalents.
    The point patterns are projected onto the $xy$ plane, orthogonal to a long diagonal of the polyhedra (the full system is shown in inset), and the structure factor is accordingly in the $k_z = 0$ plane, where we highlight the lowest-order, implicitly constrained, peaks, and the Ewald construction is obtained by scanning only $xy$ (azimuthal) orientations.
    The full distribution of Voronoi nearest neighbors is replaced by the distribution of bond orientations to nearest neighbors on a stereographic projection of the sphere.
    }
    \label{fig:quasiquasicrystals}
\end{figure*}

\textit{FReSCo variants --}
So far, we only constrained non-uniform point positions with uniform $k$-space constraints (NUwU), but the same approach can be used with non-uniform $k$-space constraints (NUwNU).
Furthermore, one may instead optimize the weights carried by uniform (on-grid) points, either with uniform (UwU) or non-uniform (UwNU) constraints in $k$-space.
These variants, whose gradients are derived in Methods, are illustrated with simple examples in Fig.~\ref{fig:UwUandfriends}.
In UwU, while the real and Fourier spaces contain the same number of pixels (hence we are free to constrain the whole $S(\bm{k})$), we only constrain the modulus of a subset of $k$-vectors, so that one may generate first guesses in phase retrieval problems~\cite{Bertolotti2012}, textures with suitable properties~\cite{Lagae2010}, or random fields with suitable correlations~\cite{Ma2017a}.
In both UwNU and NUwNU, one may impose Fourier constraints at any continuous value, with free boundary conditions instead of periodic ones.
In particular, we can impose constraints of the form $S_0 = \sum_p N \delta(\bm{k} - \bm{k}_p)$ on sets of wave-vectors $\bm{k}_p$, to impose Bragg-like peaks at arbitrary continuous positions.
When choosing minimal sets $\bm{k}_p$ with specific discrete rotational symmetries that are not attainable with simple crystals~\cite{Schwarzenbach1996}, we observe the emergence of a full quasicrystalline structure, which we now investigate.

\noindent\textit{Quasicrystalline structures --} 
In Fig.~\ref{fig:quasiquasicrystals}, we explore more discrete symmetries.
Instead of imposing repulsive interactions, we repeat several cycles of FReSCo minimization, removing at each iteration points that overlap exactly with others, and replacing them with new points drawn uniformly in a box $\left[-L/2; L/2\right]^d$, maintaining total occupancy $N$.
For $N\approx 10000$ points in $2d$, imposing $8$-, $10$-, or $12$-fold symmetries leads to point patterns with quasicrystalline characteristics~\cite{Levine1984}, characterized by aperiodicity in real space (second column), and a peaked structure factor (third column).
We also show that, like quasicrystals~\cite{Engel2015}, our structures display strong local bond-orientational order, through $2d$ histograms of nearest neighbor vectors (third column), that feature very narrow peaks, a sign that the long-range orientational order from our constraints reach all the way down to short ranges (see Extended Data and SI for additional data).

Likewise, we show (last two rows) that we can impose icosahedral or dodecahedral order in $3d$.
Like previously reported $3d$ quasicrystals, they respectively display $10$- and $6$-fold aperiodic orders in projected views~\cite{Engel2015} (second column), with associated peaked structure factors~\cite{Levine1984} (third column), and peaked nearest-neighbor vector distributions on the sphere, here shown as stereographic projections (fourth column).
We show additional projections for these $3d$ structures in Extended Data.
For all structures, we also show (right-most column), that we observe the expected anisotropic transmission patterns of quasicrystalline structures~\cite{Man2013}.
It is interesting that imposing only $n$ peaks of intensity $O(N)$ with the right symmetry around the origin in Fourier space is sufficient to obtain quasicrystalline order.
Indeed, as discussed in SI, imposing a peak of intensity $N$ at specific wave-vectors in $S(\bm{k})$ implies that integer-coefficient linear combinations of these wave-vectors will also have $N$-high peaks, but quasicrystalline peaks are typically only $O(N)$~\cite{Levine1984}, so that the constraint on linear combinations is much weaker.
Our optimization approach thus enables the \textit{non-deterministic} generation of aperiodic structures with custom photonic properties and free boundary conditions.
This dramatically expands the design space for aperiodic structures, heretofore mostly limited to \textit{deterministic} examples, \textit{e.g.} to promote Anderson localization~\cite{Sgrignuoli2021}.

\noindent\textit{Conclusions --}
We have demonstrated a highly efficient generative algorithm, FReSCo, that precisely embeds $k$-space features into point patterns up to previously inaccessible scales, and that can be combined to short-range interactions like excluded volume.
This paves the way to exploring novel wave transport properties, like new structurally colored coatings~\cite{Vynck2022}.
More generally, one may impose more complicated real-space interactions -- \textit{e.g.} constraints onto the real-space pair correlation function like in Reverse Monte Carlo~\cite{McGreevy2001} (see Methods), or potentials that favor local orientational order (such as 3-body terms, \textit{e.g.} Stillinger-Weber-type potentials~\cite{Stillinger1985}).
These extensions would clarify the role of local orientational order in wave transport~\cite{Liew2011, Djeghdi2022} and facilitate fabrication, as tetrahedral order is often imposed \textit{a posteriori}~\cite{Florescu2009,Liew2011,Muller2014}.
Our algorithm may be generalised to include higher-order correlations, for instance $3\text{-}$ and $4\text{-}$body correlations, that are also computable in $\mathcal{O}(N \log N)$ using FFTs~\cite{Sunseri2023}.
Furthermore, using automatic differentiation \cite{Goodrich2021}, our loss can guide the design of interactions realizing the self-assembly of spectrally-shaped structures.

Finally, our approach provides a way to improve blue-noise sampling methods~\cite{Pilleboue2015}.
To illustrate this, we extracted $\sim800$ frames from a Lumière movie~\cite{LaCiotat} and imposed them as structure factor constraints $S_0(k)$.
By using the point pattern for frame $i$ as the initial condition for the subsequent minimization at frame $i+1$, we encode the film in trajectories of $N=300,000$ 2d points (see \href{https://www.youtube.com/watch?v=2oVJO197Wmc}{\textcolor{blue}{video link}}, Extended Data, and SI for further details).
Such dynamical optimization can be used in adaptive sampling for real-time computer graphics~\cite{Pharr2018}.

\begin{acknowledgments}
\textit{Acknowledgments --} The authors are indebted to Praharsh Suryadevara for contributing some of the computational tools adopted in this work, to Libin Lu and Alex H. Barnett for help with and fixes in the fiNUFFT library, and to Shenglong Wang for assistance with computational resources.
The authors would also like to thank Rémi Carminati, Paul Chaikin, Olivier Dauchot, Johnathon Gales, David Grier, Matthieu Labousse, Dov Levine, David Pine, and Paddy Royall, for insightful comments on this work; as well as Jaeuk Kim, Peter Morse, Paul Steinhardt, and Salvatore Torquato for detailed comments on an earlier version of this manuscript.
A.S, M.C., and S.M. acknowledge the Simons Center for Computational Physical Chemistry for financial support.
This work was supported in part through the NYU IT High Performance Computing resources, services, and staff expertise.
Parts of the results in this work make use of the colormaps in the CMasher package~\cite{VanderVelden2020}.
\end{acknowledgments}

\textit{Author Contributions --}
\textbf{Aaron Shih:} contributed to the design and implementation of the algorithm, the data generation and analysis, and writing of this manuscript.
\textbf{Mathias Casiulis:} contributed to the design and implementation of the algorithm, the data generation and analysis, and writing of this manuscript.
\textbf{Stefano Martiniani:} contributed to the conceptualization, design and implementation of the algorithm,  data analysis, writing of this manuscript, and funding acquisition.

\textit{Code availability --}
FReSCo is available under the MIT license at \url{https://github.com/martiniani-lab/FReSCo}.

\bibliography{PostDoc-StefanoMartiniani}

\begin{thebibliography}{50}%
\makeatletter
\providecommand \@ifxundefined [1]{%
 \@ifx{#1\undefined}
}%
\providecommand \@ifnum [1]{%
 \ifnum #1\expandafter \@firstoftwo
 \else \expandafter \@secondoftwo
 \fi
}%
\providecommand \@ifx [1]{%
 \ifx #1\expandafter \@firstoftwo
 \else \expandafter \@secondoftwo
 \fi
}%
\providecommand \natexlab [1]{#1}%
\providecommand \enquote  [1]{``#1''}%
\providecommand \bibnamefont  [1]{#1}%
\providecommand \bibfnamefont [1]{#1}%
\providecommand \citenamefont [1]{#1}%
\providecommand \href@noop [0]{\@secondoftwo}%
\providecommand \href [0]{\begingroup \@sanitize@url \@href}%
\providecommand \@href[1]{\@@startlink{#1}\@@href}%
\providecommand \@@href[1]{\endgroup#1\@@endlink}%
\providecommand \@sanitize@url [0]{\catcode `\\12\catcode `\$12\catcode
  `\&12\catcode `\#12\catcode `\^12\catcode `\_12\catcode `\%12\relax}%
\providecommand \@@startlink[1]{}%
\providecommand \@@endlink[0]{}%
\providecommand \url  [0]{\begingroup\@sanitize@url \@url }%
\providecommand \@url [1]{\endgroup\@href {#1}{\urlprefix }}%
\providecommand \urlprefix  [0]{URL }%
\providecommand \Eprint [0]{\href }%
\providecommand \doibase [0]{http://dx.doi.org/}%
\providecommand \selectlanguage [0]{\@gobble}%
\providecommand \bibinfo  [0]{\@secondoftwo}%
\providecommand \bibfield  [0]{\@secondoftwo}%
\providecommand \translation [1]{[#1]}%
\providecommand \BibitemOpen [0]{}%
\providecommand \bibitemStop [0]{}%
\providecommand \bibitemNoStop [0]{.\EOS\space}%
\providecommand \EOS [0]{\spacefactor3000\relax}%
\providecommand \BibitemShut  [1]{\csname bibitem#1\endcsname}%
\let\auto@bib@innerbib\@empty
\bibitem [{\citenamefont {Kittel}(1953)}]{Kittel}%
  \BibitemOpen
  \bibfield  {author} {\bibinfo {author} {\bibfnamefont {C.}~\bibnamefont
  {Kittel}},\ }\href@noop {} {\emph {\bibinfo {title} {{Introduction to
  Solid-state Physics}}}}\ (\bibinfo  {publisher} {John Wiley {\&} Sons},\
  \bibinfo {address} {New York},\ \bibinfo {year} {1953})\BibitemShut {NoStop}%
\bibitem [{\citenamefont {Carminati}\ and\ \citenamefont
  {Schotland}(2021)}]{Carminati2021}%
  \BibitemOpen
  \bibfield  {author} {\bibinfo {author} {\bibfnamefont {R.}~\bibnamefont
  {Carminati}}\ and\ \bibinfo {author} {\bibfnamefont {J.~C.}\ \bibnamefont
  {Schotland}},\ }\href@noop {} {\emph {\bibinfo {title} {Principles of
  Scattering and Transport of Light}}}\ (\bibinfo  {publisher} {Cambridge
  University Press},\ \bibinfo {year} {2021})\BibitemShut {NoStop}%
\bibitem [{\citenamefont {Vynck}\ \emph {et~al.}(2023)\citenamefont {Vynck},
  \citenamefont {Pierrat}, \citenamefont {Carminati}, \citenamefont
  {Froufe-P{\'{e}}rez}, \citenamefont {Scheffold}, \citenamefont {Sapienza},
  \citenamefont {Vignolini},\ and\ \citenamefont {S{\'{a}}enz}}]{Vynck2023}%
  \BibitemOpen
  \bibfield  {author} {\bibinfo {author} {\bibfnamefont {K.}~\bibnamefont
  {Vynck}}, \bibinfo {author} {\bibfnamefont {R.}~\bibnamefont {Pierrat}},
  \bibinfo {author} {\bibfnamefont {R.}~\bibnamefont {Carminati}}, \bibinfo
  {author} {\bibfnamefont {L.~S.}\ \bibnamefont {Froufe-P{\'{e}}rez}}, \bibinfo
  {author} {\bibfnamefont {F.}~\bibnamefont {Scheffold}}, \bibinfo {author}
  {\bibfnamefont {R.}~\bibnamefont {Sapienza}}, \bibinfo {author}
  {\bibfnamefont {S.}~\bibnamefont {Vignolini}}, \ and\ \bibinfo {author}
  {\bibfnamefont {J.~J.}\ \bibnamefont {S{\'{a}}enz}},\ }\href@noop {}
  {\bibfield  {journal} {\bibinfo  {journal} {Reviews of Modern Physics}\
  }\textbf {\bibinfo {volume} {95}},\ \bibinfo {pages} {045003} (\bibinfo
  {year} {2023})}\BibitemShut {NoStop}%
\bibitem [{\citenamefont {Djeghdi}\ \emph {et~al.}(2022)\citenamefont
  {Djeghdi}, \citenamefont {Steiner},\ and\ \citenamefont
  {Wilts}}]{Djeghdi2022}%
  \BibitemOpen
  \bibfield  {author} {\bibinfo {author} {\bibfnamefont {K.}~\bibnamefont
  {Djeghdi}}, \bibinfo {author} {\bibfnamefont {U.}~\bibnamefont {Steiner}}, \
  and\ \bibinfo {author} {\bibfnamefont {B.~D.}\ \bibnamefont {Wilts}},\
  }\href@noop {} {\bibfield  {journal} {\bibinfo  {journal} {Advanced Science}\
  }\textbf {\bibinfo {volume} {9}},\ \bibinfo {pages} {2202145} (\bibinfo
  {year} {2022})}\BibitemShut {NoStop}%
\bibitem [{\citenamefont {Vynck}\ \emph {et~al.}(2022)\citenamefont {Vynck},
  \citenamefont {Pacanowski}, \citenamefont {Agreda}, \citenamefont {Dufay},
  \citenamefont {Granier},\ and\ \citenamefont {Lalanne}}]{Vynck2022}%
  \BibitemOpen
  \bibfield  {author} {\bibinfo {author} {\bibfnamefont {K.}~\bibnamefont
  {Vynck}}, \bibinfo {author} {\bibfnamefont {R.}~\bibnamefont {Pacanowski}},
  \bibinfo {author} {\bibfnamefont {A.}~\bibnamefont {Agreda}}, \bibinfo
  {author} {\bibfnamefont {A.}~\bibnamefont {Dufay}}, \bibinfo {author}
  {\bibfnamefont {X.}~\bibnamefont {Granier}}, \ and\ \bibinfo {author}
  {\bibfnamefont {P.}~\bibnamefont {Lalanne}},\ }\href@noop {} {\bibfield
  {journal} {\bibinfo  {journal} {Nature Materials}\ }\textbf {\bibinfo
  {volume} {21}},\ \bibinfo {pages} {1035} (\bibinfo {year}
  {2022})}\BibitemShut {NoStop}%
\bibitem [{\citenamefont {Florescu}\ \emph {et~al.}(2009)\citenamefont
  {Florescu}, \citenamefont {Torquato},\ and\ \citenamefont
  {Steinhardt}}]{Florescu2009}%
  \BibitemOpen
  \bibfield  {author} {\bibinfo {author} {\bibfnamefont {M.}~\bibnamefont
  {Florescu}}, \bibinfo {author} {\bibfnamefont {S.}~\bibnamefont {Torquato}},
  \ and\ \bibinfo {author} {\bibfnamefont {P.~J.}\ \bibnamefont {Steinhardt}},\
  }\href@noop {} {\bibfield  {journal} {\bibinfo  {journal} {Proceedings of the
  National Academy of Sciences of the United States of America}\ }\textbf
  {\bibinfo {volume} {106}},\ \bibinfo {pages} {20658} (\bibinfo {year}
  {2009})}\BibitemShut {NoStop}%
\bibitem [{\citenamefont {Man}\ \emph {et~al.}(2013)\citenamefont {Man},
  \citenamefont {Florescu}, \citenamefont {Matsuyama}, \citenamefont {Yadak},
  \citenamefont {Nahal}, \citenamefont {Hashemizad}, \citenamefont
  {Williamson}, \citenamefont {Steinhardt}, \citenamefont {Torquato},\ and\
  \citenamefont {Chaikin}}]{Man2013}%
  \BibitemOpen
  \bibfield  {author} {\bibinfo {author} {\bibfnamefont {W.}~\bibnamefont
  {Man}}, \bibinfo {author} {\bibfnamefont {M.}~\bibnamefont {Florescu}},
  \bibinfo {author} {\bibfnamefont {K.}~\bibnamefont {Matsuyama}}, \bibinfo
  {author} {\bibfnamefont {P.}~\bibnamefont {Yadak}}, \bibinfo {author}
  {\bibfnamefont {G.}~\bibnamefont {Nahal}}, \bibinfo {author} {\bibfnamefont
  {S.}~\bibnamefont {Hashemizad}}, \bibinfo {author} {\bibfnamefont
  {E.}~\bibnamefont {Williamson}}, \bibinfo {author} {\bibfnamefont
  {P.}~\bibnamefont {Steinhardt}}, \bibinfo {author} {\bibfnamefont
  {S.}~\bibnamefont {Torquato}}, \ and\ \bibinfo {author} {\bibfnamefont
  {P.}~\bibnamefont {Chaikin}},\ }\href@noop {} {\bibfield  {journal} {\bibinfo
   {journal} {Optics Express}\ }\textbf {\bibinfo {volume} {21}},\ \bibinfo
  {pages} {19972} (\bibinfo {year} {2013})}\BibitemShut {NoStop}%
\bibitem [{\citenamefont {Froufe-P{\'{e}}rez}\ \emph
  {et~al.}(2017)\citenamefont {Froufe-P{\'{e}}rez}, \citenamefont {Engel},
  \citenamefont {S{\'{a}}enz},\ and\ \citenamefont
  {Scheffold}}]{Froufe-Perez2017}%
  \BibitemOpen
  \bibfield  {author} {\bibinfo {author} {\bibfnamefont {L.~S.}\ \bibnamefont
  {Froufe-P{\'{e}}rez}}, \bibinfo {author} {\bibfnamefont {M.}~\bibnamefont
  {Engel}}, \bibinfo {author} {\bibfnamefont {J.~J.}\ \bibnamefont
  {S{\'{a}}enz}}, \ and\ \bibinfo {author} {\bibfnamefont {F.}~\bibnamefont
  {Scheffold}},\ }\href@noop {} {\bibfield  {journal} {\bibinfo  {journal}
  {Proceedings of the National Academy of Sciences of the United States of
  America}\ }\textbf {\bibinfo {volume} {114}},\ \bibinfo {pages} {9570}
  (\bibinfo {year} {2017})}\BibitemShut {NoStop}%
\bibitem [{\citenamefont {Monsarrat}\ \emph {et~al.}(2022)\citenamefont
  {Monsarrat}, \citenamefont {Pierrat}, \citenamefont {Tourin},\ and\
  \citenamefont {Goetschy}}]{Monsarrat2022}%
  \BibitemOpen
  \bibfield  {author} {\bibinfo {author} {\bibfnamefont {R.}~\bibnamefont
  {Monsarrat}}, \bibinfo {author} {\bibfnamefont {R.}~\bibnamefont {Pierrat}},
  \bibinfo {author} {\bibfnamefont {A.}~\bibnamefont {Tourin}}, \ and\ \bibinfo
  {author} {\bibfnamefont {A.}~\bibnamefont {Goetschy}},\ }\href@noop {}
  {\bibfield  {journal} {\bibinfo  {journal} {Physical Review Research}\
  }\textbf {\bibinfo {volume} {4}},\ \bibinfo {pages} {033246} (\bibinfo {year}
  {2022})}\BibitemShut {NoStop}%
\bibitem [{\citenamefont {Yan}\ \emph {et~al.}(2015)\citenamefont {Yan},
  \citenamefont {Guo}, \citenamefont {Wang}, \citenamefont {Zhang},\ and\
  \citenamefont {Wonka}}]{Yan2015}%
  \BibitemOpen
  \bibfield  {author} {\bibinfo {author} {\bibfnamefont {D.~M.}\ \bibnamefont
  {Yan}}, \bibinfo {author} {\bibfnamefont {J.~W.}\ \bibnamefont {Guo}},
  \bibinfo {author} {\bibfnamefont {B.}~\bibnamefont {Wang}}, \bibinfo {author}
  {\bibfnamefont {X.~P.}\ \bibnamefont {Zhang}}, \ and\ \bibinfo {author}
  {\bibfnamefont {P.}~\bibnamefont {Wonka}},\ }\href@noop {} {\bibfield
  {journal} {\bibinfo  {journal} {Journal of Computer Science and Technology}\
  }\textbf {\bibinfo {volume} {30}},\ \bibinfo {pages} {439} (\bibinfo {year}
  {2015})}\BibitemShut {NoStop}%
\bibitem [{\citenamefont {Pharr}\ \emph {et~al.}(2018)\citenamefont {Pharr},
  \citenamefont {Jakob},\ and\ \citenamefont {Humphreys}}]{Pharr2018}%
  \BibitemOpen
  \bibfield  {author} {\bibinfo {author} {\bibfnamefont {M.}~\bibnamefont
  {Pharr}}, \bibinfo {author} {\bibfnamefont {W.}~\bibnamefont {Jakob}}, \ and\
  \bibinfo {author} {\bibfnamefont {G.}~\bibnamefont {Humphreys}},\ }\href@noop
  {} {\emph {\bibinfo {title} {{Physically based rendering: From Theory to
  Implementation}}}},\ \bibinfo {edition} {3rd}\ ed.\ (\bibinfo  {publisher}
  {Morgan Kaufmann},\ \bibinfo {year} {2018})\BibitemShut {NoStop}%
\bibitem [{\citenamefont {Pilleboue}\ \emph {et~al.}(2015)\citenamefont
  {Pilleboue}, \citenamefont {Singh}, \citenamefont {Coeurjolly}, \citenamefont
  {Kazhdan},\ and\ \citenamefont {Ostromoukhov}}]{Pilleboue2015}%
  \BibitemOpen
  \bibfield  {author} {\bibinfo {author} {\bibfnamefont {A.}~\bibnamefont
  {Pilleboue}}, \bibinfo {author} {\bibfnamefont {G.}~\bibnamefont {Singh}},
  \bibinfo {author} {\bibfnamefont {D.}~\bibnamefont {Coeurjolly}}, \bibinfo
  {author} {\bibfnamefont {M.}~\bibnamefont {Kazhdan}}, \ and\ \bibinfo
  {author} {\bibfnamefont {V.}~\bibnamefont {Ostromoukhov}},\ }\href@noop {}
  {\bibfield  {journal} {\bibinfo  {journal} {ACM Transactions on Graphics}\
  }\textbf {\bibinfo {volume} {34}},\ \bibinfo {pages} {124} (\bibinfo {year}
  {2015})}\BibitemShut {NoStop}%
\bibitem [{\citenamefont {van Gogh}(1889)}]{StarryNight}%
  \BibitemOpen
  \bibfield  {author} {\bibinfo {author} {\bibfnamefont {V.}~\bibnamefont {van
  Gogh}},\ }\href@noop {} {\enquote {\bibinfo {title} {{Starry Night}},}\ }
  (\bibinfo {year} {1889})\BibitemShut {NoStop}%
\bibitem [{\citenamefont {Torquato}(2018)}]{Torquato2018}%
  \BibitemOpen
  \bibfield  {author} {\bibinfo {author} {\bibfnamefont {S.}~\bibnamefont
  {Torquato}},\ }\href@noop {} {\bibfield  {journal} {\bibinfo  {journal}
  {Physics Reports}\ }\textbf {\bibinfo {volume} {745}},\ \bibinfo {pages} {1}
  (\bibinfo {year} {2018})}\BibitemShut {NoStop}%
\bibitem [{\citenamefont {Klatt}\ \emph {et~al.}(2022)\citenamefont {Klatt},
  \citenamefont {Steinhardt},\ and\ \citenamefont {Torquato}}]{Klatt2022}%
  \BibitemOpen
  \bibfield  {author} {\bibinfo {author} {\bibfnamefont {M.~A.}\ \bibnamefont
  {Klatt}}, \bibinfo {author} {\bibfnamefont {P.~J.}\ \bibnamefont
  {Steinhardt}}, \ and\ \bibinfo {author} {\bibfnamefont {S.}~\bibnamefont
  {Torquato}},\ }\href@noop {} {\bibfield  {journal} {\bibinfo  {journal}
  {Proceedings of the National Academy of Sciences of the United States of
  America}\ }\textbf {\bibinfo {volume} {119}},\ \bibinfo {pages} {e2213633119}
  (\bibinfo {year} {2022})}\BibitemShut {NoStop}%
\bibitem [{\citenamefont {Haberko}\ \emph {et~al.}(2020)\citenamefont
  {Haberko}, \citenamefont {Froufe-P{\'{e}}rez},\ and\ \citenamefont
  {Scheffold}}]{Haberko2020}%
  \BibitemOpen
  \bibfield  {author} {\bibinfo {author} {\bibfnamefont {J.}~\bibnamefont
  {Haberko}}, \bibinfo {author} {\bibfnamefont {L.~S.}\ \bibnamefont
  {Froufe-P{\'{e}}rez}}, \ and\ \bibinfo {author} {\bibfnamefont
  {F.}~\bibnamefont {Scheffold}},\ }\href@noop {} {\bibfield  {journal}
  {\bibinfo  {journal} {Nature Communications}\ }\textbf {\bibinfo {volume}
  {11}},\ \bibinfo {pages} {4867} (\bibinfo {year} {2020})}\BibitemShut
  {NoStop}%
\bibitem [{\citenamefont {Uche}\ \emph {et~al.}(2004)\citenamefont {Uche},
  \citenamefont {Stillinger},\ and\ \citenamefont {Torquato}}]{Uche2004}%
  \BibitemOpen
  \bibfield  {author} {\bibinfo {author} {\bibfnamefont {O.~U.}\ \bibnamefont
  {Uche}}, \bibinfo {author} {\bibfnamefont {F.~H.}\ \bibnamefont
  {Stillinger}}, \ and\ \bibinfo {author} {\bibfnamefont {S.}~\bibnamefont
  {Torquato}},\ }\href@noop {} {\bibfield  {journal} {\bibinfo  {journal}
  {Physical Review E - Statistical Physics, Plasmas, Fluids, and Related
  Interdisciplinary Topics}\ }\textbf {\bibinfo {volume} {70}},\ \bibinfo
  {pages} {046122} (\bibinfo {year} {2004})}\BibitemShut {NoStop}%
\bibitem [{\citenamefont {Uche}\ \emph {et~al.}(2006)\citenamefont {Uche},
  \citenamefont {Torquato},\ and\ \citenamefont {Stillinger}}]{Uche2006}%
  \BibitemOpen
  \bibfield  {author} {\bibinfo {author} {\bibfnamefont {O.~U.}\ \bibnamefont
  {Uche}}, \bibinfo {author} {\bibfnamefont {S.}~\bibnamefont {Torquato}}, \
  and\ \bibinfo {author} {\bibfnamefont {F.~H.}\ \bibnamefont {Stillinger}},\
  }\href@noop {} {\bibfield  {journal} {\bibinfo  {journal} {Physical Review E
  - Statistical, Nonlinear, and Soft Matter Physics}\ }\textbf {\bibinfo
  {volume} {74}},\ \bibinfo {pages} {031104} (\bibinfo {year}
  {2006})}\BibitemShut {NoStop}%
\bibitem [{\citenamefont {Morse}\ \emph {et~al.}(2023)\citenamefont {Morse},
  \citenamefont {Kim}, \citenamefont {Steinhardt},\ and\ \citenamefont
  {Torquato}}]{Morse2023}%
  \BibitemOpen
  \bibfield  {author} {\bibinfo {author} {\bibfnamefont {P.~K.}\ \bibnamefont
  {Morse}}, \bibinfo {author} {\bibfnamefont {J.}~\bibnamefont {Kim}}, \bibinfo
  {author} {\bibfnamefont {P.~J.}\ \bibnamefont {Steinhardt}}, \ and\ \bibinfo
  {author} {\bibfnamefont {S.}~\bibnamefont {Torquato}},\ }\href@noop {}
  {\bibfield  {journal} {\bibinfo  {journal} {Arxiv Preprint}\ ,\ \bibinfo
  {pages} {2304.09139}} (\bibinfo {year} {2023})}\BibitemShut {NoStop}%
\bibitem [{\citenamefont {Froufe-P{\'{e}}rez}\ \emph
  {et~al.}(2016)\citenamefont {Froufe-P{\'{e}}rez}, \citenamefont {Engel},
  \citenamefont {Damasceno}, \citenamefont {Muller}, \citenamefont {Haberko},
  \citenamefont {Glotzer},\ and\ \citenamefont {Scheffold}}]{Froufe-Perez2016}%
  \BibitemOpen
  \bibfield  {author} {\bibinfo {author} {\bibfnamefont {L.~S.}\ \bibnamefont
  {Froufe-P{\'{e}}rez}}, \bibinfo {author} {\bibfnamefont {M.}~\bibnamefont
  {Engel}}, \bibinfo {author} {\bibfnamefont {P.~F.}\ \bibnamefont
  {Damasceno}}, \bibinfo {author} {\bibfnamefont {N.}~\bibnamefont {Muller}},
  \bibinfo {author} {\bibfnamefont {J.}~\bibnamefont {Haberko}}, \bibinfo
  {author} {\bibfnamefont {S.~C.}\ \bibnamefont {Glotzer}}, \ and\ \bibinfo
  {author} {\bibfnamefont {F.}~\bibnamefont {Scheffold}},\ }\href@noop {}
  {\bibfield  {journal} {\bibinfo  {journal} {Physical Review Letters}\
  }\textbf {\bibinfo {volume} {117}},\ \bibinfo {pages} {1} (\bibinfo {year}
  {2016})}\BibitemShut {NoStop}%
\bibitem [{\citenamefont {Leseur}\ \emph {et~al.}(2016)\citenamefont {Leseur},
  \citenamefont {Pierrat},\ and\ \citenamefont {Carminati}}]{Leseur2016}%
  \BibitemOpen
  \bibfield  {author} {\bibinfo {author} {\bibfnamefont {O.}~\bibnamefont
  {Leseur}}, \bibinfo {author} {\bibfnamefont {R.}~\bibnamefont {Pierrat}}, \
  and\ \bibinfo {author} {\bibfnamefont {R.}~\bibnamefont {Carminati}},\
  }\href@noop {} {\bibfield  {journal} {\bibinfo  {journal} {Optica}\ }\textbf
  {\bibinfo {volume} {3}},\ \bibinfo {pages} {763} (\bibinfo {year}
  {2016})}\BibitemShut {NoStop}%
\bibitem [{\citenamefont {Gorsky}\ \emph {et~al.}(2019)\citenamefont {Gorsky},
  \citenamefont {Britton}, \citenamefont {Chen}, \citenamefont {Montaner},
  \citenamefont {Lenef}, \citenamefont {Raukas},\ and\ \citenamefont {{Dal
  Negro}}}]{Gorsky2019}%
  \BibitemOpen
  \bibfield  {author} {\bibinfo {author} {\bibfnamefont {S.}~\bibnamefont
  {Gorsky}}, \bibinfo {author} {\bibfnamefont {W.~A.}\ \bibnamefont {Britton}},
  \bibinfo {author} {\bibfnamefont {Y.}~\bibnamefont {Chen}}, \bibinfo {author}
  {\bibfnamefont {J.}~\bibnamefont {Montaner}}, \bibinfo {author}
  {\bibfnamefont {A.}~\bibnamefont {Lenef}}, \bibinfo {author} {\bibfnamefont
  {M.}~\bibnamefont {Raukas}}, \ and\ \bibinfo {author} {\bibfnamefont
  {L.}~\bibnamefont {{Dal Negro}}},\ }\href@noop {} {\bibfield  {journal}
  {\bibinfo  {journal} {APL Photonics}\ }\textbf {\bibinfo {volume} {4}},\
  \bibinfo {pages} {110801} (\bibinfo {year} {2019})}\BibitemShut {NoStop}%
\bibitem [{\citenamefont {Torquato}\ and\ \citenamefont
  {Kim}(2021)}]{Torquato2021}%
  \BibitemOpen
  \bibfield  {author} {\bibinfo {author} {\bibfnamefont {S.}~\bibnamefont
  {Torquato}}\ and\ \bibinfo {author} {\bibfnamefont {J.}~\bibnamefont {Kim}},\
  }\href@noop {} {\bibfield  {journal} {\bibinfo  {journal} {Physical Review
  X}\ }\textbf {\bibinfo {volume} {11}},\ \bibinfo {pages} {21002} (\bibinfo
  {year} {2021})}\BibitemShut {NoStop}%
\bibitem [{\citenamefont {Jiao}\ and\ \citenamefont
  {Torquato}(2011)}]{Jiao2011a}%
  \BibitemOpen
  \bibfield  {author} {\bibinfo {author} {\bibfnamefont {Y.}~\bibnamefont
  {Jiao}}\ and\ \bibinfo {author} {\bibfnamefont {S.}~\bibnamefont
  {Torquato}},\ }\href@noop {} {\bibfield  {journal} {\bibinfo  {journal}
  {Physical Review E - Statistical, Nonlinear, and Soft Matter Physics}\
  }\textbf {\bibinfo {volume} {84}},\ \bibinfo {pages} {041309} (\bibinfo
  {year} {2011})}\BibitemShut {NoStop}%
\bibitem [{\citenamefont {Cort{\'{e}}}\ \emph {et~al.}(2008)\citenamefont
  {Cort{\'{e}}}, \citenamefont {Chaikin}, \citenamefont {Gollub},\ and\
  \citenamefont {Pine}}]{Corte2008}%
  \BibitemOpen
  \bibfield  {author} {\bibinfo {author} {\bibfnamefont {L.}~\bibnamefont
  {Cort{\'{e}}}}, \bibinfo {author} {\bibfnamefont {P.~M.}\ \bibnamefont
  {Chaikin}}, \bibinfo {author} {\bibfnamefont {J.~P.}\ \bibnamefont {Gollub}},
  \ and\ \bibinfo {author} {\bibfnamefont {D.~J.}\ \bibnamefont {Pine}},\
  }\href@noop {} {\bibfield  {journal} {\bibinfo  {journal} {Nature Physics}\
  }\textbf {\bibinfo {volume} {4}},\ \bibinfo {pages} {420} (\bibinfo {year}
  {2008})}\BibitemShut {NoStop}%
\bibitem [{\citenamefont {Hexner}\ and\ \citenamefont
  {Levine}(2015)}]{Hexner2015}%
  \BibitemOpen
  \bibfield  {author} {\bibinfo {author} {\bibfnamefont {D.}~\bibnamefont
  {Hexner}}\ and\ \bibinfo {author} {\bibfnamefont {D.}~\bibnamefont
  {Levine}},\ }\href@noop {} {\bibfield  {journal} {\bibinfo  {journal}
  {Physical Review Letters}\ }\textbf {\bibinfo {volume} {114}},\ \bibinfo
  {pages} {110602} (\bibinfo {year} {2015})}\BibitemShut {NoStop}%
\bibitem [{\citenamefont {Levine}\ and\ \citenamefont
  {Steinhardt}(1984)}]{Levine1984}%
  \BibitemOpen
  \bibfield  {author} {\bibinfo {author} {\bibfnamefont {D.}~\bibnamefont
  {Levine}}\ and\ \bibinfo {author} {\bibfnamefont {P.~J.}\ \bibnamefont
  {Steinhardt}},\ }\href@noop {} {\bibfield  {journal} {\bibinfo  {journal}
  {Physical Review Letters}\ }\textbf {\bibinfo {volume} {53}},\ \bibinfo
  {pages} {2477} (\bibinfo {year} {1984})}\BibitemShut {NoStop}%
\bibitem [{\citenamefont {Hansen}\ and\ \citenamefont
  {McDonald}(2006)}]{Hansen2006}%
  \BibitemOpen
  \bibfield  {author} {\bibinfo {author} {\bibfnamefont {J.-P.}\ \bibnamefont
  {Hansen}}\ and\ \bibinfo {author} {\bibfnamefont {I.~R.}\ \bibnamefont
  {McDonald}},\ }\href@noop {} {\emph {\bibinfo {title} {{Theory of simple
  liquids}}}}\ (\bibinfo  {publisher} {Elsevier Academic Press},\ \bibinfo
  {year} {2006})\BibitemShut {NoStop}%
\bibitem [{\citenamefont {Schwarzenbach}(1996)}]{Schwarzenbach1996}%
  \BibitemOpen
  \bibfield  {author} {\bibinfo {author} {\bibfnamefont {D.}~\bibnamefont
  {Schwarzenbach}},\ }\href@noop {} {\emph {\bibinfo {title}
  {{Crystallography}}}}\ (\bibinfo  {publisher} {John Wiley $\backslash${\&}
  Sons},\ \bibinfo {year} {1996})\BibitemShut {NoStop}%
\bibitem [{\citenamefont {Liu}\ and\ \citenamefont {Nocedal}(1989)}]{Liu1989}%
  \BibitemOpen
  \bibfield  {author} {\bibinfo {author} {\bibfnamefont {D.~C.}\ \bibnamefont
  {Liu}}\ and\ \bibinfo {author} {\bibfnamefont {J.}~\bibnamefont {Nocedal}},\
  }\href@noop {} {\bibfield  {journal} {\bibinfo  {journal} {Mathematical
  Programming}\ }\textbf {\bibinfo {volume} {45}},\ \bibinfo {pages} {503}
  (\bibinfo {year} {1989})}\BibitemShut {NoStop}%
\bibitem [{\citenamefont {Nocedal}\ and\ \citenamefont
  {Wright}(1999)}]{Nocedal1999}%
  \BibitemOpen
  \bibfield  {author} {\bibinfo {author} {\bibfnamefont {J.}~\bibnamefont
  {Nocedal}}\ and\ \bibinfo {author} {\bibfnamefont {S.}~\bibnamefont
  {Wright}},\ }\href@noop {} {\emph {\bibinfo {title} {{Numerical
  Optimization}}}}\ (\bibinfo  {publisher} {Springer},\ \bibinfo {year}
  {1999})\BibitemShut {NoStop}%
\bibitem [{\citenamefont {Liew}\ \emph {et~al.}(2011)\citenamefont {Liew},
  \citenamefont {Yang}, \citenamefont {Noh}, \citenamefont {Schreck},
  \citenamefont {Dufresne}, \citenamefont {O'Hern},\ and\ \citenamefont
  {Cao}}]{Liew2011}%
  \BibitemOpen
  \bibfield  {author} {\bibinfo {author} {\bibfnamefont {S.~F.}\ \bibnamefont
  {Liew}}, \bibinfo {author} {\bibfnamefont {J.~K.}\ \bibnamefont {Yang}},
  \bibinfo {author} {\bibfnamefont {H.}~\bibnamefont {Noh}}, \bibinfo {author}
  {\bibfnamefont {C.~F.}\ \bibnamefont {Schreck}}, \bibinfo {author}
  {\bibfnamefont {E.~R.}\ \bibnamefont {Dufresne}}, \bibinfo {author}
  {\bibfnamefont {C.~S.}\ \bibnamefont {O'Hern}}, \ and\ \bibinfo {author}
  {\bibfnamefont {H.}~\bibnamefont {Cao}},\ }\href@noop {} {\bibfield
  {journal} {\bibinfo  {journal} {Physical Review A - Atomic, Molecular, and
  Optical Physics}\ }\textbf {\bibinfo {volume} {84}},\ \bibinfo {pages}
  {063818} (\bibinfo {year} {2011})}\BibitemShut {NoStop}%
\bibitem [{\citenamefont {Muller}\ \emph {et~al.}(2014)\citenamefont {Muller},
  \citenamefont {Haberko}, \citenamefont {Marichy},\ and\ \citenamefont
  {Scheffold}}]{Muller2014}%
  \BibitemOpen
  \bibfield  {author} {\bibinfo {author} {\bibfnamefont {N.}~\bibnamefont
  {Muller}}, \bibinfo {author} {\bibfnamefont {J.}~\bibnamefont {Haberko}},
  \bibinfo {author} {\bibfnamefont {C.}~\bibnamefont {Marichy}}, \ and\
  \bibinfo {author} {\bibfnamefont {F.}~\bibnamefont {Scheffold}},\ }\href@noop
  {} {\bibfield  {journal} {\bibinfo  {journal} {Advanced Optical Materials}\
  }\textbf {\bibinfo {volume} {2}},\ \bibinfo {pages} {115} (\bibinfo {year}
  {2014})}\BibitemShut {NoStop}%
\bibitem [{\citenamefont {Kim}\ and\ \citenamefont {Torquato}(2020)}]{Kim2020}%
  \BibitemOpen
  \bibfield  {author} {\bibinfo {author} {\bibfnamefont {J.}~\bibnamefont
  {Kim}}\ and\ \bibinfo {author} {\bibfnamefont {S.}~\bibnamefont {Torquato}},\
  }\href@noop {} {\bibfield  {journal} {\bibinfo  {journal} {Proceedings of the
  National Academy of Sciences of the United States of America}\ }\textbf
  {\bibinfo {volume} {117}},\ \bibinfo {pages} {8764} (\bibinfo {year}
  {2020})}\BibitemShut {NoStop}%
\bibitem [{\citenamefont {Vrij}(1978)}]{Vrij1978}%
  \BibitemOpen
  \bibfield  {author} {\bibinfo {author} {\bibfnamefont {A.}~\bibnamefont
  {Vrij}},\ }\href@noop {} {\bibfield  {journal} {\bibinfo  {journal} {The
  Journal of Chemical Physics}\ }\textbf {\bibinfo {volume} {69}},\ \bibinfo
  {pages} {1742} (\bibinfo {year} {1978})}\BibitemShut {NoStop}%
\bibitem [{\citenamefont {Rosenfeld}(1990)}]{Rosenfeld1990}%
  \BibitemOpen
  \bibfield  {author} {\bibinfo {author} {\bibfnamefont {Y.}~\bibnamefont
  {Rosenfeld}},\ }\href@noop {} {\bibfield  {journal} {\bibinfo  {journal}
  {Physical Review A}\ }\textbf {\bibinfo {volume} {42}},\ \bibinfo {pages}
  {5978} (\bibinfo {year} {1990})}\BibitemShut {NoStop}%
\bibitem [{\citenamefont {Bernard}\ \emph {et~al.}(2009)\citenamefont
  {Bernard}, \citenamefont {Krauth},\ and\ \citenamefont
  {Wilson}}]{Bernard2009}%
  \BibitemOpen
  \bibfield  {author} {\bibinfo {author} {\bibfnamefont {E.~P.}\ \bibnamefont
  {Bernard}}, \bibinfo {author} {\bibfnamefont {W.}~\bibnamefont {Krauth}}, \
  and\ \bibinfo {author} {\bibfnamefont {D.~B.}\ \bibnamefont {Wilson}},\
  }\href@noop {} {\bibfield  {journal} {\bibinfo  {journal} {Physical Review E
  - Statistical, Nonlinear, and Soft Matter Physics}\ }\textbf {\bibinfo
  {volume} {80}},\ \bibinfo {pages} {5} (\bibinfo {year} {2009})}\BibitemShut
  {NoStop}%
\bibitem [{\citenamefont {Bertolotti}\ \emph {et~al.}(2012)\citenamefont
  {Bertolotti}, \citenamefont {Putten}, \citenamefont {Blum}, \citenamefont
  {Lagendijk}, \citenamefont {Vos},\ and\ \citenamefont
  {Mosk}}]{Bertolotti2012}%
  \BibitemOpen
  \bibfield  {author} {\bibinfo {author} {\bibfnamefont {J.}~\bibnamefont
  {Bertolotti}}, \bibinfo {author} {\bibfnamefont {E.~G.~V.}\ \bibnamefont
  {Putten}}, \bibinfo {author} {\bibfnamefont {C.}~\bibnamefont {Blum}},
  \bibinfo {author} {\bibfnamefont {A.}~\bibnamefont {Lagendijk}}, \bibinfo
  {author} {\bibfnamefont {W.~L.}\ \bibnamefont {Vos}}, \ and\ \bibinfo
  {author} {\bibfnamefont {A.~P.}\ \bibnamefont {Mosk}},\ }\href@noop {}
  {\bibfield  {journal} {\bibinfo  {journal} {Nature}\ }\textbf {\bibinfo
  {volume} {491}},\ \bibinfo {pages} {232} (\bibinfo {year}
  {2012})}\BibitemShut {NoStop}%
\bibitem [{\citenamefont {Lagae}\ \emph {et~al.}(2010)\citenamefont {Lagae},
  \citenamefont {Lefebvre}, \citenamefont {Cook}, \citenamefont {DeRose},
  \citenamefont {Drettakis}, \citenamefont {Ebert}, \citenamefont {Lewis},
  \citenamefont {Perlin},\ and\ \citenamefont {Zwicker}}]{Lagae2010}%
  \BibitemOpen
  \bibfield  {author} {\bibinfo {author} {\bibfnamefont {A.}~\bibnamefont
  {Lagae}}, \bibinfo {author} {\bibfnamefont {S.}~\bibnamefont {Lefebvre}},
  \bibinfo {author} {\bibfnamefont {R.}~\bibnamefont {Cook}}, \bibinfo {author}
  {\bibfnamefont {T.}~\bibnamefont {DeRose}}, \bibinfo {author} {\bibfnamefont
  {G.}~\bibnamefont {Drettakis}}, \bibinfo {author} {\bibfnamefont {D.~S.}\
  \bibnamefont {Ebert}}, \bibinfo {author} {\bibfnamefont {J.~P.}\ \bibnamefont
  {Lewis}}, \bibinfo {author} {\bibfnamefont {K.}~\bibnamefont {Perlin}}, \
  and\ \bibinfo {author} {\bibfnamefont {M.}~\bibnamefont {Zwicker}},\
  }\href@noop {} {\bibfield  {journal} {\bibinfo  {journal} {Computer Graphics
  Forum}\ }\textbf {\bibinfo {volume} {29}},\ \bibinfo {pages} {2579} (\bibinfo
  {year} {2010})}\BibitemShut {NoStop}%
\bibitem [{\citenamefont {Ma}\ and\ \citenamefont {Torquato}(2017)}]{Ma2017a}%
  \BibitemOpen
  \bibfield  {author} {\bibinfo {author} {\bibfnamefont {Z.}~\bibnamefont
  {Ma}}\ and\ \bibinfo {author} {\bibfnamefont {S.}~\bibnamefont {Torquato}},\
  }\href@noop {} {\bibfield  {journal} {\bibinfo  {journal} {Journal of Applied
  Physics}\ }\textbf {\bibinfo {volume} {121}},\ \bibinfo {pages} {244904}
  (\bibinfo {year} {2017})}\BibitemShut {NoStop}%
\bibitem [{\citenamefont {Engel}\ \emph {et~al.}(2015)\citenamefont {Engel},
  \citenamefont {Damasceno}, \citenamefont {Phillips},\ and\ \citenamefont
  {Glotzer}}]{Engel2015}%
  \BibitemOpen
  \bibfield  {author} {\bibinfo {author} {\bibfnamefont {M.}~\bibnamefont
  {Engel}}, \bibinfo {author} {\bibfnamefont {P.~F.}\ \bibnamefont
  {Damasceno}}, \bibinfo {author} {\bibfnamefont {C.~L.}\ \bibnamefont
  {Phillips}}, \ and\ \bibinfo {author} {\bibfnamefont {S.~C.}\ \bibnamefont
  {Glotzer}},\ }\href@noop {} {\bibfield  {journal} {\bibinfo  {journal}
  {Nature Materials}\ }\textbf {\bibinfo {volume} {14}},\ \bibinfo {pages}
  {109} (\bibinfo {year} {2015})}\BibitemShut {NoStop}%
\bibitem [{\citenamefont {Sgrignuoli}\ and\ \citenamefont {{Dal
  Negro}}(2021)}]{Sgrignuoli2021}%
  \BibitemOpen
  \bibfield  {author} {\bibinfo {author} {\bibfnamefont {F.}~\bibnamefont
  {Sgrignuoli}}\ and\ \bibinfo {author} {\bibfnamefont {L.}~\bibnamefont {{Dal
  Negro}}},\ }\href@noop {} {\bibfield  {journal} {\bibinfo  {journal}
  {Physical Review B}\ }\textbf {\bibinfo {volume} {103}},\ \bibinfo {pages}
  {224202} (\bibinfo {year} {2021})}\BibitemShut {NoStop}%
\bibitem [{\citenamefont {McGreevy}(2001)}]{McGreevy2001}%
  \BibitemOpen
  \bibfield  {author} {\bibinfo {author} {\bibfnamefont {R.~L.}\ \bibnamefont
  {McGreevy}},\ }\href@noop {} {\bibfield  {journal} {\bibinfo  {journal}
  {Journal of Physics Condensed Matter}\ }\textbf {\bibinfo {volume} {13}},\
  \bibinfo {pages} {R877} (\bibinfo {year} {2001})}\BibitemShut {NoStop}%
\bibitem [{\citenamefont {Stillinger}\ and\ \citenamefont
  {Weber}(1985)}]{Stillinger1985}%
  \BibitemOpen
  \bibfield  {author} {\bibinfo {author} {\bibfnamefont {F.~H.}\ \bibnamefont
  {Stillinger}}\ and\ \bibinfo {author} {\bibfnamefont {T.~A.}\ \bibnamefont
  {Weber}},\ }\href@noop {} {\bibfield  {journal} {\bibinfo  {journal}
  {Physical Review B}\ }\textbf {\bibinfo {volume} {31}},\ \bibinfo {pages}
  {5262} (\bibinfo {year} {1985})}\BibitemShut {NoStop}%
\bibitem [{\citenamefont {Sunseri}\ \emph {et~al.}(2023)\citenamefont
  {Sunseri}, \citenamefont {Slepian}, \citenamefont {Portillo}, \citenamefont
  {Hou}, \citenamefont {Kahraman},\ and\ \citenamefont
  {Finkbeiner}}]{Sunseri2023}%
  \BibitemOpen
  \bibfield  {author} {\bibinfo {author} {\bibfnamefont {J.}~\bibnamefont
  {Sunseri}}, \bibinfo {author} {\bibfnamefont {Z.}~\bibnamefont {Slepian}},
  \bibinfo {author} {\bibfnamefont {S.}~\bibnamefont {Portillo}}, \bibinfo
  {author} {\bibfnamefont {J.}~\bibnamefont {Hou}}, \bibinfo {author}
  {\bibfnamefont {S.}~\bibnamefont {Kahraman}}, \ and\ \bibinfo {author}
  {\bibfnamefont {D.~P.}\ \bibnamefont {Finkbeiner}},\ }\href@noop {}
  {\bibfield  {journal} {\bibinfo  {journal} {RAS Techniques and Instruments}\
  }\textbf {\bibinfo {volume} {2}},\ \bibinfo {pages} {62} (\bibinfo {year}
  {2023})}\BibitemShut {NoStop}%
\bibitem [{\citenamefont {Goodrich}\ \emph {et~al.}(2021)\citenamefont
  {Goodrich}, \citenamefont {King}, \citenamefont {Schoenholz}, \citenamefont
  {Cubuk},\ and\ \citenamefont {Brenner}}]{Goodrich2021}%
  \BibitemOpen
  \bibfield  {author} {\bibinfo {author} {\bibfnamefont {C.~P.}\ \bibnamefont
  {Goodrich}}, \bibinfo {author} {\bibfnamefont {E.~M.}\ \bibnamefont {King}},
  \bibinfo {author} {\bibfnamefont {S.~S.}\ \bibnamefont {Schoenholz}},
  \bibinfo {author} {\bibfnamefont {E.~D.}\ \bibnamefont {Cubuk}}, \ and\
  \bibinfo {author} {\bibfnamefont {M.~P.}\ \bibnamefont {Brenner}},\
  }\href@noop {} {\bibfield  {journal} {\bibinfo  {journal} {Proceedings of the
  National Academy of Sciences of the United States of America}\ }\textbf
  {\bibinfo {volume} {118}},\ \bibinfo {pages} {e2024083118} (\bibinfo {year}
  {2021})}\BibitemShut {NoStop}%
\bibitem [{\citenamefont {Lumi{\`{e}}re}\ and\ \citenamefont
  {Lumi{\`{e}}re}(1896)}]{LaCiotat}%
  \BibitemOpen
  \bibfield  {author} {\bibinfo {author} {\bibfnamefont {A.}~\bibnamefont
  {Lumi{\`{e}}re}}\ and\ \bibinfo {author} {\bibfnamefont {L.}~\bibnamefont
  {Lumi{\`{e}}re}},\ }\href@noop {} {\enquote {\bibinfo {title} {{Arriv{\'{e}}e
  d'un Train en Gare de La Ciotat}},}\ } (\bibinfo {year} {1896})\BibitemShut
  {NoStop}%
\bibitem [{\citenamefont {van~der Velden}(2020)}]{VanderVelden2020}%
  \BibitemOpen
  \bibfield  {author} {\bibinfo {author} {\bibfnamefont {E.}~\bibnamefont
  {van~der Velden}},\ }\href@noop {} {\bibfield  {journal} {\bibinfo  {journal}
  {Journal of Open Source Software}\ }\textbf {\bibinfo {volume} {5}},\
  \bibinfo {pages} {2004} (\bibinfo {year} {2020})}\BibitemShut {NoStop}%
\bibitem [{\citenamefont {Barnett}\ \emph {et~al.}(2019)\citenamefont
  {Barnett}, \citenamefont {Magland},\ and\ \citenamefont {{Af
  Klinteberg}}}]{Barnett2019}%
  \BibitemOpen
  \bibfield  {author} {\bibinfo {author} {\bibfnamefont {A.~H.}\ \bibnamefont
  {Barnett}}, \bibinfo {author} {\bibfnamefont {J.}~\bibnamefont {Magland}}, \
  and\ \bibinfo {author} {\bibfnamefont {L.}~\bibnamefont {{Af Klinteberg}}},\
  }\href@noop {} {\bibfield  {journal} {\bibinfo  {journal} {SIAM Journal on
  Scientific Computing}\ }\textbf {\bibinfo {volume} {48}},\ \bibinfo {pages}
  {C479} (\bibinfo {year} {2019})}\BibitemShut {NoStop}%
\bibitem [{\citenamefont {Barnett}(2021)}]{Barnett2020}%
  \BibitemOpen
  \bibfield  {author} {\bibinfo {author} {\bibfnamefont {A.~H.}\ \bibnamefont
  {Barnett}},\ }\href@noop {} {\bibfield  {journal} {\bibinfo  {journal}
  {Applied and Computational Harmonic Analysis}\ }\textbf {\bibinfo {volume}
  {51}},\ \bibinfo {pages} {1} (\bibinfo {year} {2021})}\BibitemShut {NoStop}%
\end{thebibliography}%


\begin{thebibliography}{23}%
\makeatletter
\providecommand \@ifxundefined [1]{%
 \@ifx{#1\undefined}
}%
\providecommand \@ifnum [1]{%
 \ifnum #1\expandafter \@firstoftwo
 \else \expandafter \@secondoftwo
 \fi
}%
\providecommand \@ifx [1]{%
 \ifx #1\expandafter \@firstoftwo
 \else \expandafter \@secondoftwo
 \fi
}%
\providecommand \natexlab [1]{#1}%
\providecommand \enquote  [1]{``#1''}%
\providecommand \bibnamefont  [1]{#1}%
\providecommand \bibfnamefont [1]{#1}%
\providecommand \citenamefont [1]{#1}%
\providecommand \href@noop [0]{\@secondoftwo}%
\providecommand \href [0]{\begingroup \@sanitize@url \@href}%
\providecommand \@href[1]{\@@startlink{#1}\@@href}%
\providecommand \@@href[1]{\endgroup#1\@@endlink}%
\providecommand \@sanitize@url [0]{\catcode `\\12\catcode `\$12\catcode
  `\&12\catcode `\#12\catcode `\^12\catcode `\_12\catcode `\%12\relax}%
\providecommand \@@startlink[1]{}%
\providecommand \@@endlink[0]{}%
\providecommand \url  [0]{\begingroup\@sanitize@url \@url }%
\providecommand \@url [1]{\endgroup\@href {#1}{\urlprefix }}%
\providecommand \urlprefix  [0]{URL }%
\providecommand \Eprint [0]{\href }%
\providecommand \doibase [0]{http://dx.doi.org/}%
\providecommand \selectlanguage [0]{\@gobble}%
\providecommand \bibinfo  [0]{\@secondoftwo}%
\providecommand \bibfield  [0]{\@secondoftwo}%
\providecommand \translation [1]{[#1]}%
\providecommand \BibitemOpen [0]{}%
\providecommand \bibitemStop [0]{}%
\providecommand \bibitemNoStop [0]{.\EOS\space}%
\providecommand \EOS [0]{\spacefactor3000\relax}%
\providecommand \BibitemShut  [1]{\csname bibitem#1\endcsname}%
\let\auto@bib@innerbib\@empty
\bibitem [{\citenamefont {Gabrielli}\ and\ \citenamefont
  {Torquato}(2004)}]{Gabrielli2004}%
  \BibitemOpen
  \bibfield  {author} {\bibinfo {author} {\bibfnamefont {A.}~\bibnamefont
  {Gabrielli}}\ and\ \bibinfo {author} {\bibfnamefont {S.}~\bibnamefont
  {Torquato}},\ }\href@noop {} {\bibfield  {journal} {\bibinfo  {journal}
  {Physical Review E - Statistical Physics, Plasmas, Fluids, and Related
  Interdisciplinary Topics}\ }\textbf {\bibinfo {volume} {70}},\ \bibinfo
  {pages} {12} (\bibinfo {year} {2004})}\BibitemShut {NoStop}%
\bibitem [{\citenamefont {van Gogh}(1889)}]{StarryNight}%
  \BibitemOpen
  \bibfield  {author} {\bibinfo {author} {\bibfnamefont {V.}~\bibnamefont {van
  Gogh}},\ }\href@noop {} {\enquote {\bibinfo {title} {{Starry Night}},}\ }
  (\bibinfo {year} {1889})\BibitemShut {NoStop}%
\bibitem [{\citenamefont {{Morse, Philip}}\ and\ \citenamefont
  {Feshbach}(1953)}]{MorseFeshbachI}%
  \BibitemOpen
  \bibfield  {author} {\bibinfo {author} {\bibfnamefont {M.}~\bibnamefont
  {{Morse, Philip}}}\ and\ \bibinfo {author} {\bibfnamefont {H.}~\bibnamefont
  {Feshbach}},\ }\href@noop {} {\emph {\bibinfo {title} {{Methods of
  Theoretical Physics: Part I}}}}\ (\bibinfo  {publisher} {McGraw-Hill series
  in fundamentals of physics},\ \bibinfo {year} {1953})\BibitemShut {NoStop}%
\bibitem [{\citenamefont {Carminati}\ and\ \citenamefont
  {Schotland}(2021)}]{Carminati2021}%
  \BibitemOpen
  \bibfield  {author} {\bibinfo {author} {\bibfnamefont {R.}~\bibnamefont
  {Carminati}}\ and\ \bibinfo {author} {\bibfnamefont {J.~C.}\ \bibnamefont
  {Schotland}},\ }\href@noop {} {\emph {\bibinfo {title} {Principles of
  Scattering and Transport of Light}}}\ (\bibinfo  {publisher} {Cambridge
  University Press},\ \bibinfo {year} {2021})\BibitemShut {NoStop}%
\bibitem [{\citenamefont {Economou}(2006)}]{Economou2006}%
  \BibitemOpen
  \bibfield  {author} {\bibinfo {author} {\bibfnamefont {E.~N.}\ \bibnamefont
  {Economou}},\ }\href@noop {} {\emph {\bibinfo {title} {Springer Series in
  Solid-state Sciences}}}\ (\bibinfo  {publisher} {Springer},\ \bibinfo
  {address} {Berlin},\ \bibinfo {year} {2006})\BibitemShut {NoStop}%
\bibitem [{\citenamefont {Gonz{\'{a}}lez}(2010)}]{Gonzalez2010a}%
  \BibitemOpen
  \bibfield  {author} {\bibinfo {author} {\bibfnamefont {{\'{A}}.}~\bibnamefont
  {Gonz{\'{a}}lez}},\ }\href@noop {} {\bibfield  {journal} {\bibinfo  {journal}
  {Mathematical Geosciences}\ }\textbf {\bibinfo {volume} {42}},\ \bibinfo
  {pages} {49} (\bibinfo {year} {2010})}\BibitemShut {NoStop}%
\bibitem [{\citenamefont {Leseur}\ \emph {et~al.}(2016)\citenamefont {Leseur},
  \citenamefont {Pierrat},\ and\ \citenamefont {Carminati}}]{Leseur2016}%
  \BibitemOpen
  \bibfield  {author} {\bibinfo {author} {\bibfnamefont {O.}~\bibnamefont
  {Leseur}}, \bibinfo {author} {\bibfnamefont {R.}~\bibnamefont {Pierrat}}, \
  and\ \bibinfo {author} {\bibfnamefont {R.}~\bibnamefont {Carminati}},\
  }\href@noop {} {\bibfield  {journal} {\bibinfo  {journal} {Optica}\ }\textbf
  {\bibinfo {volume} {3}},\ \bibinfo {pages} {763} (\bibinfo {year}
  {2016})}\BibitemShut {NoStop}%
\bibitem [{\citenamefont {Monsarrat}\ \emph {et~al.}(2022)\citenamefont
  {Monsarrat}, \citenamefont {Pierrat}, \citenamefont {Tourin},\ and\
  \citenamefont {Goetschy}}]{Monsarrat2022}%
  \BibitemOpen
  \bibfield  {author} {\bibinfo {author} {\bibfnamefont {R.}~\bibnamefont
  {Monsarrat}}, \bibinfo {author} {\bibfnamefont {R.}~\bibnamefont {Pierrat}},
  \bibinfo {author} {\bibfnamefont {A.}~\bibnamefont {Tourin}}, \ and\ \bibinfo
  {author} {\bibfnamefont {A.}~\bibnamefont {Goetschy}},\ }\href@noop {}
  {\bibfield  {journal} {\bibinfo  {journal} {Physical Review Research}\
  }\textbf {\bibinfo {volume} {4}},\ \bibinfo {pages} {033246} (\bibinfo {year}
  {2022})}\BibitemShut {NoStop}%
\bibitem [{\citenamefont {Hansen}\ and\ \citenamefont
  {McDonald}(2006)}]{Hansen2006}%
  \BibitemOpen
  \bibfield  {author} {\bibinfo {author} {\bibfnamefont {J.-P.}\ \bibnamefont
  {Hansen}}\ and\ \bibinfo {author} {\bibfnamefont {I.~R.}\ \bibnamefont
  {McDonald}},\ }\href@noop {} {\emph {\bibinfo {title} {{Theory of simple
  liquids}}}}\ (\bibinfo  {publisher} {Elsevier Academic Press},\ \bibinfo
  {year} {2006})\BibitemShut {NoStop}%
\bibitem [{\citenamefont {Lutfalla}(2021)}]{Lutfalla2021}%
  \BibitemOpen
  \bibfield  {author} {\bibinfo {author} {\bibfnamefont {V.~H.}\ \bibnamefont
  {Lutfalla}},\ }in\ \href@noop {} {\emph {\bibinfo {booktitle} {27th IFIP WG
  1.5 International Workshop on Cellular Automata and Discrete Complex Systems
  (AUTOMATA 2021)}}},\ \bibinfo {series and number} {\bibinfo {number} {9}},\
  \bibinfo {editor} {edited by\ \bibinfo {editor} {\bibfnamefont
  {A.}~\bibnamefont {Castillo-Ramirez}}, \bibinfo {editor} {\bibfnamefont
  {P.}~\bibnamefont {Guillon}}, \ and\ \bibinfo {editor} {\bibfnamefont
  {K.}~\bibnamefont {Perrot}}}\ (\bibinfo  {publisher} {Schloss Dagstuhl –
  Leibniz-Zentrum f{\"{u}}r Informatik, Dagstuhl Publishing, Germany},\
  \bibinfo {year} {2021})\ pp.\ \bibinfo {pages} {9:1--9:12}\BibitemShut
  {NoStop}%
\bibitem [{\citenamefont {{De Bruijn}}(1981)}]{deBruijn1981}%
  \BibitemOpen
  \bibfield  {author} {\bibinfo {author} {\bibfnamefont {N.}~\bibnamefont {{De
  Bruijn}}},\ }\href@noop {} {\bibfield  {journal} {\bibinfo  {journal}
  {Indagationes Mathematicae}\ }\textbf {\bibinfo {volume} {43}},\ \bibinfo
  {pages} {39} (\bibinfo {year} {1981})}\BibitemShut {NoStop}%
\bibitem [{\citenamefont {de~Bruijn}(1986)}]{deBruijn1986}%
  \BibitemOpen
  \bibfield  {author} {\bibinfo {author} {\bibfnamefont {N.}~\bibnamefont
  {de~Bruijn}},\ }\href {\doibase 10.1016/s1385-7258(86)80002-6} {\bibfield
  {journal} {\bibinfo  {journal} {Indagationes Mathematicae (Proceedings)}\
  }\textbf {\bibinfo {volume} {89}},\ \bibinfo {pages} {123} (\bibinfo {year}
  {1986})}\BibitemShut {NoStop}%
\bibitem [{\citenamefont {Engel}\ \emph {et~al.}(2015)\citenamefont {Engel},
  \citenamefont {Damasceno}, \citenamefont {Phillips},\ and\ \citenamefont
  {Glotzer}}]{Engel2015}%
  \BibitemOpen
  \bibfield  {author} {\bibinfo {author} {\bibfnamefont {M.}~\bibnamefont
  {Engel}}, \bibinfo {author} {\bibfnamefont {P.~F.}\ \bibnamefont
  {Damasceno}}, \bibinfo {author} {\bibfnamefont {C.~L.}\ \bibnamefont
  {Phillips}}, \ and\ \bibinfo {author} {\bibfnamefont {S.~C.}\ \bibnamefont
  {Glotzer}},\ }\href@noop {} {\bibfield  {journal} {\bibinfo  {journal}
  {Nature Materials}\ }\textbf {\bibinfo {volume} {14}},\ \bibinfo {pages}
  {109} (\bibinfo {year} {2015})}\BibitemShut {NoStop}%
\bibitem [{\citenamefont {Levine}\ and\ \citenamefont
  {Steinhardt}(1986)}]{Levine1986}%
  \BibitemOpen
  \bibfield  {author} {\bibinfo {author} {\bibfnamefont {D.}~\bibnamefont
  {Levine}}\ and\ \bibinfo {author} {\bibfnamefont {P.~J.}\ \bibnamefont
  {Steinhardt}},\ }\href@noop {} {\bibfield  {journal} {\bibinfo  {journal}
  {Physical Review B}\ }\textbf {\bibinfo {volume} {34}},\ \bibinfo {pages}
  {596} (\bibinfo {year} {1986})}\BibitemShut {NoStop}%
\bibitem [{\citenamefont {Steinhardt}\ \emph {et~al.}(1983)\citenamefont
  {Steinhardt}, \citenamefont {Nelson},\ and\ \citenamefont
  {Ronchetti}}]{Steinhardt1983}%
  \BibitemOpen
  \bibfield  {author} {\bibinfo {author} {\bibfnamefont {P.~J.}\ \bibnamefont
  {Steinhardt}}, \bibinfo {author} {\bibfnamefont {D.~R.}\ \bibnamefont
  {Nelson}}, \ and\ \bibinfo {author} {\bibfnamefont {M.}~\bibnamefont
  {Ronchetti}},\ }\href@noop {} {\bibfield  {journal} {\bibinfo  {journal}
  {Physical Review B}\ }\textbf {\bibinfo {volume} {28}},\ \bibinfo {pages}
  {784} (\bibinfo {year} {1983})}\BibitemShut {NoStop}%
\bibitem [{\citenamefont {Nelson}\ and\ \citenamefont
  {Halperin}(1979)}]{Nelson1979}%
  \BibitemOpen
  \bibfield  {author} {\bibinfo {author} {\bibfnamefont {D.~R.}\ \bibnamefont
  {Nelson}}\ and\ \bibinfo {author} {\bibfnamefont {B.~I.}\ \bibnamefont
  {Halperin}},\ }\href@noop {} {\bibfield  {journal} {\bibinfo  {journal}
  {Physical Review B}\ }\textbf {\bibinfo {volume} {19}},\ \bibinfo {pages}
  {2457} (\bibinfo {year} {1979})}\BibitemShut {NoStop}%
\bibitem [{\citenamefont {Li}\ and\ \citenamefont {Ciamarra}(2020)}]{Li2020}%
  \BibitemOpen
  \bibfield  {author} {\bibinfo {author} {\bibfnamefont {Y.-W.}\ \bibnamefont
  {Li}}\ and\ \bibinfo {author} {\bibfnamefont {M.~P.}\ \bibnamefont
  {Ciamarra}},\ }\href@noop {} {\bibfield  {journal} {\bibinfo  {journal}
  {Physical Review Letters}\ }\textbf {\bibinfo {volume} {124}},\ \bibinfo
  {pages} {218002} (\bibinfo {year} {2020})}\BibitemShut {NoStop}%
\bibitem [{\citenamefont {Lumi{\`{e}}re}\ and\ \citenamefont
  {Lumi{\`{e}}re}(1896)}]{LaCiotat}%
  \BibitemOpen
  \bibfield  {author} {\bibinfo {author} {\bibfnamefont {A.}~\bibnamefont
  {Lumi{\`{e}}re}}\ and\ \bibinfo {author} {\bibfnamefont {L.}~\bibnamefont
  {Lumi{\`{e}}re}},\ }\href@noop {} {\enquote {\bibinfo {title} {{Arriv{\'{e}}e
  d'un Train en Gare de La Ciotat}},}\ } (\bibinfo {year} {1896})\BibitemShut
  {NoStop}%
\bibitem [{\citenamefont {Fienup}(1982)}]{Fienup1982}%
  \BibitemOpen
  \bibfield  {author} {\bibinfo {author} {\bibfnamefont {J.~R.}\ \bibnamefont
  {Fienup}},\ }\href@noop {} {\bibfield  {journal} {\bibinfo  {journal}
  {Applied Optics}\ }\textbf {\bibinfo {volume} {21}},\ \bibinfo {pages} {2758}
  (\bibinfo {year} {1982})}\BibitemShut {NoStop}%
\bibitem [{\citenamefont {Bertolotti}\ \emph {et~al.}(2012)\citenamefont
  {Bertolotti}, \citenamefont {Putten}, \citenamefont {Blum}, \citenamefont
  {Lagendijk}, \citenamefont {Vos},\ and\ \citenamefont
  {Mosk}}]{Bertolotti2012}%
  \BibitemOpen
  \bibfield  {author} {\bibinfo {author} {\bibfnamefont {J.}~\bibnamefont
  {Bertolotti}}, \bibinfo {author} {\bibfnamefont {E.~G.~V.}\ \bibnamefont
  {Putten}}, \bibinfo {author} {\bibfnamefont {C.}~\bibnamefont {Blum}},
  \bibinfo {author} {\bibfnamefont {A.}~\bibnamefont {Lagendijk}}, \bibinfo
  {author} {\bibfnamefont {W.~L.}\ \bibnamefont {Vos}}, \ and\ \bibinfo
  {author} {\bibfnamefont {A.~P.}\ \bibnamefont {Mosk}},\ }\href@noop {}
  {\bibfield  {journal} {\bibinfo  {journal} {Nature}\ }\textbf {\bibinfo
  {volume} {491}},\ \bibinfo {pages} {232} (\bibinfo {year}
  {2012})}\BibitemShut {NoStop}%
\bibitem [{\citenamefont {Lagae}\ \emph {et~al.}(2010)\citenamefont {Lagae},
  \citenamefont {Lefebvre}, \citenamefont {Cook}, \citenamefont {DeRose},
  \citenamefont {Drettakis}, \citenamefont {Ebert}, \citenamefont {Lewis},
  \citenamefont {Perlin},\ and\ \citenamefont {Zwicker}}]{Lagae2010}%
  \BibitemOpen
  \bibfield  {author} {\bibinfo {author} {\bibfnamefont {A.}~\bibnamefont
  {Lagae}}, \bibinfo {author} {\bibfnamefont {S.}~\bibnamefont {Lefebvre}},
  \bibinfo {author} {\bibfnamefont {R.}~\bibnamefont {Cook}}, \bibinfo {author}
  {\bibfnamefont {T.}~\bibnamefont {DeRose}}, \bibinfo {author} {\bibfnamefont
  {G.}~\bibnamefont {Drettakis}}, \bibinfo {author} {\bibfnamefont {D.~S.}\
  \bibnamefont {Ebert}}, \bibinfo {author} {\bibfnamefont {J.~P.}\ \bibnamefont
  {Lewis}}, \bibinfo {author} {\bibfnamefont {K.}~\bibnamefont {Perlin}}, \
  and\ \bibinfo {author} {\bibfnamefont {M.}~\bibnamefont {Zwicker}},\
  }\href@noop {} {\bibfield  {journal} {\bibinfo  {journal} {Computer Graphics
  Forum}\ }\textbf {\bibinfo {volume} {29}},\ \bibinfo {pages} {2579} (\bibinfo
  {year} {2010})}\BibitemShut {NoStop}%
\bibitem [{\citenamefont {Dong}\ \emph {et~al.}(2020)\citenamefont {Dong},
  \citenamefont {Liu}, \citenamefont {Yao}, \citenamefont {Chantler},
  \citenamefont {Qi}, \citenamefont {Yu},\ and\ \citenamefont
  {Jian}}]{Dong2020}%
  \BibitemOpen
  \bibfield  {author} {\bibinfo {author} {\bibfnamefont {J.}~\bibnamefont
  {Dong}}, \bibinfo {author} {\bibfnamefont {J.}~\bibnamefont {Liu}}, \bibinfo
  {author} {\bibfnamefont {K.}~\bibnamefont {Yao}}, \bibinfo {author}
  {\bibfnamefont {M.}~\bibnamefont {Chantler}}, \bibinfo {author}
  {\bibfnamefont {L.}~\bibnamefont {Qi}}, \bibinfo {author} {\bibfnamefont
  {H.}~\bibnamefont {Yu}}, \ and\ \bibinfo {author} {\bibfnamefont
  {M.}~\bibnamefont {Jian}},\ }\href@noop {} {\bibfield  {journal} {\bibinfo
  {journal} {Sensors (Switzerland)}\ }\textbf {\bibinfo {volume} {20}},\
  \bibinfo {pages} {1135} (\bibinfo {year} {2020})}\BibitemShut {NoStop}%
\bibitem [{\citenamefont {Ma}\ and\ \citenamefont {Torquato}(2017)}]{Ma2017a}%
  \BibitemOpen
  \bibfield  {author} {\bibinfo {author} {\bibfnamefont {Z.}~\bibnamefont
  {Ma}}\ and\ \bibinfo {author} {\bibfnamefont {S.}~\bibnamefont {Torquato}},\
  }\href@noop {} {\bibfield  {journal} {\bibinfo  {journal} {Journal of Applied
  Physics}\ }\textbf {\bibinfo {volume} {121}},\ \bibinfo {pages} {244904}
  (\bibinfo {year} {2017})}\BibitemShut {NoStop}%
\end{thebibliography}%

\appendix
\setcounter{figure}{0}
\renewcommand{\figurename}{Extended Data Fig.}
\renewcommand{\thefigure}{\arabic{figure}}
\setcounter{equation}{0}
\newtagform{M}{(M})
\usetagform{M}

\section{Methods}
\textit{Algorithm parameters --}
The size of the domain in which the optimization is performed, $\mathcal{N}_K = |\mathcal{K}|$, counted in number of discrete wave-vectors, is limited in practice by the number of degrees of freedom, $dN$, for $N$ points embedded in $d$-dimensional space.
This constraint often leads to defining a ratio $\chi \equiv \mathcal{N}_K / (2 d N)$~\cite{Uche2004,Uche2006,Monsarrat2022,Morse2023}, where the 2 stems from Friedel's law. 
Optimizations with $\chi < 1$ are in theory possible, while those with $\chi \geq 1$ are overconstrained and cannot necessarily be achieved.
In practice, past work has reported good achievability with other algorithms using $\chi \lesssim 0.5$~\cite{Uche2004} in $2d$ space.
In this work, we always set $\chi \approx 0.4$, so that the number of constrained $k$-space features scales linearly with $N$.

The bulk of the computations consists of Fourier transforms between uniform and non-uniform spaces (Figure \ref{fig:Sketch}$(a)$).
We use the Flatiron Institute Non-Uniform Fast Fourier Transform (FINUFFT) framework which provides transforms of three types and is highly optimized for multithreaded CPU computations~\cite{Barnett2019,Barnett2020}.
Type-1 refers to a non-uniform to uniform transform (e.g. real space points to a $k$-space grid as used in the calculation of the structure factor in Eq.~\ref{eq:structure_factor}).
Type-2 refers to a uniform to non-uniform transform (e.g. $k$-space grid to known points in real space as used in the calculation of the loss gradient in Eq.~\ref{eq:gradient}).
Type-3 refers to a non-uniform to non-uniform transform (e.g. real space points to specific points in $k$-space, with free boundary conditions, as used in the calculation of the Ewald sphere or in NUwNU, see below).

The termination criterion for optimization throughout the text was a threshold value of $10^{-39}$ on the gradient.
In the special case of stealthy hyperuniform systems, this criterion achieves low-$k$ values, $S \sim 10^{-25}-10^{-20}$.
Lower values such as those reported in Ref.~\onlinecite{Morse2023} can be attained within the powerful framework of FReSCo using a different termination value and higher precision arithmetics, but such small values are not realistic in any practical realization (see SI for a detailed discussion of this point).

\textit{Analytical gradients --}
We here show how one may write the gradient of the structure factor loss, $\mathcal{L}_S$, analytically as a Fourier series for all four constructions UwU, UwNU, NUwU, NUwNU. 
In Eq.~\ref{eq:structure_factor} we introduced the structure factor
\begin{align*}
    S(\bm{k}) &\equiv \frac{|\widehat{\rho}\,(\bm{k})|^2}{N}
\end{align*}
associated to a $d$-dimensional density field describing $N$ points with complex-valued weights $(c_1, \ldots, c_N)$ at positions $(\bm{r}_1, \ldots \bm{r}_N)$ in $\mathbb{R}^d$,
\begin{align}
    \rho(\bm{r}) = \sum\limits_{n=1}^{N} c_n \delta\left(\bm{r} - \bm{r}_n\right),
\end{align}
through its Fourier transform
\begin{align}
    \widehat{\rho}(\bm{k}) &= \sum_{n=1}^{N}c_n \exp(i\bm{k}\cdot\bm{r}_n).
    \label{eq:ft}
\end{align}
For simplicity, in the following, we absorb the normalization by $S_0$ in Eq.~\ref{eq:loss} into the weighting function $W(\bm{k}) = w(\bm{k}) / S_0(\bm{k})^2$ if $S_0(\bm{k}) \neq 0$ and $W(\bm{k}) = w(\bm{k})$ otherwise, so that at every point we can rewrite the loss, Eq.~\ref{eq:SF_loss}, as
\begin{align}
    \mathcal{L}_S = \sum_{\bm{k}\in \mathcal{K}} W(\bm{k})\left(S(\bm{k})-S_0(\bm{k})\right)^2.
\end{align}

We first restrict ourselves to the case of real-valued weights, and write the gradient of this loss with respect to one of the weights, $c_n \in \mathbb{R}$, corresponding to the optimization of a real-valued field at fixed mesh positions,
\begin{align}
    \frac{\partial\mathcal{L}_S}{\partial c_n} = \sum_{\bm{k}\in\mathcal{K}} 2W(\bm{k})\left(S(\bm{k})-S_0(\bm{k})\right)\frac{\partial S(\bm{k})}{\partial c_n}.
    \label{eq:weightgrad}
\end{align}
The corresponding derivative of the Fourier transform of the density field, Eq.~\ref{eq:ft}, then reads
\begin{align}
\frac{\partial \widehat{\rho}\,(\bm{k})}{\partial c_n} = e^{i\bm{k}\cdot\bm{r}_n},
\end{align}
Recalling that, in general, the structure factor can be written as 
\begin{align}
    S(\bm{k}) = |\widehat{\rho}\,(\bm{k})|^2 / \rho_0 =\widehat{\rho}\,(\bm{k})\widehat{\rho}\,^\dagger(\bm{k})/\rho_0,
\end{align}
where $\rho_0 = \sum_j |c_j|^2$, we may write the gradient components of $S$ as:
\begin{align}
    \frac{\partial S}{\partial c_n}= \frac{1}{\rho_0} \left(\widehat{\rho}\,^\dagger(\bm{k})e^{i\bm{k}\cdot\bm{r}_n} + \widehat{\rho}\,(\bm{k})e^{-i\bm{k}\cdot\bm{r}_n}\right) - \frac{2c_n|\widehat{\rho}(\bm{k})|^2}{\rho_0^2}.
\end{align}
The first two terms in this expression may be simplified by noticing that they are the sum of a number with its conjugate, so that 
\begin{align}
    \frac{\partial S}{\partial c_n}= \frac{2}{\rho_0} Re \left[ \widehat{\rho}\,(\bm{k})e^{-i\bm{k}\cdot\bm{r}_n} \right]- \frac{2c_n}{\rho_0}S(\bm{k}). \label{eq:weightgradient}
\end{align}
Injecting this expression into Eq.~\ref{eq:weightgrad}, one may find that the gradient can be calculated by taking the real part of a Fourier Transform:
\begin{align}
    \frac{\partial\mathcal{L}_S}{\partial c_n} &= Re \left[\sum_{\bm{k}\in\mathcal{K}} \bm{C}_{c_n}(\bm{k})e^{-i\bm{k}\cdot\bm{r}_n} \right]- \bm{F}_{c_n}\nonumber\\
    &=Re\left[\widehat{\bm{C}}_{c_n}(\bm{r}_n)\right]- \bm{F}_{c_n}
    \label{eq:lossweightgradient}
\end{align}
where $\bm{C}_{c_n}(\bm{k})=4W(\bm{k})\left(S(\bm{k})-S_0(\bm{k})\right)\widehat{\rho}\,(\bm{k})/\rho_0$ are coefficients of a Fourier series, $\widehat{\bm{C}}_{c_n}$ is the Fourier transform of $\bm{C}_{c_n}$, and $\bm{F}_{c_n} = \frac{4c_n}{\rho_0}\sum_{\bm{k}\in\mathcal{K}}W(\bm{k})S(\bm{k})(S(\bm{k})-S_0(\bm{k}))$.

Likewise, we write the gradient of the loss with respect to one of positions, $\bm{r}_n$, corresponding to the optimization of continuous positions with fixed complex weights,
\begin{align}
    \frac{\partial\mathcal{L}_S}{\partial \bm{r}_n} = \sum_{\bm{k}\in\mathcal{K}} 2W(\bm{k})\left(S(\bm{k})-S_0(\bm{k})\right)\frac{\partial S(\bm{k})}{\partial \bm{r}_n}.
    \label{eq:positiongrad}
\end{align}
Using the definition of $S$ again, coupled with the derivative of the Fourier transform of the density field:
\begin{align}
\frac{\partial \widehat{\rho}\,(\bm{k})}{\partial \bm{r}_n} = i\bm{k} c_n e^{i\bm{k}\cdot\bm{r}_n},
\end{align}
the gradient components of $S$ can be recast as
\begin{align}
    \frac{\partial S}{\partial \bm{r}_n}=\frac{i\bm{k} }{N}\left( c_n\widehat{\rho}\,^\dagger(\bm{k})e^{i\bm{k}\cdot\bm{r}_n}- c^\dagger_n\widehat{\rho}\,(\bm{k})e^{-i\bm{k}\cdot\bm{r}_n}\right).
\end{align}
Once again, this expression may be simplified by noticing that it is the sum of a number with its conjugate, so that 
\begin{align}
    \frac{\partial S}{\partial \bm{r}_n}= 2 Re \left[ - \frac{i\bm{k}c^\dagger_n}{N}  \widehat{\rho}\,(\bm{k})e^{-i\bm{k}\cdot\bm{r}_n} \right]. \label{eq:Sgradient}
\end{align}
Injecting this expression into Eq.~\ref{eq:positiongrad}, one recovers Eq.~\ref{eq:gradient} of the main text:
\begin{align}
    \frac{\partial \mathcal{L}_S}{\partial \bm{r}_{n}} &= Re\left[\sum_{\bm{k}}\bm{C}_{\bm{r}_n}(\bm{k}) c^\dagger_n \exp(-i\bm{k}\cdot\bm{r}_n)\right] \nonumber\\ 
    &= Re\left[c^\dagger_n\widehat{\bm{C}}_{\bm{r}_n}(\bm{r}_n)\right]
    \label{eq:losspositiongradient}
\end{align}
where $\bm{C}_{\bm{r}_n}(\bm{k})=-4i\bm{k} w(\bm{k})\left[S(\bm{k})-S_0(\bm{k})\right]\widehat{\rho}\,(\bm{k})/N$ are coefficients of a Fourier series, and $\widehat{\bm{C}}_{\bm{r}_n}$ is the Fourier transform of $\bm{C}_{\bm{r}_n}$.

From Eqs.~\ref{eq:lossweightgradient} and~\ref{eq:losspositiongradient}, we define 4 variants of our optimization algorithm depending on the structure of $k$-space constraints and of the density field that are considered.
While it is in principle possible to simultaneously optimize weights and non-uniform positions, we leave that possibility for future work, and only consider optimization of either weights or positions.

\textit{UwU: Uniform real space with uniform k-space constraints --}
First, consider the case of a density field constrained to a square grid, also called a \textit{uniform} sampling of space~\cite{Pharr2018}.
In that case, one may optimize the weights at each point, so as to obtain a square-grid meshing of a field with desirable correlations.
In particular, if the system is defined with periodic boundary conditions, one may impose features at $\bm{k}$-vectors also lying on a grid, or in other words use a \textit{uniform} constraint.
We therefore call this variant UwU, Uniform real space density with Uniform k-space constraints.
Since both the real-space density field and the Fourier-space constraints are uniform, this is the only scenario in which one may compute both the loss and its gradient with usual discrete Fourier transform algorithms.

\textit{UwNU: Uniform real space with non-uniform k-space constraints --}
Now, consider the case of a uniform density field, but with free boundary conditions, so that one may impose non-uniform Fourier constraints at arbitrary continuous positions, or UwNU.
This case (UwNU) utilizes one FINUFFT Type 2 transformation (uniform to non-uniform) to calculate $S(\bm{k})$ and one FINUFFT Type 1 transformation (non-uniform to uniform) to calculate the gradient of the loss.

\textit{NUwU: Non-uniform real space with uniform k-space constraints --}
If we now optimize real-space positions, resulting in a non-uniform set of real positions with periodic boundary conditions, and therefore use uniform Fourier-space constraints, we get NUwU.
This case (NUwU) utilizes one FINUFFT Type 1 transformation (non-uniform to uniform) to calculate $S(\bm{k})$ and $d$ FINUFFT Type 2 transformations (uniform to non-uniform) to calculate the gradient of the loss (one for each dimension of $\bm{k}$).

\textit{NUwNU: Non-uniform real space with non-uniform k-space constraints --}
Finally, we may optimize non-uniform real-space positions, but this time with free boundary conditions, leading to arbitrary non-uniform Fourier-space constraints, and to NUwNU.
This case (NUwNU) utilizes one FINUFFT Type 3 transformation (non-uniform to non-uniform) to calculate $S(\bm{k})$ and $d$ additional FINUFFT Type 3 transformations to calculate the gradient of the loss (one for each dimension of $\bm{k}$).

\textit{Pair correlation function optimization --}
Our optimization strategy can be generalized to optimize the real-space pair correlation function $g(\bm{r})$.
For a real-valued density field generated by a spatially uniform process inside a cubic box with sidelength $L$, the pair correlation function is defined~\cite{Hansen2006} as
\begin{align}
g(\bm{r}) \equiv L^d \frac{\int d^d \bm{r}_1 d^d \bm{r}_2 \rho(\bm{r}_1) \rho(\bm{r}_2) \delta(\bm{r} - \bm{r}_{12})}{\left(\int d^d \bm{r}_1 \rho(\bm{r}_1)\right) \left( \int d^d \bm{r}_2 \rho(\bm{r}_2)\right)},
\end{align}
which tends to $1$ as the density fields at positions $\bm{r}_1$ and $\bm{r}_1 + \bm{r}$ become independent, usually as $r \to \infty$.
This expression can be simplified using the definition of the density field of a point pattern, and excluding the $i = j$ point from the sum per the usual convention~\cite{Hansen2006}, yielding
\begin{align}
g(\bm{r}) &= \frac{L^d}{N^2 \overline{c_n}^2} \sum_{i \neq j} c_i c_j \delta(\bm{r} - \bm{r}_{ij}),
\end{align}
where we defined the arithmetic average of the weights, $\overline{c_n} \equiv \sum_n c_n / N$.
This last expression can be written as an inverse Fourier transform of $S(\bm{k})-1$, namely
\begin{align}
    g(\bm{r})= \frac{1}{n_0 \overline{c_n}^2}\int_{\mathcal{F}} \frac{d^d\bm{k}}{(2\pi)^d}(S(\bm{k})-1)e^{-i\bm{k}\cdot\bm{r}}, \label{eq:gdef}
\end{align}
where the integral is computed over the whole Fourier domain $\mathcal{F}$, and $n_0$ is the spatially averaged number density, $n_0 = N /L^d$.
Note that in the standard setting of liquid theory, $\forall n, c_n = 1$ so that the prefactor simply becomes $1/n_0$, yielding the more usual relation between $S$ and $g$~\cite{Hansen2006}.
For compactness, we henceforth define $\rho_0 = n_0 \overline{c_n}^2$.

The associated loss term can be written as a sum over a discrete set $\mathcal{R}$ of constrained real-space distances $\bm{x}$ instead of reciprocal-space wave-vectors,
\begin{align}
\mathcal{L}_g \equiv \sum_{\bm{x} \in \mathcal{R}}W_g(\bm{x})\left(g(\bm{x})-g_0(\bm{x})\right)^2,
\end{align}
where $W_g$ is a weight function that can for instance select short-range order to be jointly optimized with some longer-range property in $S$.
The gradient of this loss term with respect to the position  $\bm{r}_n$ of particle $n$ can be expressed as
\begin{align}
    \frac{\partial \mathcal{L}_g}{\partial \bm{r}_n} &= \sum_{\bm{x} \in \mathcal{R}}2W_g(\bm{x})\left(g(\bm{x})-g_0(\bm{x})\right)\frac{\partial g}{\partial \bm{r}_n}(\bm{x}).
\end{align}

Since we optimize structures in finite periodic boxes, the integral in Eq.~\ref{eq:gdef} reduces to a discrete Fourier transform,
\begin{align}
    g(\bm{r})= Re\left[\frac{V_k}{\rho_0}\sum_{\bm{k}}(S(\bm{k})-1)e^{-i\bm{k}\cdot\bm{r}}\right],
    \label{eq:gDFT}
\end{align}
where $V_k = (1/L)^d$ is the discretization volume used when switching to a discrete Fourier transform. 
As a result, one may express the gradient of $g$ with respect to the position $\bm{r}_n$ of particle $n$ as
\begin{align}
    \frac{\partial g}{\partial \bm{r}_n} (\bm{x}) & = Re\left[\frac{V_k}{\rho_0}\sum_{\bm{k}}\frac{\partial S(\bm{k})}{\partial \bm{r}_n}e^{-i\bm{k}\cdot\bm{x}}\right].
\end{align}
The gradient of $\mathcal{L}_g$ can then be expressed as
\begin{align}
    \frac{\partial \mathcal{L}_g}{\partial \bm{r}_n} = \frac{2 V_k}{\rho_0}Re\left[\sum_{\bm{k}}\frac{\partial S(\bm{k})}{\partial \bm{r}_n}\sum_{\bm{x} \in \mathcal{R}}W_g(\bm{x})\left(g(\bm{x})-g_0(\bm{x})\right)e^{-i\bm{k}\cdot\bm{x}}\right].
\end{align}
Finally, one may define
\begin{align}
G(\bm{k}) \equiv 2\sum_{\bm{r} \in \mathcal{R}} W_g(\bm{x})\left(g(\bm{x})-g_0(\bm{x})\right)e^{-i\bm{k}\cdot\bm{x}}
\end{align}
such that
\begin{align}
    \frac{\partial \mathcal{L}_g}{\partial \bm{r}_n} = \frac{V_k}{\rho_0}Re\left[\sum_{\bm{k}}\frac{\partial S(\bm{k})}{\partial \bm{r}_n}G(\bm{k})\right].
\end{align}
All in all, introducing the weight $c_n$ of each point again,
\begin{align}
    \frac{\partial \mathcal{L}_g}{\partial \bm{r}_n} = - \frac{2V_k }{\rho_0 N}Re\left[\sum_{\bm{k}}i\bm{k}c^\dagger_n\rho(\bm{k})G(\bm{k})e^{-i\bm{k}\cdot\bm{r}_n}\right]. 
    \label{eq:dgDFT}
\end{align}
This last expression can be evaluated using two Fourier transforms. As Equations \ref{eq:gDFT} and \ref{eq:dgDFT} can be evaluated using regular FFTs, the loss minimization in $g(\bm{r})$ may thus be performed in $\mathcal{O}(N \log N)$ time as well (per iteration).

\begin{figure*}
    \centering
    \includegraphics[width=0.96\textwidth]{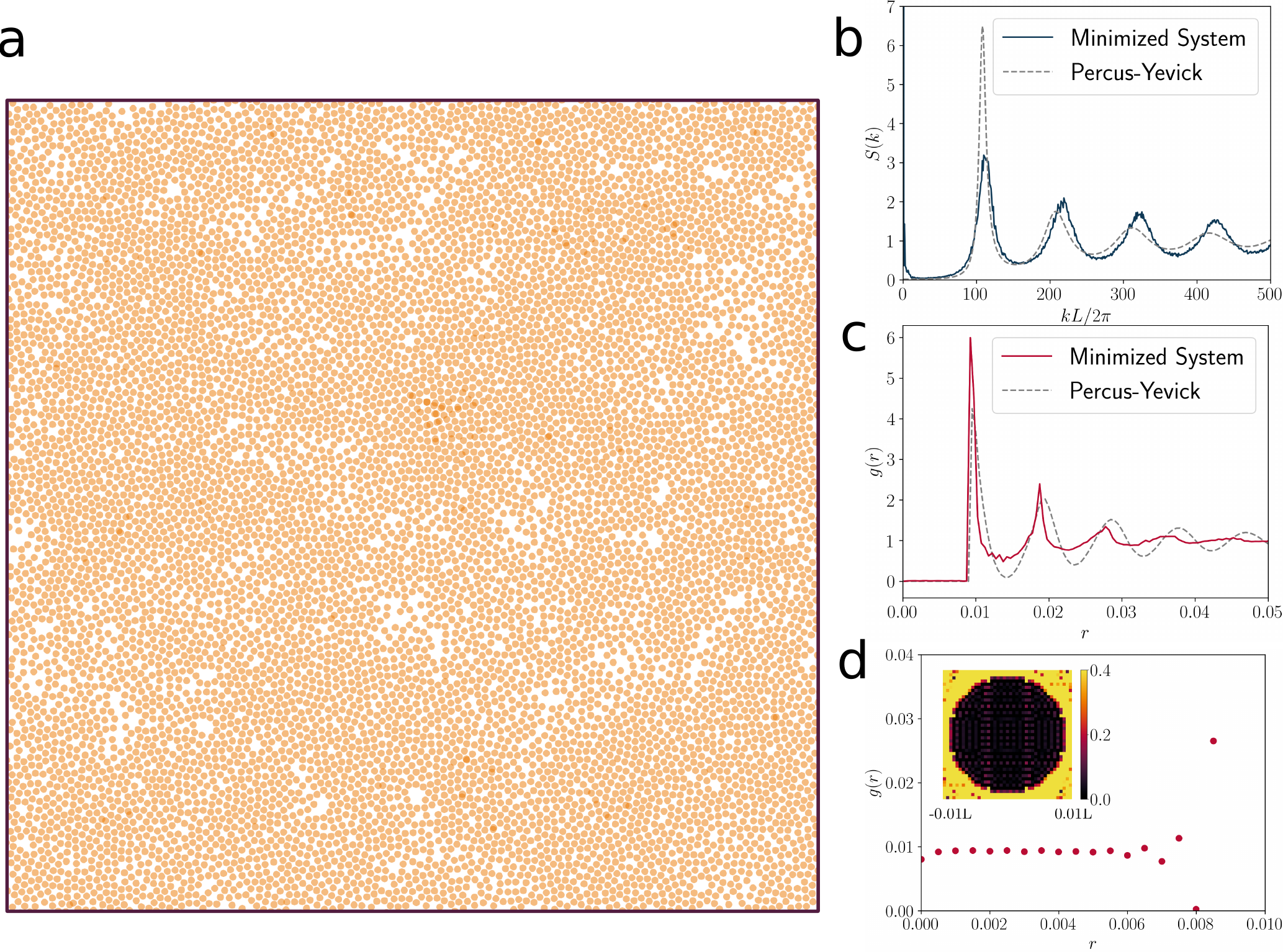}
    \caption{\textbf{Constraining the pair correlation function $g(r)$ using modified NUwU.}
    (a) A point pattern obtained by constraining $g_0(r<\sigma)=0$ for an exclusion diameter $\sigma$ such that $\phi=0.7$.
    Disks are drawn with diameter $\sigma$ with transparency to highlight overlaps remaining in the final system.
    (b) The radial structure factor of the point pattern (solid line) compared to the structure factor of the Percus-Yevick solution for ideal hard disks (dashed line).
    (c) The pair correlation function of the point pattern calculated using radial binning, compared with the Percus-Yevick prediction.
    (d) A zoomed in portion of the pair correlation function used during minimization (calculated using the Fourier transform of the structure factor).
    Due to the Fourier transform using only a finite number of modes, the $g(r)$ that is used to minimize exhibits aliasing, resulting in imperfect minimization.
    }
    \label{fig:grconstraint}
\end{figure*}

In Extended Data Fig.~\ref{fig:grconstraint}, we show an example output of this strategy, using $g_0(r\leq \sigma) = 0$ as a target and $2001 \times 2001$ Fourier modes in an $N= 10000$ point pattern, with $\sigma$ an exclusion diameter.
The result does exhibit a hard-disk-like structure factor, panel $(b)$, and pair correlation function, panel $(c)$, although it is less ordered than equilibrium hard disks, as highlighted by the comparison to Percus-Yevick curves.
This indicates that our optimization does \textit{not} sample hard disk configurations uniformly like equilibrium simulations would.
Note however that, due to the finite number of modes, the evaluation of $g$ via an inverse Fourier transformed displays aliasing errors: a perfect step function cannot be represented with any finite number of Fourier modes, leading to rippling~\cite{Pharr2018}.
In panel $(d)$, we highlight this by showing a zoom onto the early values of the final $g(r)$, which is not exactly zero (this can also be seen in panel $(a)$ in the form of a small number of overlaps).
In the inset, we show the corresponding zoom onto the center of the $2d$ $g(\bm{r})$, which shows that these overlaps actually lie on a discrete grid due to aliasing.
Note that this issue is likely to be worse in sharp features like the hardcore repulsion we show here, but should not be as much of an issue when imposing smoother features, $i.e.$ features with less high-frequency content (ideally band-limited features).

\textit{Additional data --}
We here show extended figures from the main text.

\begin{figure*}
    \centering
    \includegraphics[width=0.96\textwidth]{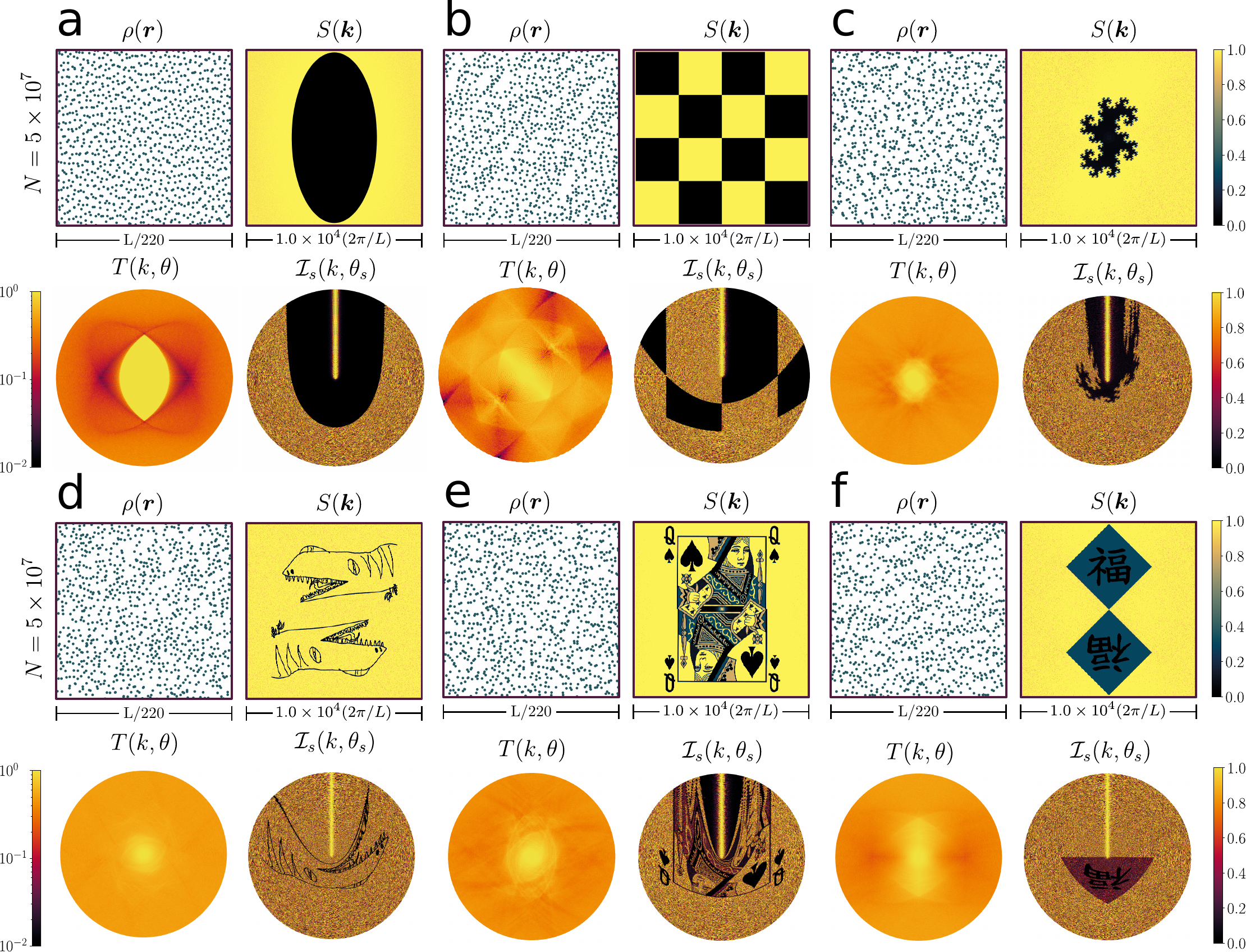}
    \caption{\textbf{Additional examples of output point patterns and associated structure factors and transmission spectra.} All point patterns contain a total of $N=5\times 10^7$ points.
    (a) Elliptical stealthy hyperuniform.
    (b) Checkerboard structure factor.
    (c) A twin dragon fractal.
    (d) A sketch of a Yangchuanosaurus by Leonardo S. Martiniani.
    (e) A queen of spades playing card.
    (f) The Chinese character `fu' on a square.}
    \label{fig:Bonus}
\end{figure*}

In Extended Data Fig.~\ref{fig:Bonus}, which may be seen as an extension of Fig.~\ref{fig:Sketch}$(c)-(e)$ we show a few more examples of outputs of the NUwU variant.
Panel $(a)$ was obtained by imposing an elliptical hyperuniform domain, which may be used to create orientation-dependent colors in backscattered light~\cite{Leseur2016}.
Panel $(b)$ was obtained by imposing a checkerboard of alternated ones and zeros around the origin of Fourier space, showing a minimal example of non-radial and anisotropic structure factor.
Panel $(c)$ was obtained by imposing a twin dragon fractal into the structure factor, showing that fine resolution may be achieved.
This point is pushed even further in panels $(d)-(f)$, in which we respectively impose a line drawing, a playing card, and an ideogram. 

\begin{figure*}
    \centering
    \includegraphics[width=0.96\textwidth]{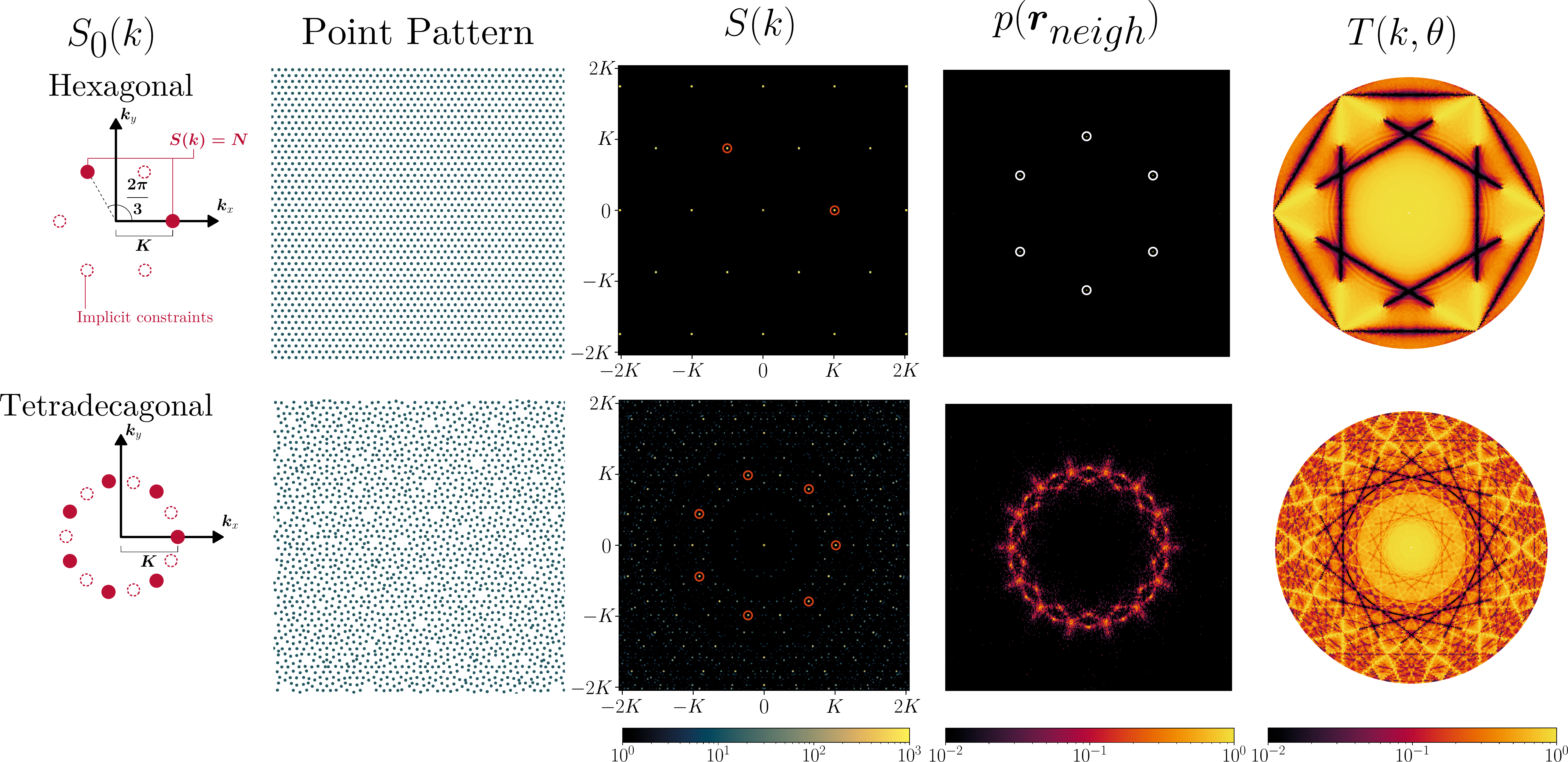}
    \caption{\textbf{Additional 2d systems with special symmetries generated using NUwNU.}
    Top: Hexagonal order, resulting in a typical triangular lattice.
    As the peaks in the nearest neighbor plot are so narrow, we indicate their locations by circling them in green.
    Bottom: Tetradecagonal order.
    }
    \label{fig:quasiquasicrystals_ext2d}
\end{figure*}

In Extended Data Fig.~\ref{fig:quasiquasicrystals_ext2d}, we show two additional discrete rotational symmetries we imposed in $2d$: hexagonal and tetradecagonal order, resulting respectively in a triangular lattice and in a quasicrystalline structure with 14-fold symmetry.

\begin{figure*}
    \centering
    \includegraphics[width=0.96\textwidth]{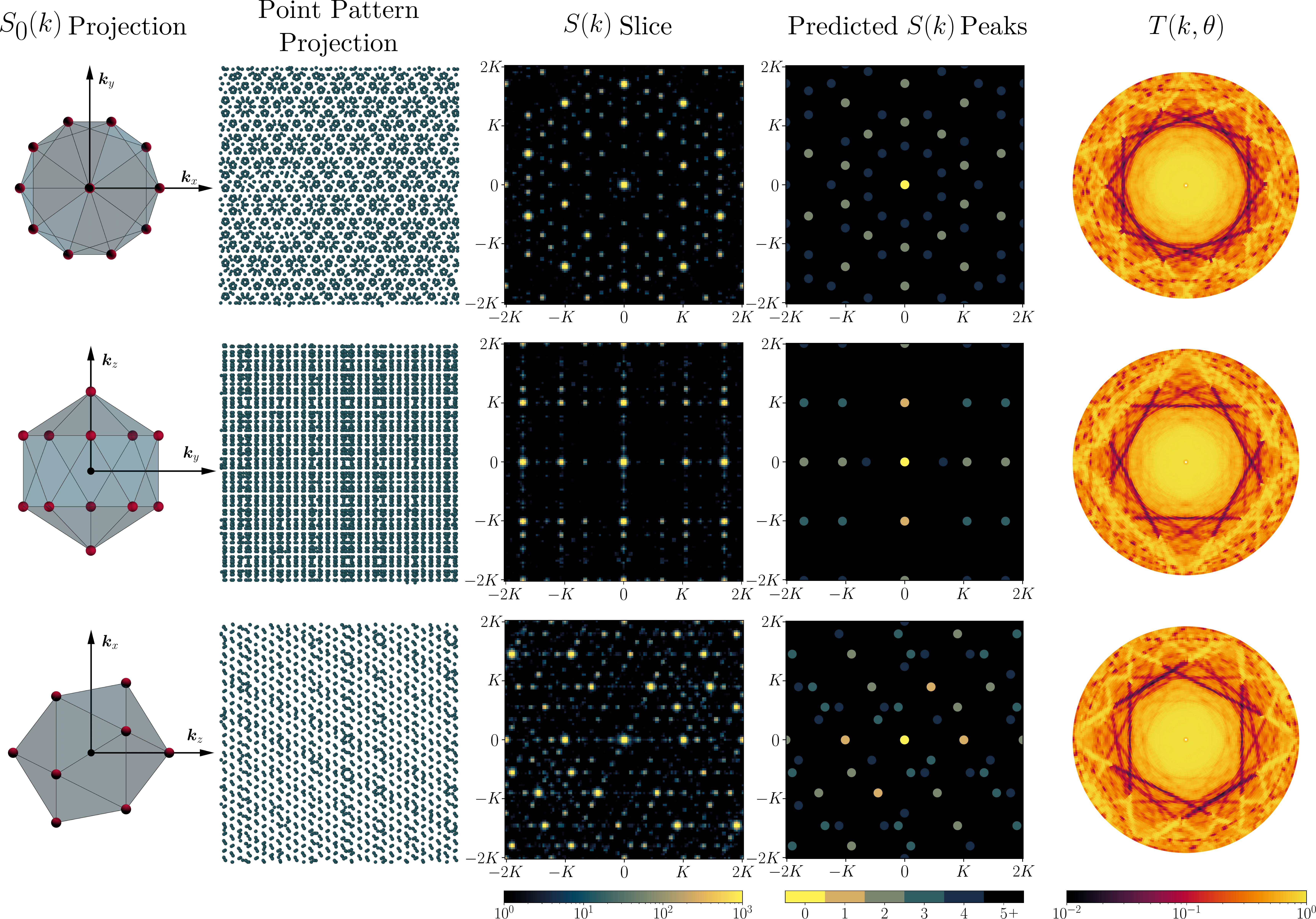}
    \caption{\textbf{Icosahedral order in a system generated using NUwNU.}
    The first column depicts the 3d constrained peaks $S_0(k)$ in various projections.
    The second column depicts the generated 3d point pattern in various projections.
    The third column depicts a slice of the 3d structure factor through the origin along the appropriate directions.
    The fourth column plots predicted peak locations on the slice calculated as linear combinations of the constrained vectors in $S_0(k)$.
    The predicted peaks are colored by the minimum number of constrained vectors required to form that linear combination.
    The fifth column depicts the transmission spectrum calculated using the Ewald sphere method.
    }
    \label{fig:quasiquasicrystals_extico}
\end{figure*}

\begin{figure*}
    \centering
    \includegraphics[width=0.96\textwidth]{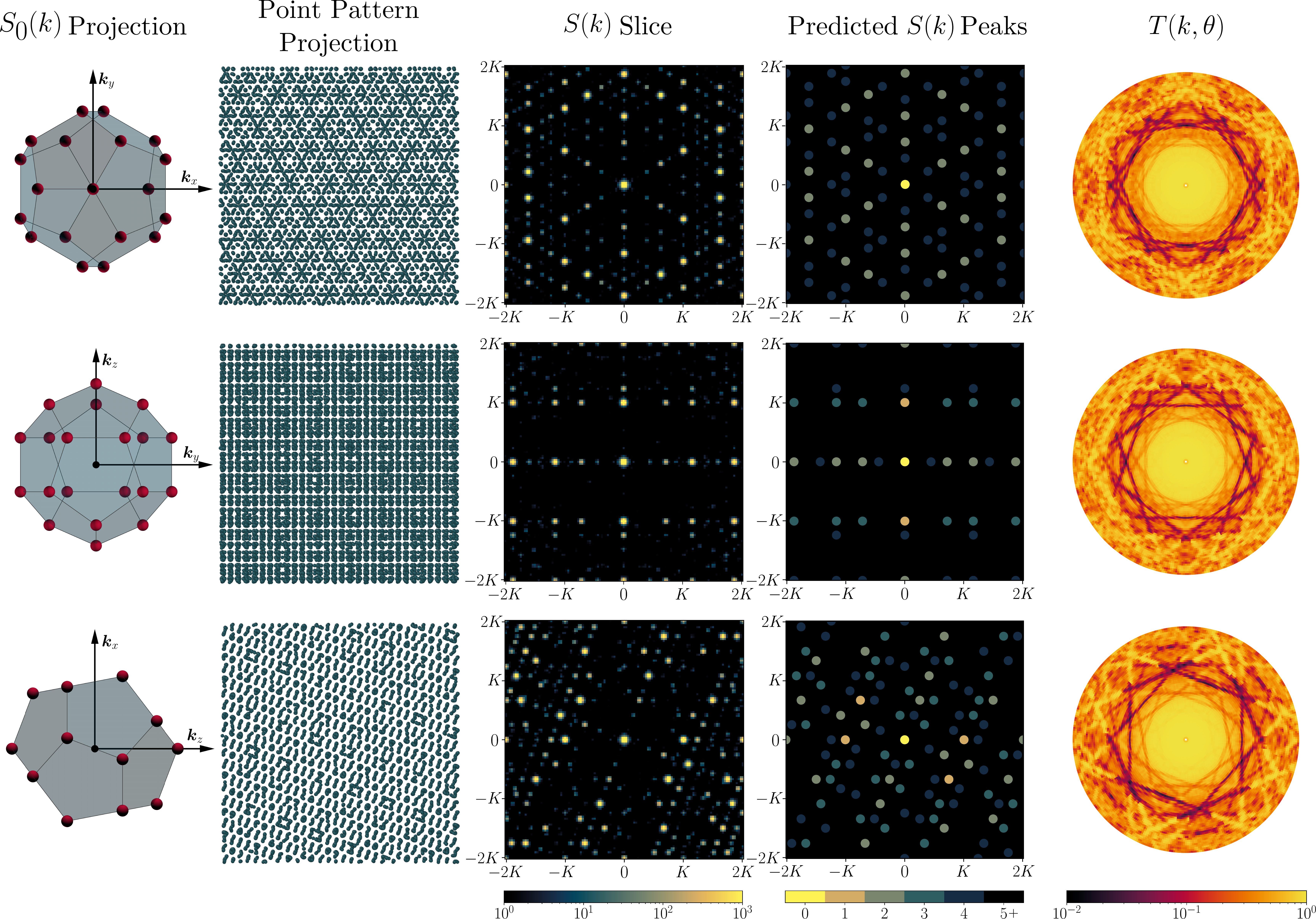}
    \caption{\textbf{Dodecahedral order in a system generated using NUwNU.}
    The first column depicts the 3d constrained peaks $S_0(k)$ in various projections.
    The second column depicts the generated 3d point pattern in various projections.
    The third column depicts a slice of the 3d structure factor through the origin along the appropriate directions.
    The fourth column plots predicted peak locations on the slice calculated as linear combinations of the constrained vectors in $S_0(k)$.
    The predicted peaks are colored by the minimum number of constrained vectors required to form that linear combination.
    The fifth column depicts the transmission spectrum calculated using the Ewald sphere method.
    }
    \label{fig:quasiquasicrystals_extdodec}
\end{figure*}

In Extended Data Fig.~\ref{fig:quasiquasicrystals_extico}, we show additional data on the icosahedral quasicrystal of Fig.~\ref{fig:quasiquasicrystals}.
Each row is obtained by changing the plane onto which real-space positions are projected (second column) and the slicing plane used both in reciprocal space (third column) and in the transmission plots (right-most column).
To complete Fig.~\ref{fig:quasiquasicrystals}, in lieu of nearest-neighbor histogram, the fourth column represents the location of the integer linear combinations of the peaks we imposed in the slice of Fourier space we are considering in each row.
In each direction, the FReSCo output matches the expected results for an icosahedral quasicrystal~\cite{Engel2015}.
We show the same data for our dodecahedral quasicrystal in Extended Data Fig.~\ref{fig:quasiquasicrystals_extdodec}.

\end{document}


\preprint{APS/123-QED}

\title{Supplementary Information for ``Fast Generation of Spectrally-Shaped Disorder''}

\author{Aaron Shih}
\thanks{Equal contribution.}
\affiliation{\nyucourant}
\affiliation{\nyuphysics}
\author{Mathias Casiulis}
\thanks{Equal contribution.}
\affiliation{\nyuphysics}
\affiliation{\nyusimons}
\author{Stefano Martiniani}
\email{sm7683@nyu.edu}
\affiliation{\nyucourant}
\affiliation{\nyuphysics}
\affiliation{\nyusimons}

\date{\today}

\maketitle

\renewcommand{\figurename}{FIG.}
\renewcommand{\thefigure}{S\arabic{figure}}
\renewcommand{\thetable}{S\arabic{figure}}
\newtagform{S}{(S})
\usetagform{S}

\begin{figure}
    \centering
    \includegraphics[width=0.96\textwidth]{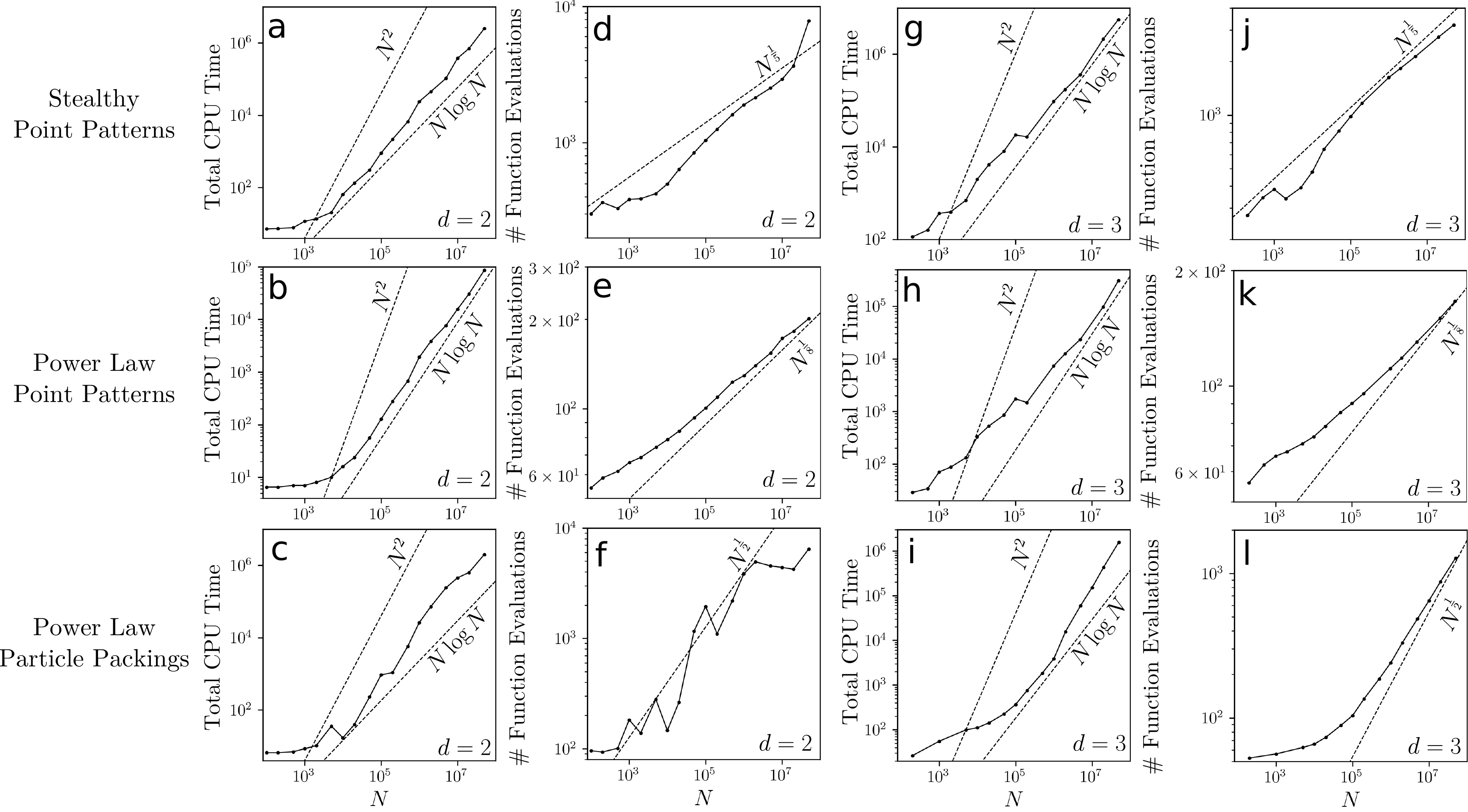} 
   
    \caption{\textbf{Performance benchmarks.} Total CPU time utilized for full minimization plotted against system size $N$, in a log-log scale in 2d $(a-c)$ and 3d $(g-i)$.
    Alongside, $(d-f),(j-l)$ are plots of the total number of function evaluations used to achieve the termination condition for minimization.
    Systems generated are stealthy disordered hyperuniform point patterns $(a,d),(g,j)$, power law hyperuniform point patterns $(b,e),(h,k)$, and power law hyperuniform disk/sphere packings $(c,f),(i,l)$.
    On each plot of CPU time, a reference line representing $\mathcal{O}(N\log N)$ scaling is plotted for comparison, as well as the best-case scaling per iteration of previous methods $\mathcal{O}(N^2)$ (note that this does not take into account the scaling of number of evaluations in previous methods).
    On each plot of the number of function evaluations, an overestimate of the large-$N$ scaling is plotted as a guide.
    }
    \label{fig:Benchmarks}
\end{figure}

\section{Benchmark of $\text{FReSCo}$}

We collect data for full minimization times of stealthy disordered hyperuniform point patterns, power law hyperuniform point patterns, and power law hyperuniform disk/sphere packings in 2d and 3d, all starting from independent Poisson point processes (Fig. \ref{fig:Benchmarks}).
Point patterns were minimized to a root mean square error of $10^{-39}$ on the gradient, while particle packings were minimized to a root mean square error of $10^{-10}$ on the gradient.
All systems demonstrate large-$N$ scaling very near $N\log N$ for total times (dashed black lines) as the system size increases.
Note that this scaling is observed for full minimization procedures, which is a stronger result than the scaling of individual iterations, that is $N \log N$ by definition.

To emphasize this point, we also show the number of required iterations as a function of $N$. Since a single minimization step is guaranteed to scale like $\mathcal{O}(N\log N)$, the overall scaling is that of the number of function evaluations multiplied by $N\log N$.
Therefore, we want to show that the number of function evaluations, $n_{ev}$, is sublinear in $N$ to demonstrate that FReSCo overall outperforms previous $N^2$ or $N^3$ algorithms.
We show that the scalings are systematically sublinear for all tested systems, with scalings $n_{ev} \lessapprox {o}(N^{\frac{1}{5}})$ for stealthy hyperuniform point patterns, $n_{ev} \lessapprox {o}(N^{\frac{1}{8}})$ for power law hyperuniform point patterns, and $n_{ev} \lessapprox {o}(N^{\frac{1}{2}})$ for power law hyperuniform particle packings.
All minimizations were here performed on a single CPU node, parallelized over multiple cores.
The number of cores used for all systems of Fig.~\ref{fig:Benchmarks} are given in detail in Table~\ref{tab:CPUUsage}.

\begin{table}
\centering
\begin{tabular}{ |c|c|c|c|c|} 
 \hline
 $N$ & \# cores & \# cores & \# cores & \# cores \\
  & (2d Points)& (2d Packings)& (3d Points)& (3d Packings)\\
 \hline
 \hline
 $10^2$         &  2 &  2 &  2 &  2\\
 $2\times 10^2$ &  2 &  2 &  2 &  2\\
 $5\times 10^2$ &  2 &  2 &  2 &  2\\
 \hline
 $10^3$         &  2 &  2 &  4 &  4\\
 $2\times 10^3$ &  2 &  2 &  4 &  4\\
 $5\times 10^3$ &  2 &  2 &  4 &  4\\
 \hline
 $10^4$         &  4 &  4 &  8 &  4\\
 $2\times 10^4$ &  4 &  6 &  8 &  4\\
 $5\times 10^4$ &  8 &  8 &  8 &  4\\
 \hline
 $10^5$         & 12 & 12 & 12 &  4\\
 $2\times 10^5$ & 12 & 16 & 12 &  4\\
 $5\times 10^5$ & 12 & 24 & 12 &  4\\
 \hline
 $10^6$         & 24 & 32 & 24 &  4\\
 $2\times 10^6$ & 24 & 32 & 24 &  4\\
 $5\times 10^6$ & 24 & 48 & 24 &  6\\
 \hline
 $10^7$         & 24 & 48 & 48 &  8\\
 $2\times 10^7$ & 24 & 48 & 48 & 12\\
 $5\times 10^7$ & 48 & 48 & 48 & 24\\
 \hline
\end{tabular}
\caption{Number of cores used for each system shown in the benchmark of Fig.~\ref{fig:Benchmarks}.
\label{tab:CPUUsage}
}
\end{table}

\section{Size matters: $S$ and number fluctuation behaviour in small systems}

\begin{figure}
    \centering
    \includegraphics[width=0.96\textwidth]{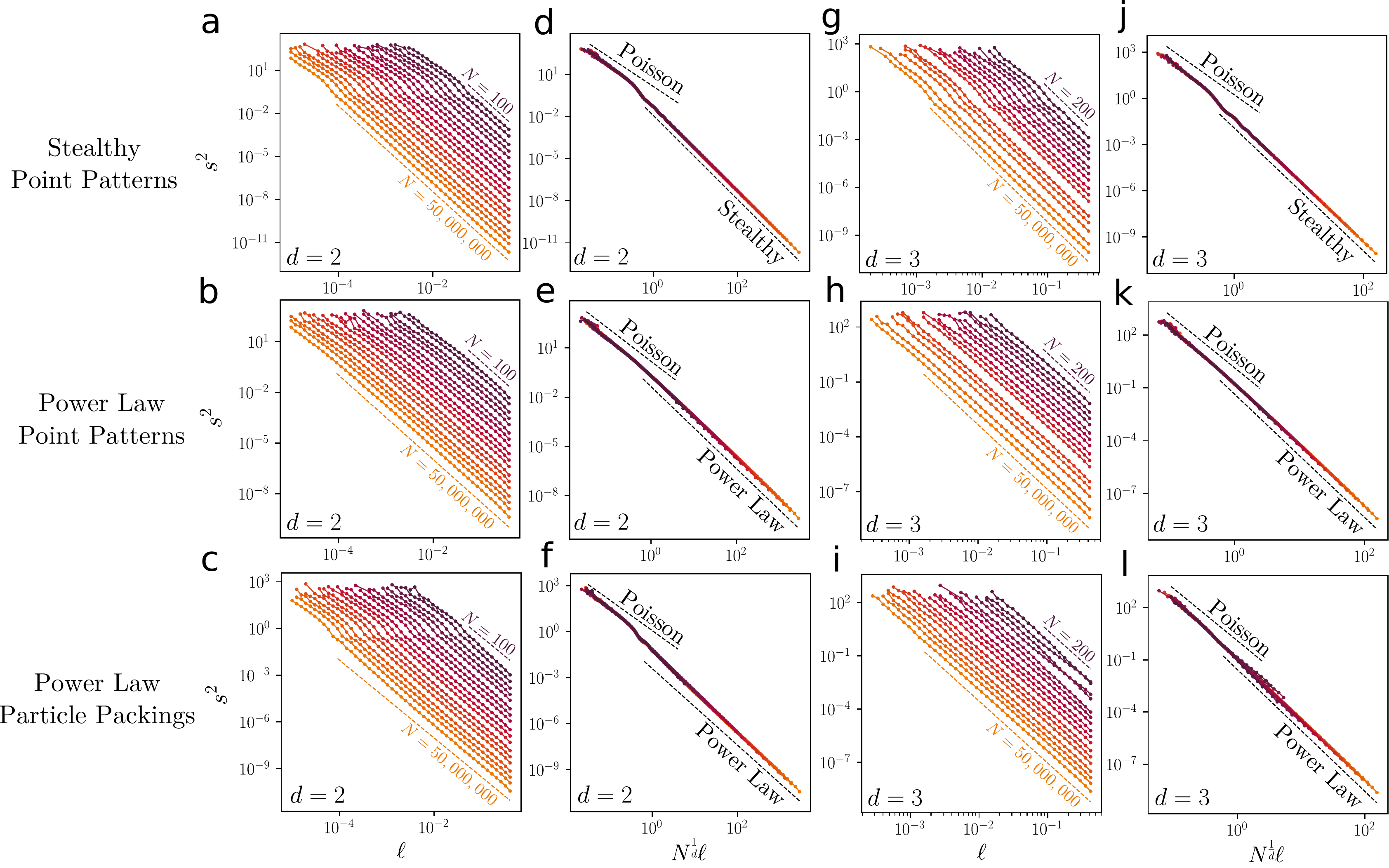} 
   
    \caption{\textbf{Hyperuniform scaling across system sizes in 2d and 3d.}
    Reduced number variance plots for stealthy disordered hyperuniform point patterns (top), power law hyperuniform point patterns (middle), and power law hyperuniform sphere packings (bottom).
    Plots on the left are plotted against the radius of the sample circle as a fraction of box length, showing a much narrower range of hyperuniform scaling in smaller systems.
    Plots on the right depict the data rescaled by $N^{\frac{1}{d}}$, showing the Poissonian scaling at short length scales transitioning to hyperuniform scaling at large length scales.}
    \label{fig:Variances}
\end{figure}

From each of the systems minimized for the benchmark in Fig. \ref{fig:Benchmarks}, we evaluate the reduced number variance $s^2 \equiv \left\langle N^2\right\rangle / \left\langle N\right\rangle^2 - 1$, with averages performed over a set of circular/spherical sample volumes with radii $\ell \in [10^{-5}L,0.5L]$ for 2d systems (Fig.~\ref{fig:Variances} a-f) and $\ell \in [10^{-4}L,0.5L]$ for 3d systems (Fig.~\ref{fig:Variances} g-l).
We observe hyperuniform scaling for nearly four decades in 2d and at least two decades in 3d for systems containing $N=5\times 10^7$ points.
In contrast, for our smallest systems, we only observe about one decade of hyperuniform scaling in 2d ($N=100$) and less than one decade in 3d ($N=200$).
We note that the power law scaling in the hyperuniform disk packing in Fig.~\ref{fig:Variances}c,f appears to exhibit a broader crossover region, \text{i.e.} that follows neither of the asymptotic scalings, between the Poissonian scaling and the large-scale scaling imposed by the power law in $S(k)$.
We believe this to be an effect of the treatment of the points as particles with pair repulsion at a somewhat high packing density $\phi=0.6$, as we do not observe any deviation in the corresponding 3d case for a comparatively dilute packing fraction $\phi=0.25$, Fig. ~\ref{fig:Variances} $(i,l)$.

\section{Robustness of stealthy hyperuniformity against noise\label{sec:Robustness}}

\begin{figure}
    \centering
    \includegraphics[width=0.96\textwidth]{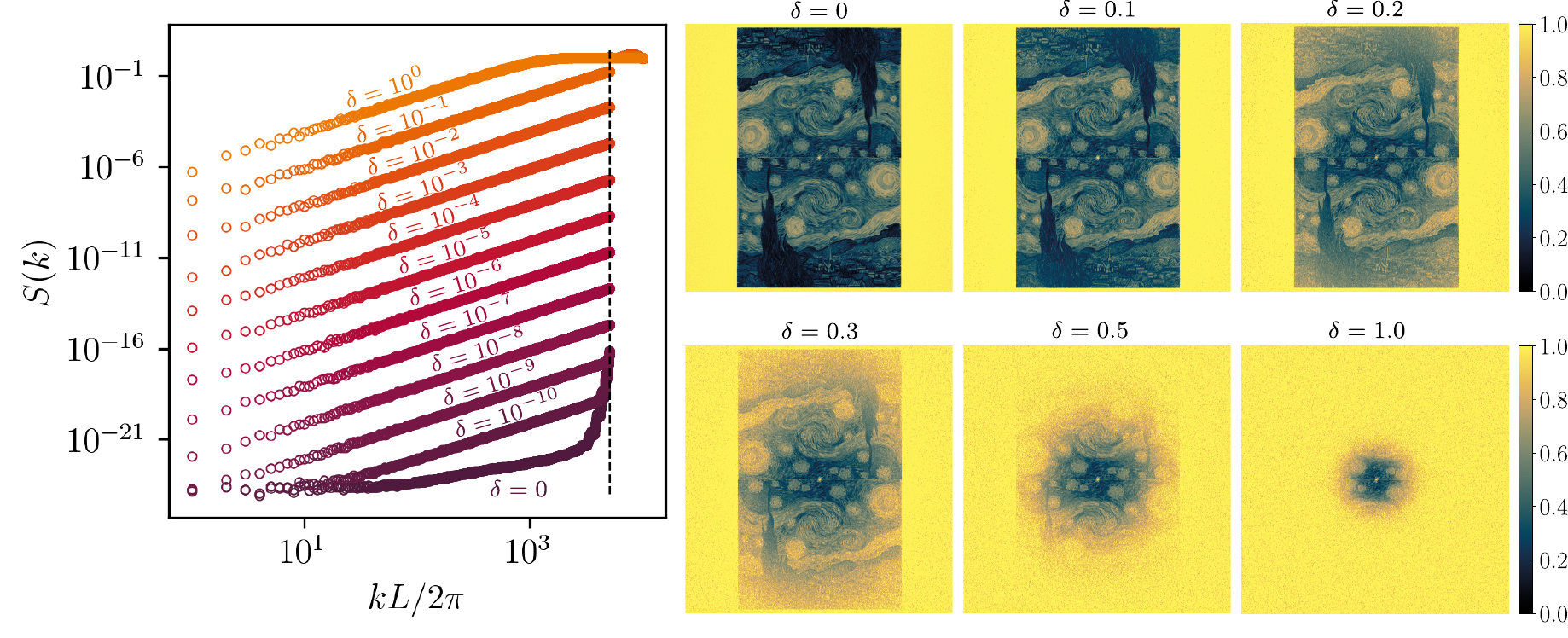} 
   
    \caption{\textbf{Effect of Gaussian random noise on structure factor.}
    Left: Angular averaged structure factors (log scale) of a $N=5\times 10^7$ stealthy hyperuniform point pattern subjected to varying degrees of Gaussian noise ($\sigma = L\delta/N^{\frac{1}{d}}$). Right: 2d structure factor (linear scale) of a  $N=5\times 10^7$ starry night structure factor point pattern subjected to varying degrees of Gaussian noise ($\sigma = L\delta/N^{\frac{1}{d}}$).}
    \label{fig:Noise}
\end{figure}

One of the main motivations for using disordered material structures in applications such as photonics where crystalline structures already perform well is that disordered structures can be more resilient to defects.
Here, we take minimized point patterns ($N=5\times 10^7$) and displace each point using a Gaussian random normal distribution of mean $\mu = 0$ and standard deviation $\sigma = L\delta /N^{\frac{1}{d}}$, so that $\delta$ is a fraction of the average inter-particle distance (Fig. \ref{fig:Noise}).
Doing so, one expects the new low-$k$ structure factor to be, at every point, the maximum of the power spectrum of the noise, which here grows like $k^2$ and of the original structure factor~\cite{Gabrielli2004}.
In the stealthy hyperuniform case, we show that, although we are able to minimize the structure factor down to a magnitude of $\sim 10^{-24}$ at its minimum value, it only takes a noise corresponding to $\delta=10^{-9}$ to raise that value.
For the case of a $N=5\times 10^7$ point pattern, this $\delta$ value results in a standard deviation of $\sigma \approx 1.4\times10^{-13}L$, which implies that the noise level would be equivalent to sub-Angstrom displacements in a $L=1$ km size device with typical inter-particle spacing $d \approx 10$ cm.
In practice, the difference in $S(k)$ magnitude between, say $10^{-6}$ and $10^{-24}$ is irrelevant for any realistic fabrication process.
Furthermore, most practical applications likely do not need such high precision in $S(k)$ in practice.
We demonstrate this by showing the effect of Gaussian noise on a highly-detailed structure factor (in linear intensity scale), namely a Starry Night~\cite{StarryNight} structure factor point pattern ($N=5\times 10^7$).
We observe that long-range correlations are well-preserved up to high values of the noise, while short-range correlations are washed out earlier on.
The effect of noise for $\delta=0.1$ is only somewhat noticeable, corresponding to a standard deviation of $\sigma \approx 1.4\times10^{-5}L$.
This implies that a $L=1$ mm device with $d \approx 100$ nm distance between particles would be resilient to $10$-nanometer-scale defects.

\section{Single-scattering approximation}

\subsection{Ewald circle construction \label{sec:SS_theory}}

We here remind of how to reach a single-scattering level steady-state description of system of small dielectric scatterers analytically, and how the result relates to the Ewald sphere construction.
A monochromatic electric field $\bm{E}(\bm{r}; \omega)$ at pulsation $\omega$ propagating in $3d$ space through a relative dielectric constant field $\varepsilon(\bm{r})$ obeys the monochromatic Maxwell-Helmholtz equation~\cite{MorseFeshbachI,Carminati2021},
\begin{align}
    \bm{\nabla}\times\bm{\nabla}\times \bm{E}(\bm{r}; \omega) - \frac{\omega^2}{c^2} \varepsilon(\bm{r}; \omega) \bm{E}(\bm{r};\omega) = i \mu_0 \omega \bm{j}_{ext}(\bm{r};\omega),
\end{align}
where $c$ is the speed of light in vacuum, $\mu_0$ is the magnetic permeability of vacuum and $\bm{j}_{ext}$ is the externally imposed charge current density, that results in light sources.
Following usual conventions~\cite{Carminati2021}, we define the incident field $\bm{E}_{inc}(\bm{r})$ as the solution of the wave equation for the same sources, but in vacuum,
\begin{align}
    \bm{\nabla}\times\bm{\nabla}\times \bm{E}_{inc}(\bm{r}; \omega) - \frac{\omega^2}{c^2}\bm{E}_{inc}(\bm{r};\omega) = i \mu_0 \omega \bm{j}_{ext}(\bm{r};\omega),\label{eq:incidentE}
\end{align}
as well as the scattered field $\bm{E}_s = \bm{E} - \bm{E}_{inc}$.
The latter obeys the equation
\begin{align}
    \bm{\nabla}\times\bm{\nabla}\times \bm{E}_s(\bm{r}; \omega) - \frac{\omega^2}{c^2} \bm{E}_s(\bm{r};\omega) = \frac{\omega^2}{c^2} \delta\varepsilon(\bm{r}; \omega) \bm{E}(\bm{r};\omega), \label{eq:scatteredE}
\end{align}
in which we introduced the relative dielectric contrast $\delta\varepsilon = \varepsilon - 1$.
To go further, we introduce the \textit{dyadic Green's function}~\cite{Economou2006,Carminati2021}, $\overline{\overline{G}}_0(\bm{r},\bm{r'}; \omega)$ , associated to propagation in free space.
This $3\times 3$ rank-$2$ tensor is defined as the solution of the free-space Maxwell-Helmholtz equation when the source is replaced by a Dirac delta in each direction,
\begin{align}
    \bm{\nabla}\times\bm{\nabla}\times \overline{\overline{G}}_0(\bm{r},\bm{r'}; \omega) - \frac{\omega^2}{c^2}\overline{\overline{G}}_0(\bm{r},\bm{r'}; \omega) = \delta(\bm{r} - \bm{r}') \overline{\overline{I}},
\end{align}
with $\overline{\overline{I}}$ the identity tensor.
Defining the Green tensor enables to write any field propagating in vacuum as an integral equation over the source term.
For instance, the incident field defined in Eq.~\ref{eq:incidentE} verifies
\begin{align}
    \bm{E}_{inc}(\bm{r}; \omega) = i \mu_0 \omega \int d^3\bm{r}'\, \overline{\overline{G}}_0(\bm{r},\bm{r'}; \omega) \bm{j}_{ext}(\bm{r}'; \omega).
\end{align}
More importantly, writing the analogue equation for the scattered field, Eq.~\ref{eq:scatteredE}, leads to the Lippman-Schwinger equation~\cite{Carminati2021},
\begin{align}
    \bm{E}(\bm{r}; \omega) &= \bm{E}_{inc}(\bm{r}; \omega) + \frac{\omega^2}{c^2} \int d^3\bm{r}'\, \overline{\overline{G}}_0(\bm{r},\bm{r'}; \omega) \delta\varepsilon(\bm{r}'; \omega) \bm{E}(\bm{r}'; \omega).
    \label{eq:Lippman-Schwinger}
\end{align}
In this paper, we study propagation through media composed of $N$ discrete scatterers, each with homogeneous dielectric contrasts $\delta\varepsilon_i$, placed at positions $\bm{r}_i$, in a homogeneous medium that can be assumed to be vacuum, leading to
\begin{align}
    \bm{E}_s(\bm{r}; \omega) &= \frac{\omega^2}{c^2} \sum\limits_{i=1}^N \delta\varepsilon_i(\omega) \int_{\mathcal{V}_i} d^3\bm{r}'\, \overline{\overline{G}}_0(\bm{r},\bm{r'}; \omega)  \bm{E}(\bm{r}'; \omega),
\end{align}
with $\mathcal{V}_i$ the domain occupied by scatterer $i$.
Up until this point, all equations are exact.

Since the Lippman-Schwinger equation is implicit for $\bm{E}$, it may instead be expanded into a series in powers of $\bm{E_{inc}}$, known as the Born series~\cite{Carminati2021}.
In that series, the number of integrations indicates a number of scattering events so that, in particular, one can truncate the series at first order (perform the Born approximation) to describe single scattering:
\begin{align}
    \bm{E}_s(\bm{r}; \omega) &\approx \frac{\omega^2}{c^2} \sum\limits_{i=1}^N \delta\varepsilon_i(\omega) \int_{\mathcal{V}_i} d^3\bm{r}'\, \overline{\overline{G}}_0(\bm{r},\bm{r'}; \omega)  \bm{E}_{inc}(\bm{r}'; \omega),
\end{align}
which is now a closed-form integral equation for $\bm{E}_s$.
We then assume that each scatterer is smaller than a wavelength, $\omega a /(2 \pi c) \ll 1$ with $a$ the typical size of scatterers.
In this regime, the field component at $\omega$ is well approximated by the field at the center of the scatterer, so that
\begin{align}
    \bm{E}_s(\bm{r}; \omega) &\approx \frac{\omega^2}{c^2}  \sum\limits_{i=1}^N \left[ \int_{\mathcal{V}_i} d^3\bm{r}'\, \overline{\overline{G}}_0(\bm{r},\bm{r'}; \omega)\right] \delta\varepsilon_i(\omega)  \bm{E}_{inc}(\bm{r}_i; \omega) .
\end{align}
Finally, we assume that the scattered field is measured in the far field~\cite{Carminati2021} of the system, $r \ll L$ where $L$ is the linear size of the medium enclosing all scatterers, and $r \ll L^2/\lambda$.
In this regime, the Green's function of free space can be approximated by its far-field expression~\cite{MorseFeshbachI,Carminati2021},
\begin{align}
    \overline{\overline{G}}_0(\bm{r},\bm{r}'; \omega) \approx \frac{e^{i k_0 r}}{4 \pi r} e^{i \bm{k}_{sca}\cdot\bm{r}'} \left( \overline{\overline{I}} - \hat{\bm{r}}\otimes\hat{\bm{r}} \right) ,\label{eq:FFG3d}
\end{align}
where we introduced $k_0 = \omega/c$ the wave-vector associated to $\omega$ in vacuum, and $\bm{k}_{sca} = k_0 \hat{\bm{r}}$ the scattered wave vector observed at $\bm{r}$ in direction $\hat{\bm{r}} = \bm{r}/r$.
As a result, assuming that scatterers are balls with volumes $V_i$, the scattered field can be rewritten
\begin{align}
    \bm{E}_s(\bm{r}; \omega) &\approx {k_0}^2 \frac{e^{i k_0 r}}{4 \pi r} \sum\limits_{i=1}^N \alpha_i(\omega) \left[ e^{i \bm{k}_{sca}\cdot\bm{r}_i} \left( \overline{\overline{I}} - \hat{\bm{r}}\otimes\hat{\bm{r}} \right) \bm{E}_{inc}(\bm{r}_i; \omega)\right],
\end{align}
with $\alpha_i = V_i \delta\varepsilon_i$ the polarizability of scatterer $i$.
Finally, we assume the incident field to be a plane wave with incident wave-vector $\bm{k}_{inc} = k_0 \hat{\bm{e}}(\theta)$, so that
\begin{align}
\bm{E}_{inc}(\bm{r}_i; \omega) = \bm{E}_0 e^{- i \bm{k}_{inc}\cdot \bm{r_i}}.
\end{align}

Altogether, introducing the scattered intensity $I_s \equiv \bm{E}_s \cdot \bm{E}_s^{\dagger}$, one finds
\begin{align}
    I_s(\bm{r}; \omega) \approx {k_0}^4 |\alpha(\omega)|^2 \frac{I_0^{\perp}(\hat{\bm{r}}; \omega)}{16 \pi^2 r^2} \left|\sum\limits_{i=1}^N  \left[ e^{i (\bm{k}_{sca} - \bm{k}_{inc})\cdot\bm{r}_i} \right] \right|^2,
\end{align}
where we defined the transverse component of the intensity of the plane wave
\begin{align}
    I_0^{\perp}(\hat{\bm{r}}; \omega) \equiv \left|\left( \overline{\overline{I}} - \hat{\bm{r}}\otimes\hat{\bm{r}} \right) \bm{E}_{0}\right|^2.
\end{align}
Therefore, recalling that the Fourier representation of a point pattern reads
\begin{align}
    \widehat{\rho}\, (\bm{k}) = \sum\limits_{i=1}^{N} e^{i \bm{k}\cdot \bm{r}_i} \nonumber
\end{align}
and introducing $\bm{q} \equiv \bm{k}_{sca} - \bm{k}_{inc} $, the far-field scattered intensity in the direction of $\bm{k}_{sca}$, in the limits of small scatterers and in the single-scattering regime, is explicitly proportional to the structure factor evaluated at $\bm{q}$,
\begin{align}
    I_s(\bm{r}; \omega) \approx N {k_0}^4 |\alpha(\omega)|^2 \frac{I_0^{\perp}(\hat{\bm{r}}; \omega)}{16 \pi^2 r^2} S(\bm{q}),
\end{align}
where, for a fixed incident field, $\bm{q} = k_0 (\hat{\bm{r}} - \hat{\bm{k}}_inc)$ is only a function of $\hat{\bm{r}}$ and $\omega$ (see Fig.2 of the main text for a sketch).
In particular, one may normalize this intensity, measured at some distance $R$, by the total intensity scattered over the whole sphere $\mathcal{S}_R$ with radius $R$, so as to recover a purely angular function,
\begin{align}
    \mathcal{I}_s(\hat{\bm{r}}; \omega) = \frac{I_0^{\perp}(\hat{\bm{r}}; \omega) S(\bm{q})}{\oint\limits_{\mathcal{S}_R} d^{2} \hat{\bm{r}}\, I_0^{\perp}(\hat{\bm{r}}; \omega) S(\bm{q})}.
\end{align}
In that expression, the $I_0^{\perp}$ dependence simply ensures that the measured field is transverse in the far field, and for any one choice of source leads to a slow modulation of the scattered intensity compared to the effect of $S$.
In practice, in many cases, only a scalar wave level of description is needed to approach even optical problems~\cite{Carminati2021}, so that polarization is simply averaged over, and
\begin{align}
    \mathcal{I}_s^{scalar}(\hat{\bm{r}}; \omega) \approx \frac{ S(\bm{q})}{\oint\limits_{\mathcal{S}_R} d^{2} \hat{\bm{r}}\, S(\bm{q})}.
\end{align}
This approximation is well justified for isotropic media, where the system is not expected to interact differently with any specific polarization.
In the main text, we integrate this intensity over a half-sphere $\mathcal{F}$ around the incoming wave-vector to generate the forward-scattered transmission $T$,
\begin{align}
        T(\bm{k}_{inc}) \underset{3d}{=} \frac{\int_{\mathcal{F} \setminus 0} S\left[k  \hat{\bm{e}}(\theta + \vartheta, \phi + \varphi) - k \hat{\bm{e}}(\theta, \phi )\right]d\vartheta d\varphi}{\oint_{\mathcal{S} \setminus 0}S\left[k  \hat{\bm{e}}(\theta + \vartheta, \phi + \varphi) - k \hat{\bm{e}}(\theta, \phi )\right]d\vartheta d\varphi},
\end{align}
where the integration is here explicitly written in terms of the deviation in the two spherical coordinates, and where we dropped the $R$ subscript from the integration domains since this quantity is independent of $R$.
The integration is performed numerically using a Fibonacci lattice~\cite{Gonzalez2010a} with $N_s = 2000$ points, evenly spaced over the sphere.

In the case of $2d$ point patterns, the derivation above remains mostly unchanged if one considers that scatterers are now slender, circular-section rods.
In that case, assuming that the incoming wave-vector $\bm{k}_{inc}$ lies in the plane perpendicular to rods, the propagation problem can be split into two independent problems: one for in-plane polarization components (transverse electric, or TE), and one for the out-of-plane polarization component (transverse magnetic, or TM).
The two propagators for these problems can be written in the far field (see for instance Refs.~\cite{MorseFeshbachI,Leseur2016,Monsarrat2022}) as
\begin{align}
    \overline{\overline{G}}_0^{TE}(\bm{r},\bm{r}'; \omega) &\approx \frac{i(1-i)}{4} \frac{e^{i k_0 r}}{\sqrt{\pi k_0 r}} e^{i k_0 \hat{\bm{r}}\cdot\bm{r}'} \left(\overline{\overline{I}} - \hat{\bm{r}}\otimes\hat{\bm{r}}\right),\label{eq:FFG2d_TE} \\
    G_{0}^{TM}(\bm{r},\bm{r}'; \omega) &\approx  \frac{i(1-i)}{4} \frac{e^{i k_0 r}}{\sqrt{\pi k_0 r}} e^{i k_0 \hat{\bm{r}}\cdot\bm{r}'}. \label{eq:FFG2d_TM}
\end{align}
Using the same derivation as in $3d$ with these expressions, defining $\theta$ the angle that parameterizes the direction of $\hat{\bm{r}}$, we find
\begin{align}
    I_s^{TE}(\bm{r}; \omega) &\approx \frac{N {k_0}^3 |\alpha(\omega)|^2}{8 \pi r} I_0^{TE,\perp}(\bm{r}; \omega) S(\bm{q}), \\
    I_s^{TM}(\bm{r}; \omega) &\approx \frac{N {k_0}^3 |\alpha(\omega)|^2}{8 \pi r} I_0^{TM}(\omega) S(\bm{q}),
\end{align}
where 
\begin{align}
    I_0^{TE,\perp} &\equiv \left| \left( \overline{\overline{I}} - \hat{\bm{r}}\otimes\hat{\bm{r}} \right) \bm{E}_0^{TE}  \right|^2, \\ 
    I_0^{TM} &\equiv \left| \bm{E}_0^{TM}  \right|^2.
\end{align}
Notice that, by construction, the TM mode is always transverse, so that no modulation by a projection is needed.
All in all, for $2d$ systems, the scattered intensities normalized by the total scattered intensity on the circle $\mathcal{C}_R$ read
\begin{align}
    \mathcal{I}_s^{TE}(\theta; \omega) &\approx \frac{I_0^{TE,\perp}(\theta; \omega) S(\bm{q})}{\oint\limits_{\mathcal{C}_R} d \theta \, I_0^{TE,\perp}(\theta; \omega) S(\bm{q})}, \\
    \mathcal{I}_s^{TM}(\theta; \omega) &\approx \frac{ S(\bm{q})}{\oint\limits_{\mathcal{C}_R} d \theta \, S(\bm{q})}.
\end{align}
Notice that the TM mode behaves like a scalar wave, as expected from the out-of-plane polarization.
In the main text, we integrate the $TM$ expression over the full forward half-circle $\mathcal{F}$ to generate the forward-scattered transmission $T$,
\begin{align}
    T(\bm{k}_{inc}) \underset{2d}{=} \frac{\int_{\mathcal{F} \setminus 0} S\left[k  \hat{\bm{e}}(\theta + \vartheta) - k \hat{\bm{e}}(\theta )\right]d\vartheta}{\oint_{\mathcal{C} \setminus 0}S\left[k  \hat{\bm{e}}(\theta + \vartheta) - k\hat{\bm{e}}(\theta )\right]d\vartheta},
\end{align}
where $\vartheta$ is the deviation angle.
The integration is performed using $N_s = 360$ points evenly spaced over the circle.

\subsection{Comparison between Born approximation and Ewald circle results \label{sec:SS_numerics}}

To justify the use of the Ewald circle construction, we here show that it yields the same results as computing the intensity at a finite but long range in the Born approximation.
In the latter, we put the circle-cut system with diameter $L$ (see Fig. 2$(a)$ of the main text) such that its center lies at $\bm{r} = \bm{0}$, and set the source to be a perfectly collimated Gaussian beam~\cite{Carminati2021},
\begin{align}
\bm{E}_{inc}(\bm{r}; \omega) = \bm{E}_0 e^{- i \bm{k}_{inc}\cdot\bm{r}} e^{-r^2/w^2},  
\end{align}
with $w = 10 L$ so as to approximate a plane wave but still have a finite-width source.
We set the refractive index of scatterers to be $n = 1.5$, their radius to be such that the packing fraction is $0.1$, and assume the surrounding medium to be vacuum.
Furthermore, in order to approximate the far-field regime, we evaluate the scattered field within the Born approximation on a sphere with radius $R_{meas} = 10L$, so that $kR_{meas} \gg 1$ at all tested values.

\begin{figure}
    \centering
    \includegraphics[width=0.96\textwidth]{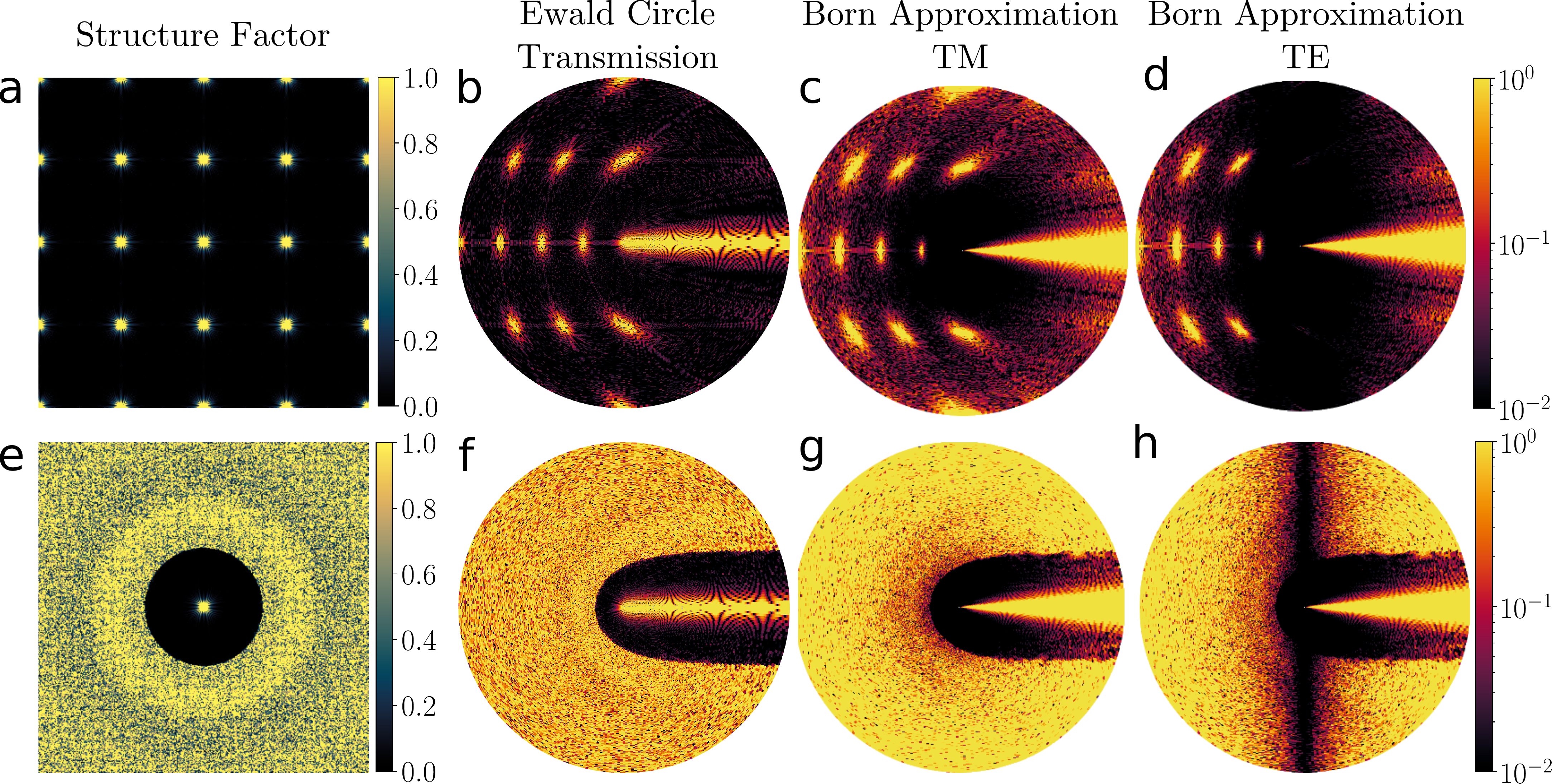} 
    \caption{\textbf{Comparison between Born approximation and Ewald circle.}
    Top line: example of an $N = 10000$ square lattice system.
    $(a)$ Structure factor, $(b)$ Ewald circle output for the normalized scattered intensity $\mathcal{I}_s$ when $\bm{k} = k \hat{\bm{e}}_x$, $(c)-(d)$ Corresponding Born approximation results for a TM, $\bm{E}_0 = E_0 \hat{\bm{e}}_z$ and a TE, $\bm{E}_0 = E_0 \hat{\bm{e}}_y$ beam, respectively.
    For $(b)-(d)$, results are plotted up to $kL/2\pi = 200$.
    Bottom line: same plots for a stealthy hyperuniform system.
    }
    \label{fig:BornEwald}
\end{figure}

Results on a couple of example systems are shown in Fig.~\ref{fig:BornEwald}.
In Fig.~\ref{fig:BornEwald}$(a)-(d)$, we show the example of a square-lattice structure of $N=10000$ points, the structure factor of which is shown in panel $(a)$.
We show that the Ewald circle construction, Fig.~\ref{fig:BornEwald}$(b)$, matches the Born approximation result in TM polarization, Fig.~\ref{fig:BornEwald}$(c)$, even considering a finite-size system, a finite-distance measurement, and a finite-width source, as well as arbitrary values for the refractive index and scatterer radius.
The only notable difference is a slow $k$ dependence of the width of the forward-scattering feature near $\vartheta = 0$ corresponding to the central peak, which stems from the fact that the measurement is not strictly performed in the far field.
Furthermore, notice that the TE polarization results obtained through the Born approximation, Fig.~\ref{fig:BornEwald}$(d)$, are simply the result of modulating the TM or Ewald-circle results by a projection of the incoming polarization onto the transverse direction in each direction, as expected.

Finally, in Fig.~\ref{fig:BornEwald}$(e)-(h)$, we also show results for a $2d$ disordered stealthy hyperuniform system with $N = 10000$, $\chi = 0.4$, leading to a disk $S(k< K) = 0$ with $K \approx 71$.
Once again, the Ewald construction captures single-scattering properties as expected.

\subsection{Effect of detection width on perceived transmission gaps}

Throughout the text, we show results for $T$ obtained by integrating over a full half-circle or half-sphere.
In practice, any experimental result would be performed using a smaller detection width, so that the integration domain $\mathcal{F}$ in the definition of $T$ should be replaced by a finite angular region.
We argue that this has a particular importance in the context of stealthy hyperuniform systems.
To do so while preserving axial symmetry in the detection, we introduce a variant of $T$,
\begin{align}
        T_\psi(\bm{k}_{inc}) &\underset{3d}{=} \frac{\int_{\mathcal{F}_\psi \setminus 0} S\left[k  \hat{\bm{e}}(\theta + \vartheta, \phi + \varphi) - k \hat{\bm{e}}(\theta, \phi )\right]d\vartheta d\varphi}{\oint_{\mathcal{S} \setminus 0}S\left[k  \hat{\bm{e}}(\theta + \vartheta, \phi + \varphi) - k \hat{\bm{e}}(\theta, \phi )\right]d\vartheta d\varphi}, \\ 
        T_\psi(\bm{k}_{inc}) &\underset{2d}{=} \frac{\int_{\mathcal{F}_\psi \setminus 0} S\left[k  \hat{\bm{e}}(\theta + \vartheta) - k \hat{\bm{e}}(\theta )\right]d\vartheta}{\oint_{\mathcal{C} \setminus 0}S\left[k  \hat{\bm{e}}(\theta + \vartheta) - k\hat{\bm{e}}(\theta )\right]d\vartheta},
\end{align}
where $\mathcal{F}_\psi$ is a forward cone with half-aperture angle $\psi$.
In the limit $\psi = \pi/2$, in both $2d$ and $3d$, the cone becomes a half circle or sphere, and the previous definition is recovered.

\begin{figure}
    \centering
    \includegraphics[width=0.96\textwidth]{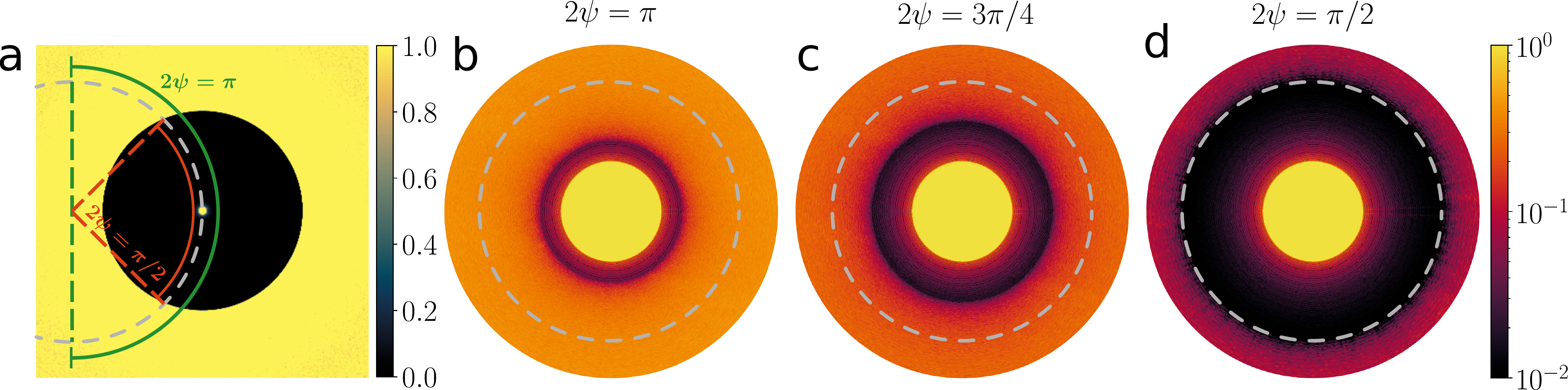}
    \caption{\textbf{Effects of varying aperture in stealthy hyperuniform systems.}
    $(a)$ Sketch of the effect of aperture in $2d$.
    Depending on the aperture angle $0 \leq 2\psi \leq \pi$ used for detection, the onset of forward scattering, and therefore the end of the transmission trough is observed at different frequencies.
    $(b)-(d)$ Illustration: $T$ obtained for a stealthy hyperuniform system ($N = 4\times 10^7$, $K = 3000$), plotted up to $k_{max} = 5000$, for $2\psi = \pi$ $(b)$, $2\psi = 3\pi/4$ $(c)$, and $2\psi = \pi / 2$ $(d)$.
    The dashed gray circle in panels $(b)-(d)$ represents the same frequency as the one represented as a dashed gray circle in $(a)$.
    }
    \label{fig:Aperture}
\end{figure}

We sketch the effect of the aperture $\psi$ in Fig.~\ref{fig:Aperture}$(a)$: as the aperture angle is reduced from $\pi/2$, the smalles frequency at which forward scattering reaches the detector is pushed to higher and higher frequencies.
More precisely, the onset is expected at $k_f(\psi) = K / \sqrt{2 (1 - \cos \psi)}$, so that $k_f(\pi /2) = K /\sqrt{2}$ as in the main text, and $k_f(0) \to \infty$, so that the width of the observed trough is unbounded from above.
To illustrate this, we show intensity maps for this quantity, for a few values of $\psi$, in Fig.~\ref{fig:Aperture}$(b)-(d)$, in a stealthy hyperuniform system of $N=4\times 10^7$ particles with $K = 3000$.
We indeed observe that switching from $\psi = \pi/2$ to $\psi = \pi/4$ leads to a roughly $2-$fold change in the value of $k_f$, leading to a large change in the perceived width of the transmission trough.

\section{Characterization of quasicrystalline order\label{sec:Quasicrystals}}

In the main text, we present evidence that constraining a finite set of peaks to high, order $N$ values is sufficient to non-deterministically generate structures with quasicrystalline order.
The main text focuses on two aspects of the characterization: the distribution of vectors of Voronoi nearest-neighbors and the structure factor of the output configuration.
In this section, we present additional data to support the observation that the systems indeed have quasicrystalline order, focusing on the $2d$ case.

\subsection{Full $2d$ pair correlation function}

\begin{figure}
    \centering
    \includegraphics[width=0.46\textwidth]{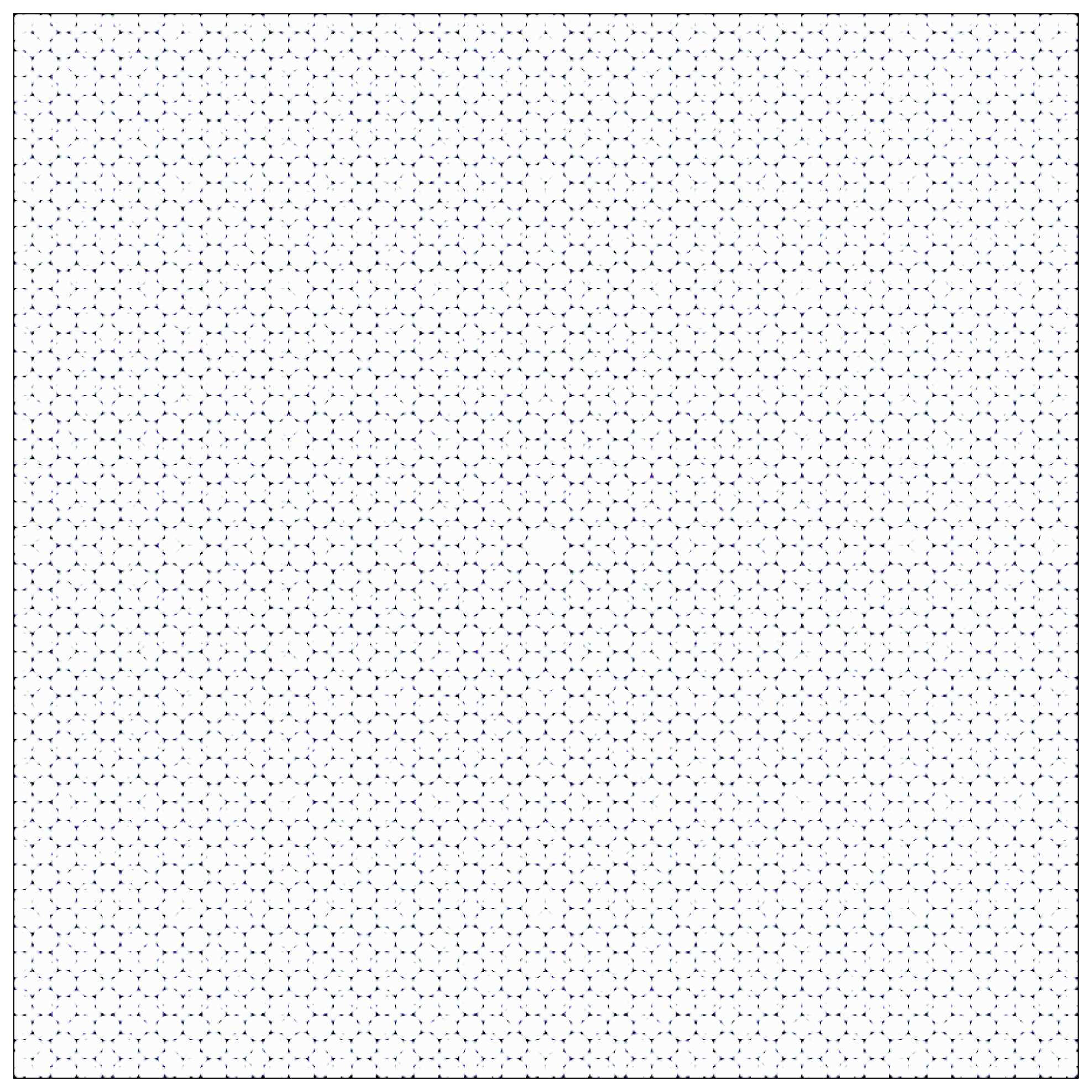}
    \includegraphics[width=0.46\textwidth]{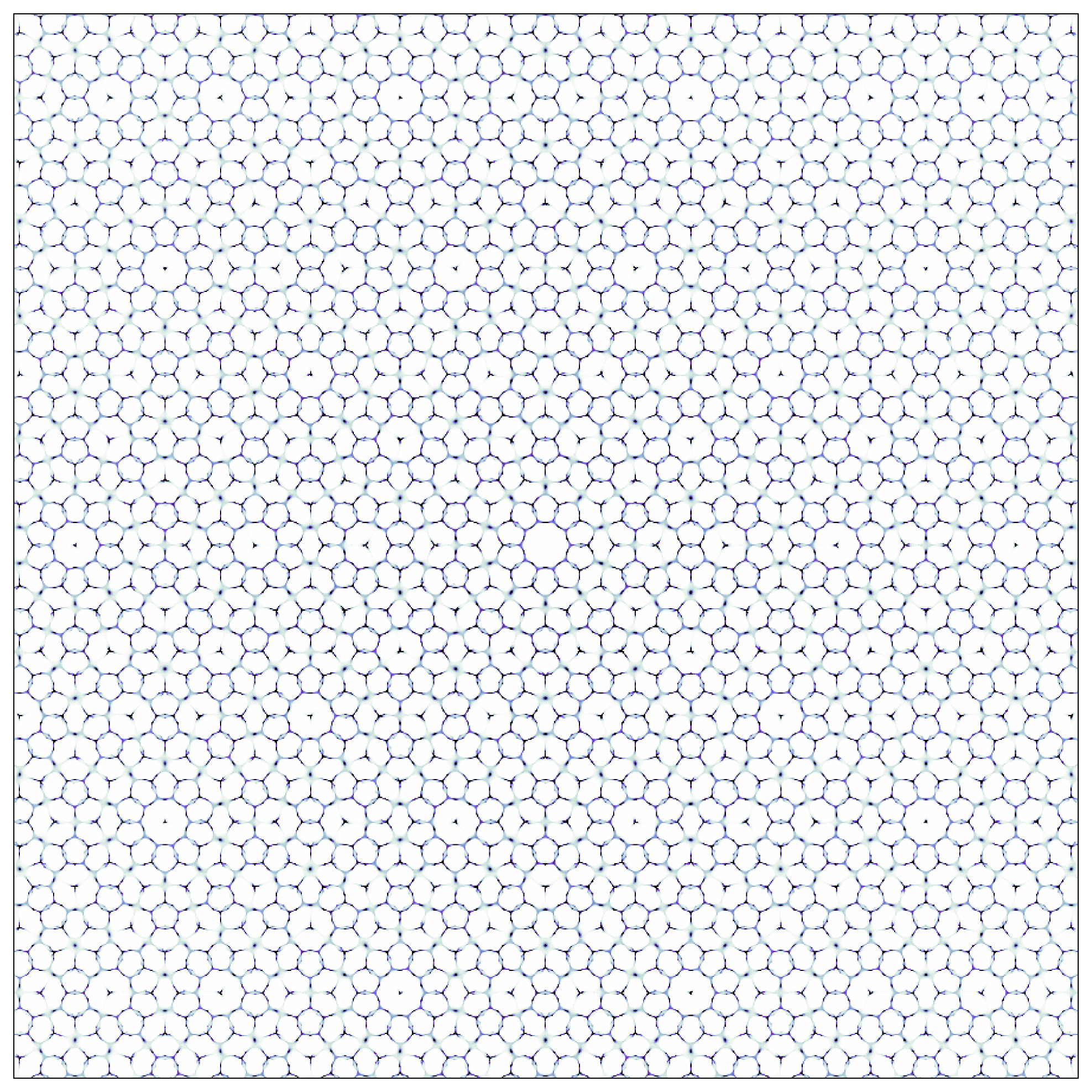}
    \includegraphics[height=0.46\textwidth]{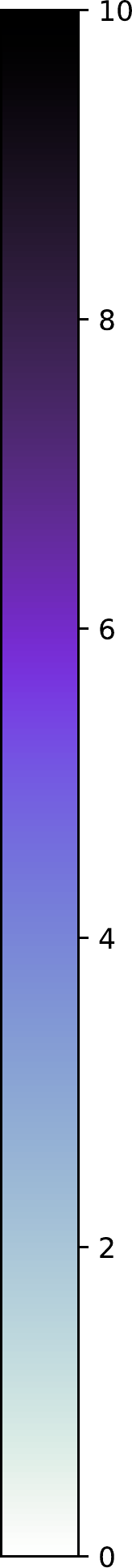}\\ 
    \includegraphics[width=0.46\textwidth]{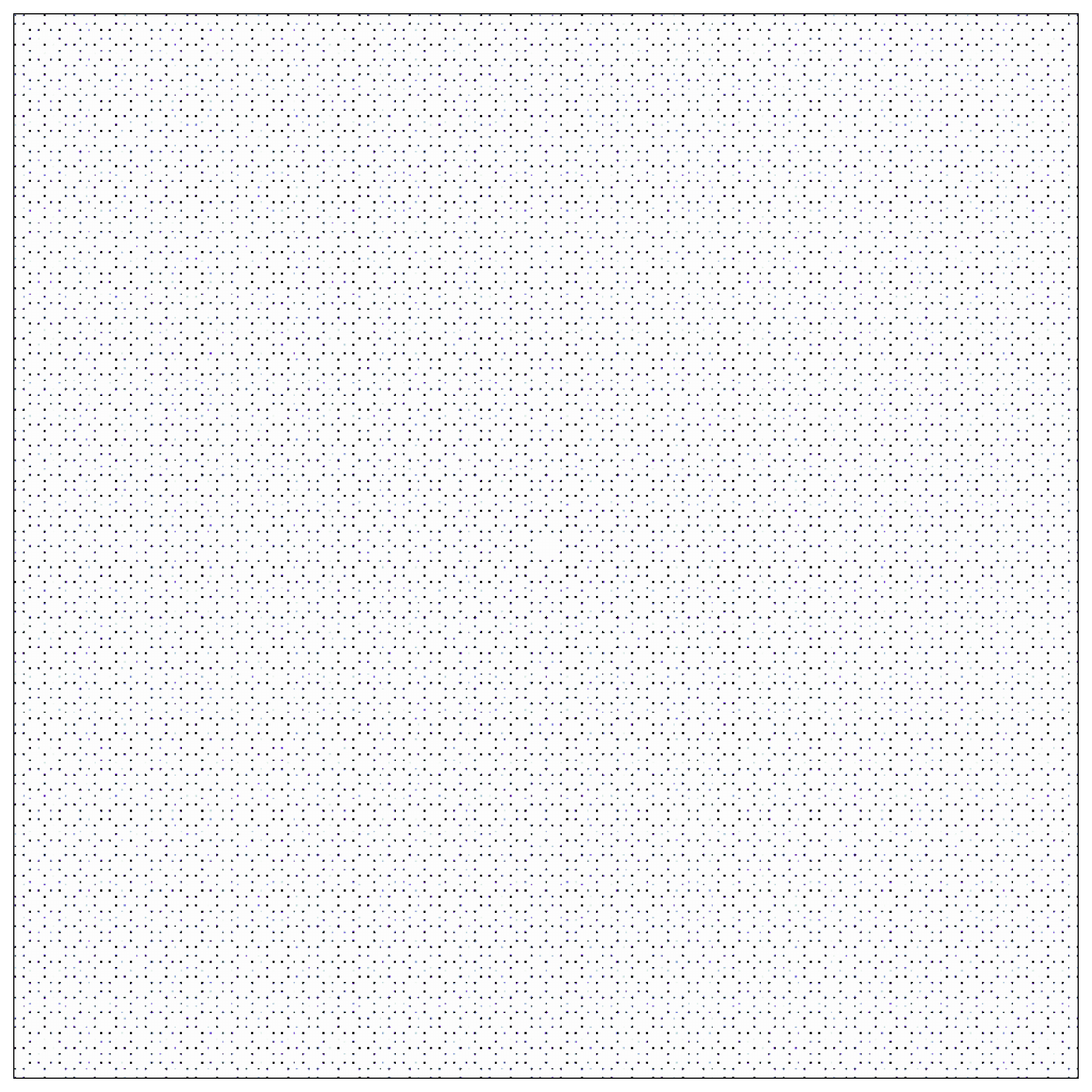}
    \includegraphics[width=0.46\textwidth]{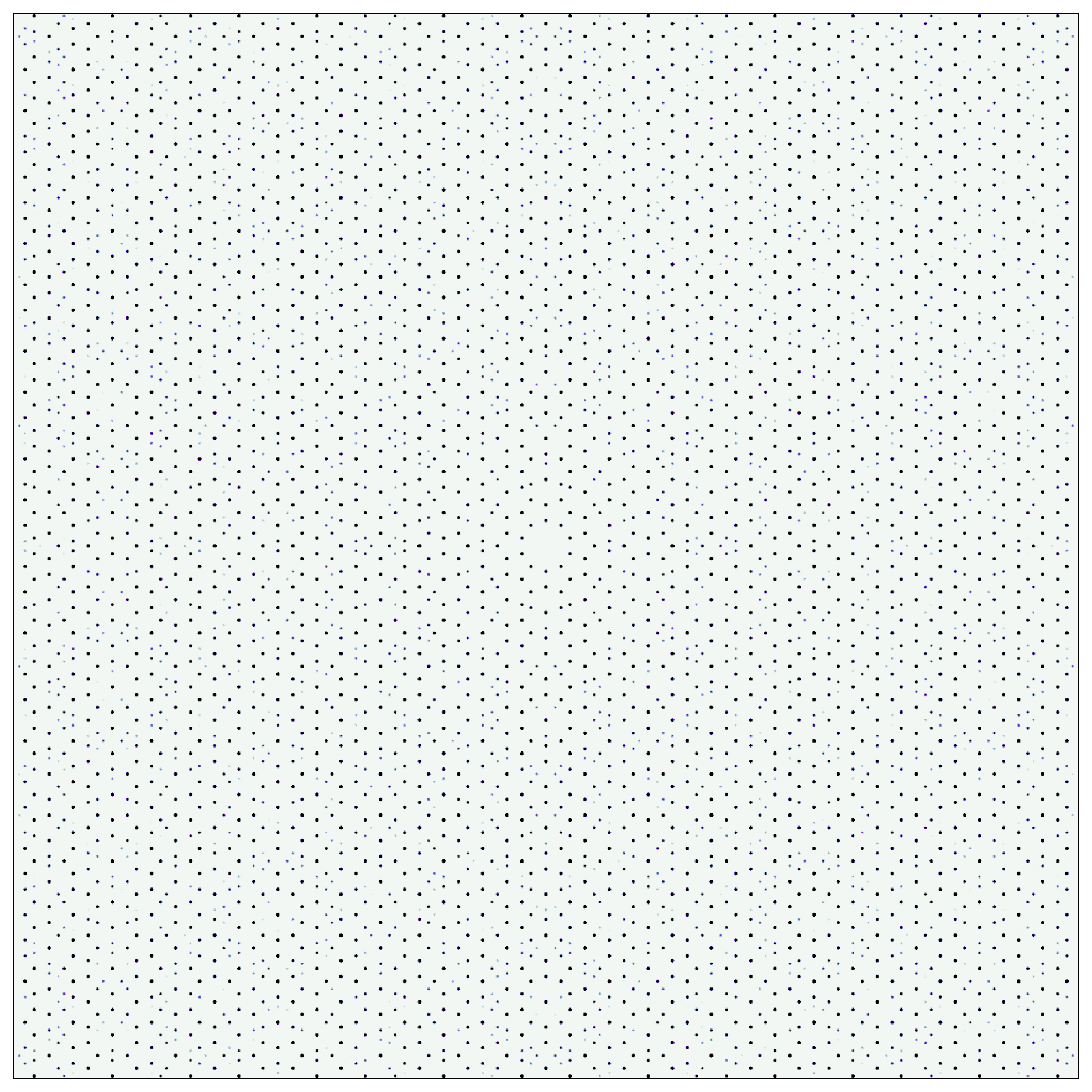}
    \includegraphics[height=0.46\textwidth]{Figures_SI/Quasicrystal_2dgr/amethystcbar.pdf}
    \caption{\textbf{Quasicrystalline Pair Correlation Functions: Octagonal, Decagonal.}
    Top row: Density plots of $2d$ $g(\bm{r})$ of single FReSCo NUwNU quasicrystalline output configurations for $n = 8$- (top left), and $10$--fold (top-right) symmetries.
    The plots run from $-L/4$ to $L/4$ with $L$ the sidelength of the system (see extra files for full range).
    Bottom row: same outputs for sets of vertices of rhombic tilings of perfect quasicrystals with the same symmetries.
    }
    \label{fig:Quasicrystal_2d_g(r)_octagonal_decagonal}
\end{figure}

\begin{figure}
    \centering
    \includegraphics[width=0.46\textwidth]{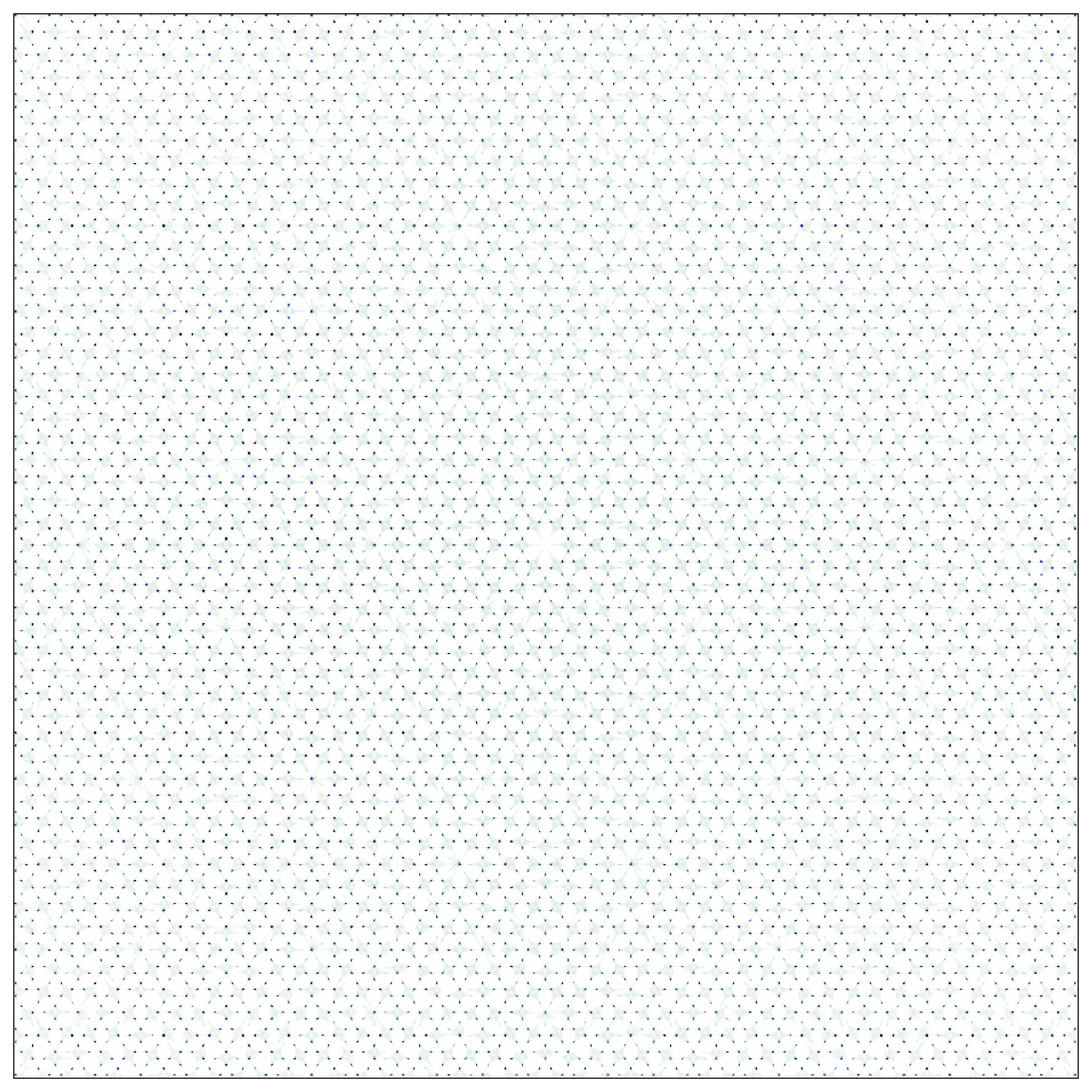}
    \includegraphics[width=0.46\textwidth]{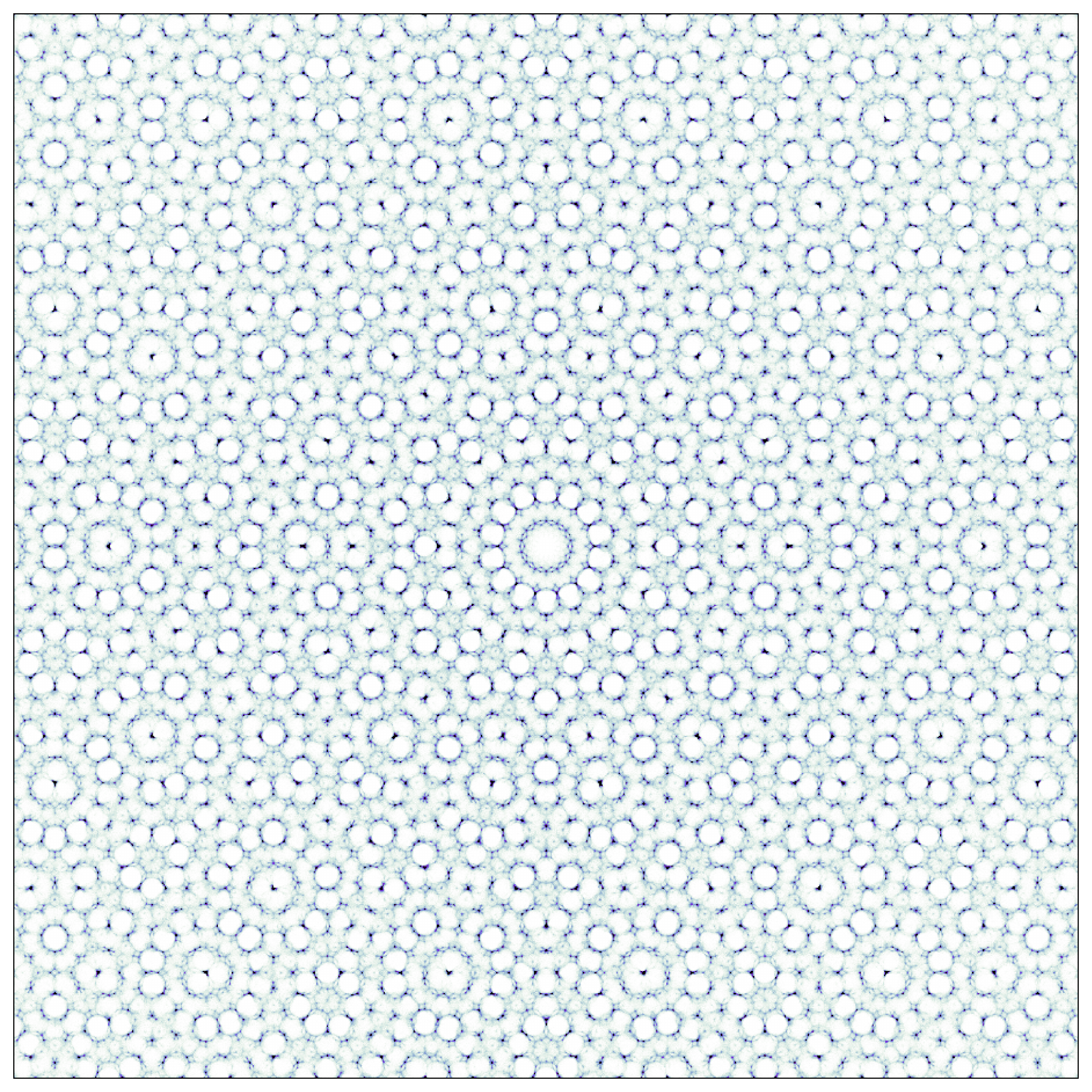}
    \includegraphics[height=0.46\textwidth]{Figures_SI/Quasicrystal_2dgr/amethystcbar.pdf}\\ 
    \includegraphics[width=0.46\textwidth]{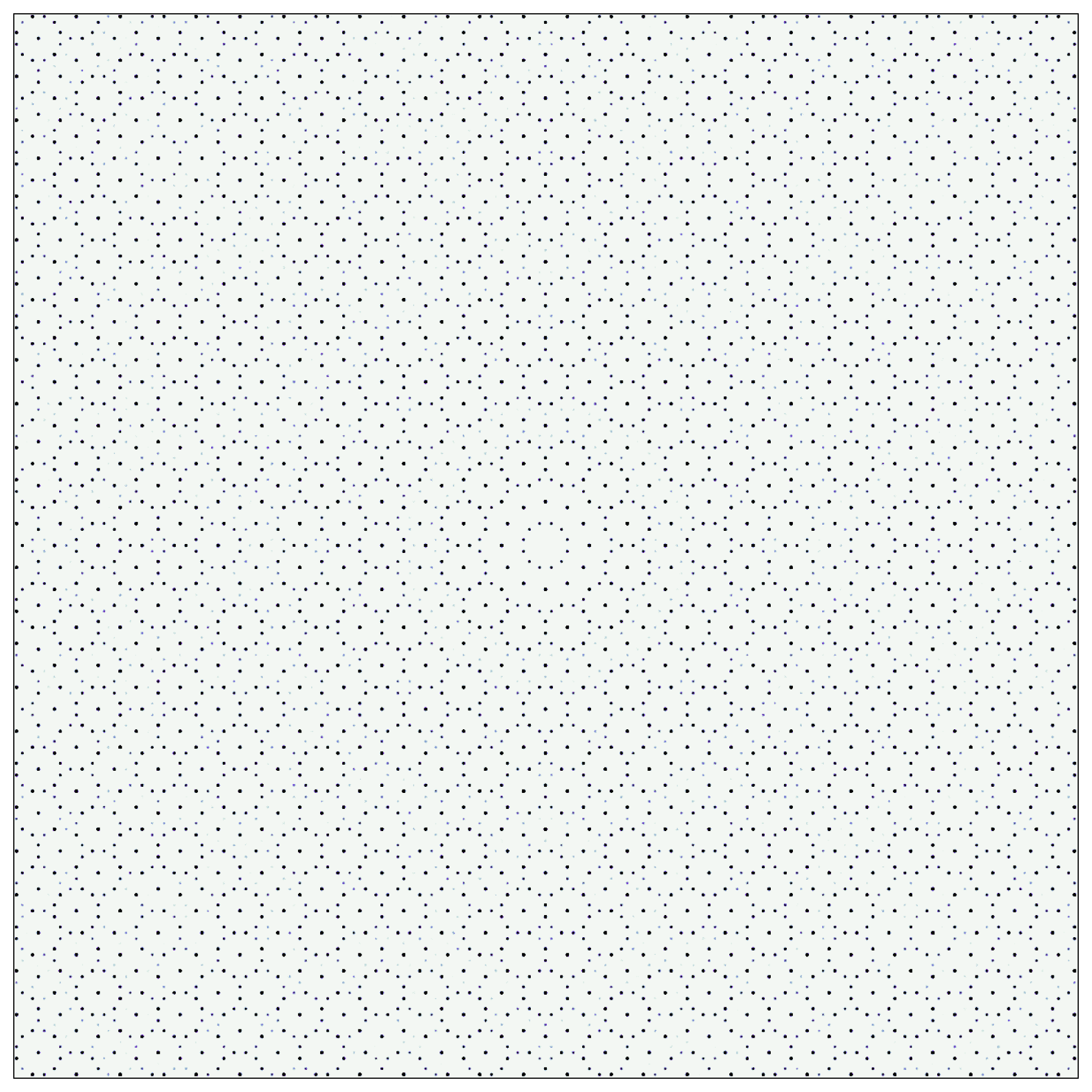}
    \includegraphics[width=0.46\textwidth]{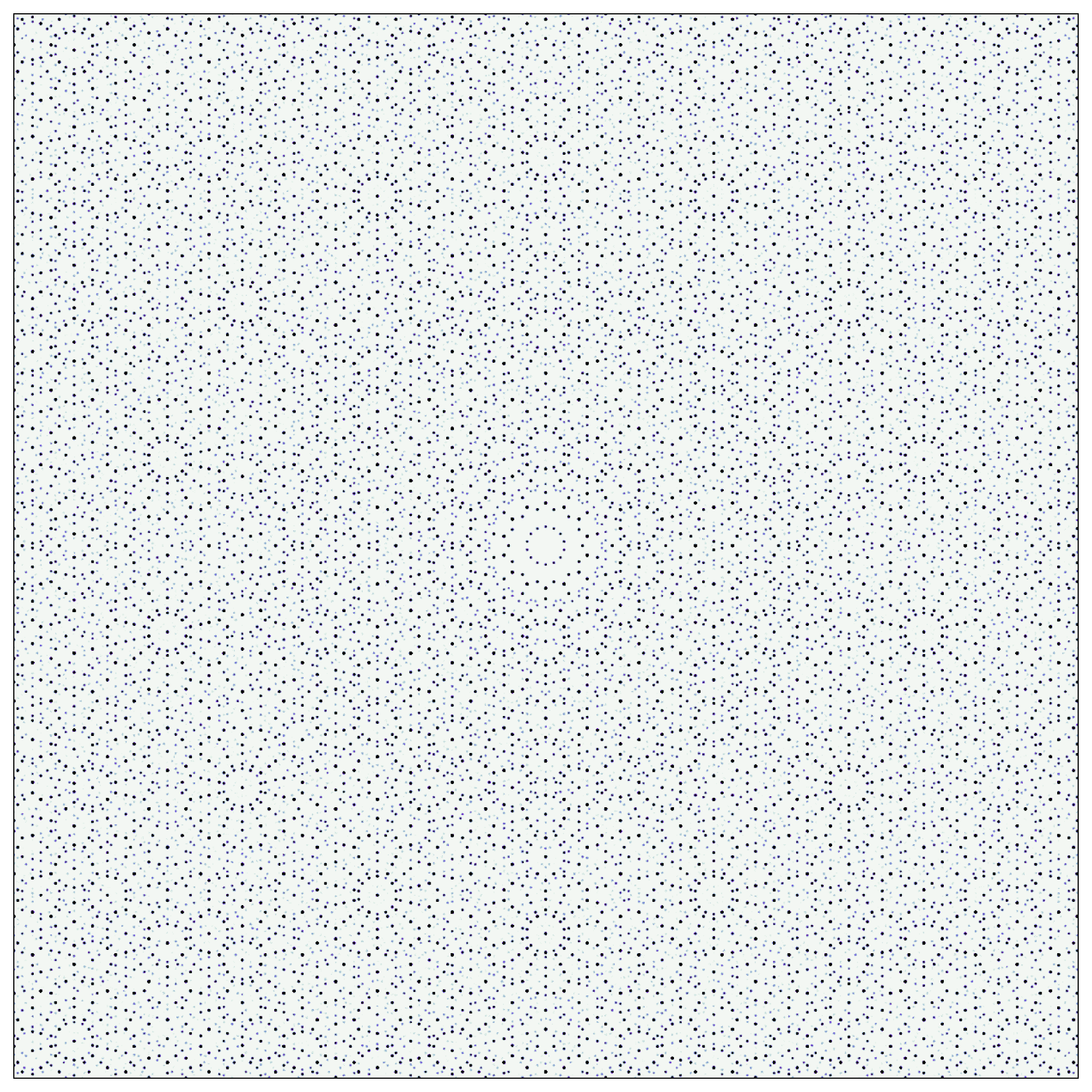}
    \includegraphics[height=0.46\textwidth]{Figures_SI/Quasicrystal_2dgr/amethystcbar.pdf}
    \caption{\textbf{Quasicrystalline Pair Correlation Functions: Dodecagonal, Tetradecagonal.}
    Top row: Density plots of $2d$ $g(\bm{r})$ of single FReSCo NUwNU quasicrystalline output configurations for $n = 12$- (top left), and $14$--fold (top-right) symmetries.
    The plots run from $-L/4$ to $L/4$ with $L$ the sidelength of the system (see extra files for full range).
    Bottom row: same outputs for sets of vertices of rhombic tilings of perfect quasicrystals with the same symmetries.
    }
    \label{fig:Quasicrystal_2d_g(r)_dodecagonal_tetradecagonal}
\end{figure}

First, we compute the full, two-dimensional pair correlation function~\cite{Hansen2006},
\begin{align}
    g(\bm{r}) \equiv \frac{\langle \rho(\bm{r}_0) \rho(\bm{r}_0 + \bm{r}) \rangle_{\bm{r}_0}}{\langle\rho(\bm{r_0})\rangle_{\bm{r}_0}^2},
\end{align}
where $\langle \cdot \rangle_{\bm{r}_0}$ represents a spatial averaging over point positions (origin), and the denominator reduces to $\rho_0^2$, the square of the average point density.
This function represents the averaged two-point correlation of the point density at all ranges, and is (up to normalization) identical to the nearest-neighbor vector distribution from the main text at short range.
In practice, we compute $g$ by directly binning distance vectors between the $N(N-1)$ particles pairs in $2d$.
We show the result for $N\approx10000$ in the top row of Figs.~\ref{fig:Quasicrystal_2d_g(r)_octagonal_decagonal} and~\ref{fig:Quasicrystal_2d_g(r)_dodecagonal_tetradecagonal} for all explored $2d$ quasicrystals.
In each panel, we show that our systems preserve peaked positional order, as well as the right discrete rotational symmetry breaking, at all ranges.
For comparison, in the bottom row, we show the same figures for sets of $N\approx 10000$ points corresponding to the vertices of quasicrystalline rhombic tilings obtained via a dualing map~\cite{Lutfalla2021} applied to the intersections of de Bruijn's multigrid method~\cite{deBruijn1981,deBruijn1986}, onto which a small random kick (uniform in both directions with amplitude $\delta = 10^{-4}L$).
This second row stresses the point that these two features, peaked and system-wide discrete rotational symmetry in $g$, are characteristic of quasicrystals, supporting our claim of quasicrystalline order.
Notice that, save for the 8-fold case, the point patterns generated by FReSCo are \textit{not} simply the vertices of the rhombic tiling, but a different decoration of the tiling, as discussed in the main text.
In particular, notice that the same $N$ in our FReSCo outputs and in an exact dualed deBruijn construction do not lead to the same feature scales: the features look larger in our outputs, which highlights that our systems look like \textit{decorations} of usual quasicrystals, meaning that a rhombic tiling superimposed on our system would contain more than one point per rhombus, much like a body-centered cubic lattice is a decoration of the primitive cubic lattice with more than one atom per cell.

\subsection{Rhombic tilings}

\begin{figure}[b]
    \centering
    \includegraphics[width=0.90\textwidth]{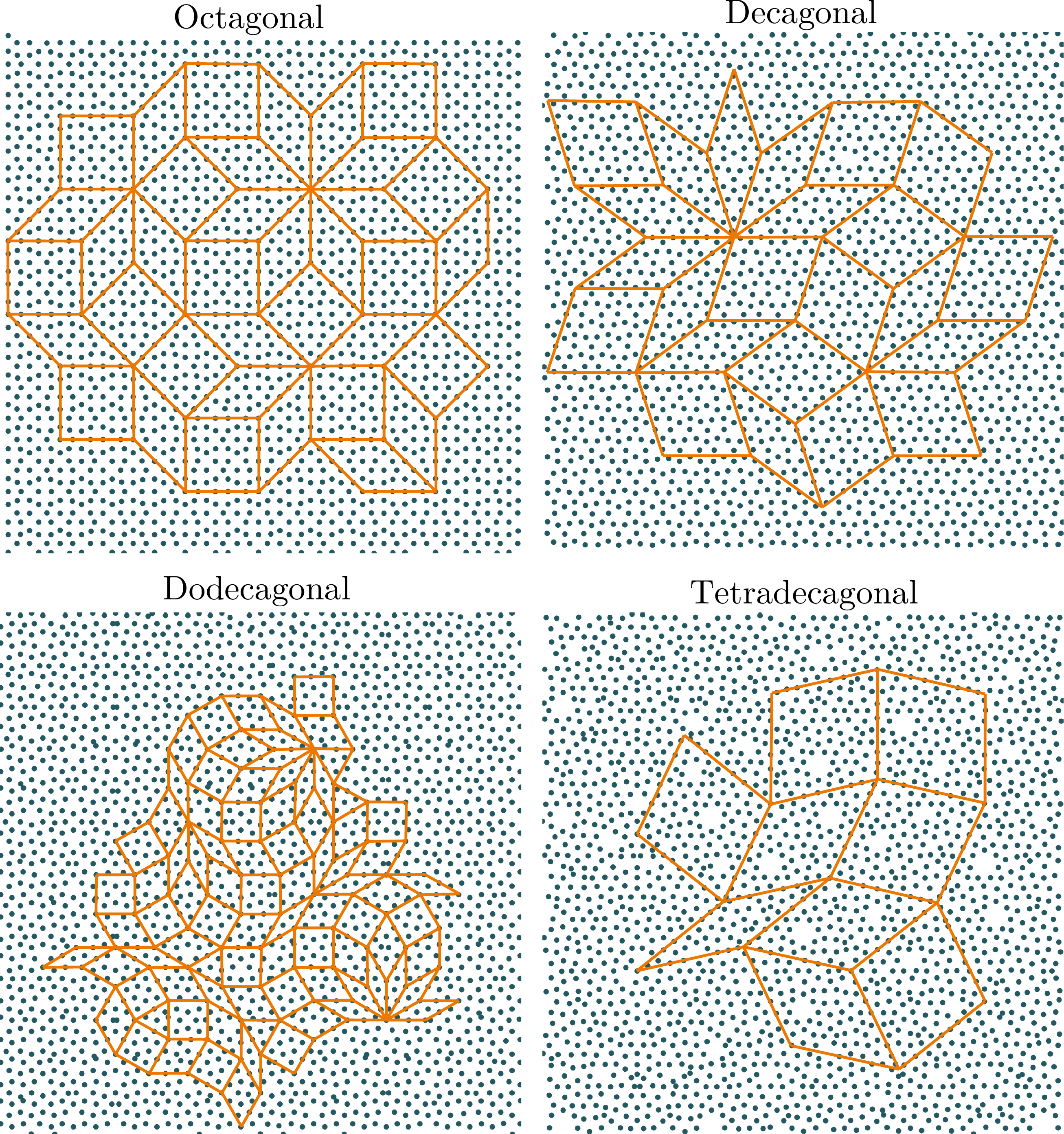}
    \caption{\textbf{Rhombic tilings drawn on systems generated using NUwNU.}
    Points depicted are $L/2 \times L/2$ subsections of the full minimized point pattern.}
    \label{fig:tilings}
\end{figure}
This is illustrated more clearly in Fig.~\ref{fig:tilings}, where we show parts of quasicrystalline FReSCo outputs, and draw possible rhombic tilings onto them.
To draw these, following Ref.~\onlinecite{Engel2015}, we use centers of regular $n$-gons of particles as candidates for rhombic vertices.
This criterion allows to draw consistent rhombic tilings that obey usual matching rules~\cite{Levine1986}, and may be merged together or divided into smaller units to generate tilings at other scales, down to a minimal scale limited by the smallest distance between the regular $n$-gons.
Notice that in the octagonal and dodecagonal symmetry systems, the centers of these polygons tend to lie on points in the point pattern.
In the decagonal and tetradecagonal symmetry systems, however, these centers tend to lie between points, here at Voronoi vertices.
The tiles we draw are decorated by points inside of them, so that the whole system can be covered by a finite set of tiles with decorations at all scales.
Also, the smallest scale coincides with rhombic vertices using strictly all points in the point pattern and nothing else only in the octagonal case.
In all other cases, the decoration inside the smallest set of tiles gets more and more intricate as the order of symmetry increases.

\subsection{Bond-orientational Order Parameters (BOOPs)}
\begin{figure}[b]
    \centering
    \includegraphics[width=0.96\textwidth]{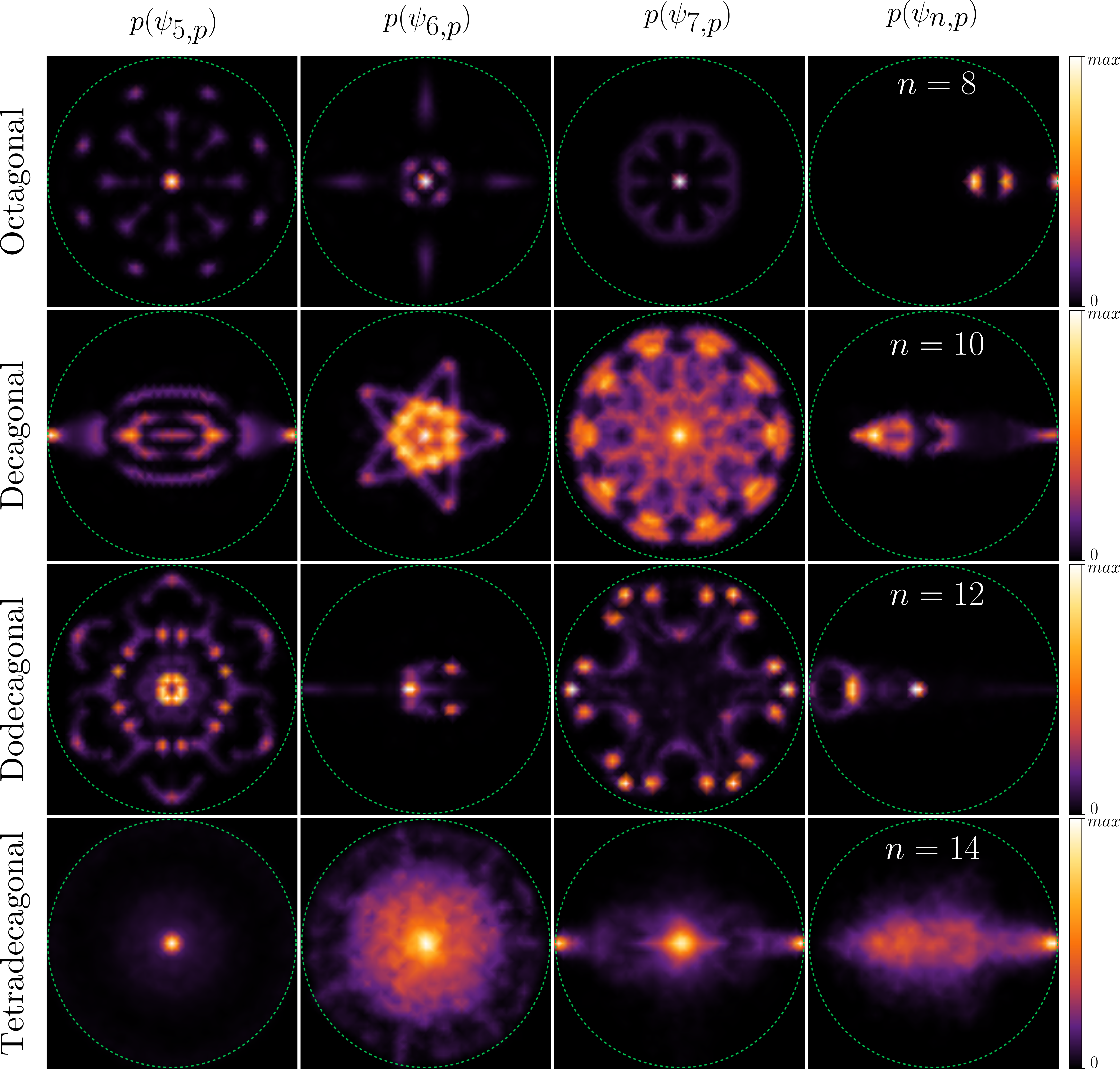}
    \caption{\textbf{Bond-Orientational Order Parameters: FReSCo outputs.}
    From left to right in each row, density maps of the $2d$ probability density functions of pentatic $\psi_5$, hexatic $\psi_6$, heptatic $\psi_7$, and $n$-atic order, for FReSCo outputs with $n = 8$- (top row), $n = 10$- (second row), $n = 12$- (third row), and $n=14$-fold symmetry. 
    Dashed green circles indicate the end of the unit disk.
    }
    \label{fig:Quasicrystal_BOOPs}
\end{figure}
\begin{figure}[b]
    \centering
    \includegraphics[width=0.96\textwidth]{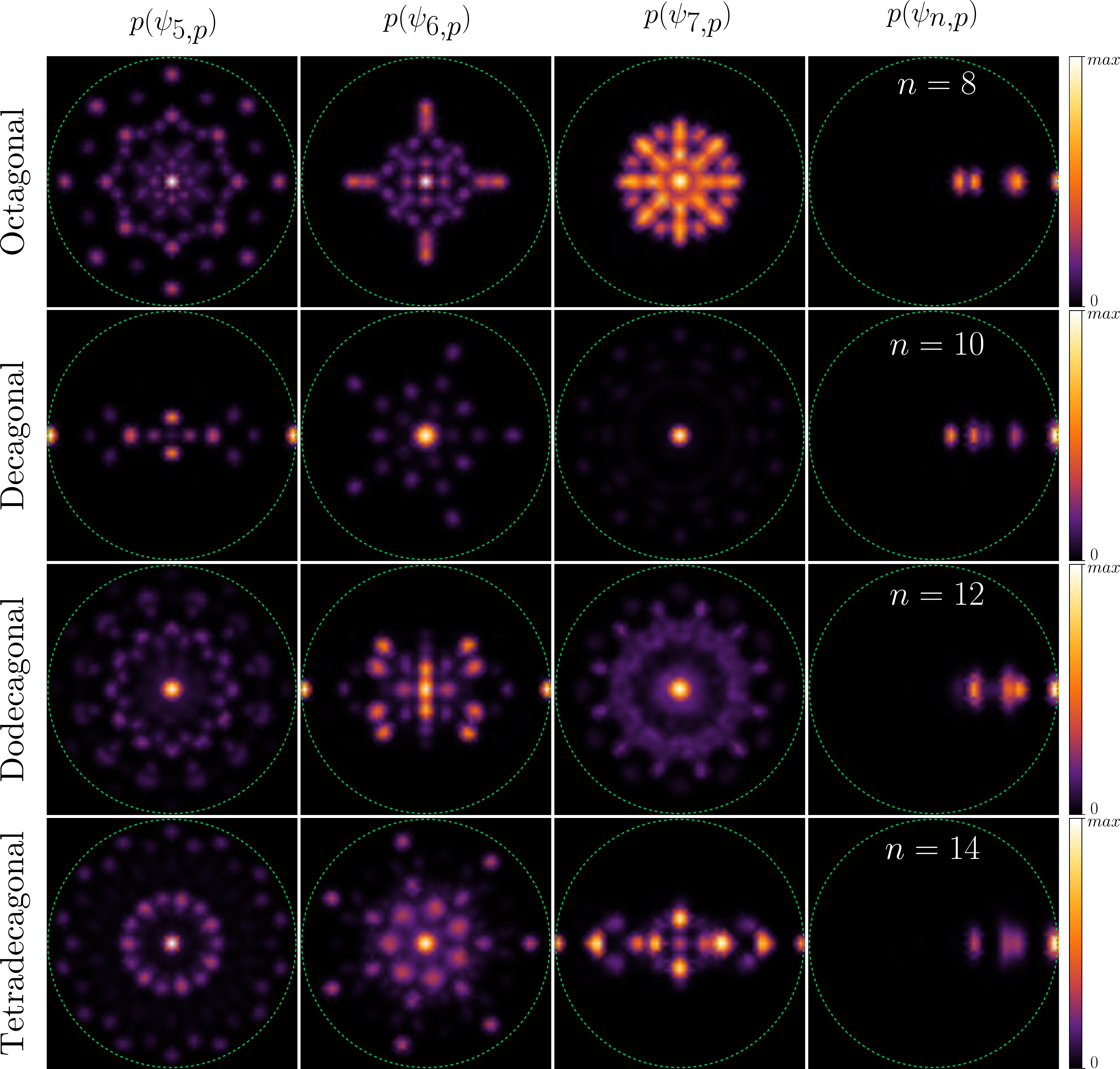}
    \caption{\textbf{Bond-Orientational Order Parameters: Quasicrystals.}
    From left to right in each row, density maps of the $2d$ probability density functions of pentatic $\psi_5$, hexatic $\psi_6$, heptatic $\psi_7$, and $n$-atic order, for weakly noisy quasicrystals with $n = 8$- (top row), $n = 10$- (second row), $n = 12$- (third row), and $n=14$-fold symmetry. 
    Dashed green circles indicate the end of the unit disk.
    }
    \label{fig:DeBruijn_BOOPs}
\end{figure}
Following Ref.~\onlinecite{Engel2015}, it is possible to argue about quasicrystallinity using Steinhardt's bond-orientational order parameters (BOOPs)~\cite{Steinhardt1983}.
In $2d$, these BOOPs are usually defined via local complex-valued observables associated to each particle,
\begin{align}
    \psi_{n,p} = \frac{1}{N_{neigh}} \sum\limits_{q = 1}^{N_{neigh}} e^{i n \theta_{pq}}
\end{align}
where $p$ denotes a particle in the system, $q$ runs over its $N_{neigh}$ Voronoi nearest-neighbors, and $n$ is the order of discrete rotational symmetry being checked for.
The most famous examples of these parameters are the tetratic, $n=4$, and hexatic, $n = 6$, order parameters~\cite{Nelson1979,Li2020}, but the same definition can be used for any $n$.
These complex numbers can equivalently be thought of as vector order parameters $\bm{\psi}_{n,p}$ with lengths bounded from above by 1.
If all neighbors of a particle lie on a subset of the vertices of the regular $n$-gon, $\psi_{n,p}$ has unit modulus, and $arg(\psi_{n,p})/n$ reflects the orientation of the polygon within the $[0; 2\pi/n)$ interval.
A system with long-range orientational order is characterized by an average BOOP, $\Psi_n = N^{-1} \sum_p \psi_{n,p}$ with high modulus, which implies that individual particles all have polygonal environments with the same orientation.
In a polycrystalline system, while individual particles have high $\psi_{n,p}$ moduli, the distribution of BOOP vectors will contain different orientations averaging to a small average BOOP $\Psi_n$.
Any deviation from perfect local polygonal environments lowers the modulus of $\psi_{n,p}$, and also leads to lower $\Psi_n$.
In practice, it is therefore much more informative to look at the full distribution of local BOOPs to characterize the orientational order in a system.
Since Steinhardt's BOOPs can by definition only exist in the unit disk, these distributions are convenient to show as density plots within $[-1;1]^2$.

In Fig.~\ref{fig:Quasicrystal_BOOPs} we show such density plots at a few relevant values of $n$ for our quasicrystalline structures.
In this figure, each row corresponds to a given system, from top to bottom $n = 8$-, $10$-, $12$-, and $14$-fold symmetric outputs of FReSCo.
On each row, the first three plots correspond to density plots of the distribution of the microscopic pentatic $\psi_{5,p}$, hexatic $\psi_{6,p}$, and heptatic $\psi_{7,p}$ orders; while the last panel corresponds to the relevant $\psi_{n,p}$.
The last panel of each row consistently shows a very anisotropic distribution, skewed towards a very clear orientation of the BOOP, with a strong peak at unit modulus and a few lower-modulus peaks.
This means that the system indeed does have strong BOOP in a consistent orientation across the system, but that all particles are not strictly equivalent, a result consistent with the idea that these configurations correspond to decorations of the usual quasicrystalline rhombic tilings.
Furthermore, the more usual pentatic, hexatic, and heptatic BOOPs are surprisingly informative about the orientational order in the system.
Indeed, for an $n$-fold order, the $(n/2)$-fold BOOP tends to be very strongly peaked at opposite values, meaning that there are, for instance, a lot of locally 5-fold structures in the $10$-fold quasicrystalline system, but with two competing orientations.
Likewise, incommensurate BOOPs also indicate the order of the discrete rotational symmetry: the hexatic order in the $10$-fold system, for instance, is weak on average, but has a strongly 5- and,therefore, 10-fold symmetric distribution.

For comparison, in Fig.~\ref{fig:DeBruijn_BOOPs}, we show the same distributions in the weakly noisy configurations used for the bottom row of Figs.~\ref{fig:Quasicrystal_2d_g(r)_octagonal_decagonal} and~\ref{fig:Quasicrystal_2d_g(r)_dodecagonal_tetradecagonal}.
While the configurations are different in absolute terms, they share their main features: strong orientational order at the required $n$, symmetric peaks at order $n/2$, and peaks reflecting the symmetry of the system at other values.
Note that the differences between these reference systems and the output of FReSCo become more noticeable as $n$ increases: this is likely due to the fact that the FReSCo outputs then correspond to more and more ``decorated'' rhombic tiles (see main text).

Altogether, gathering information from the final point patterns, their structure factors, the statistics of nearest neighbor vectors, the full $2d$ $g(r)$, and BOOP distributions, we have shown that FReSCo outputs display quasicrystalline order, although they are not trivially identical to vertices from a rhombic tiling or any such simple exact construction.

\section{Additional examples of imposed structure factors\label{sec:Additional_examples}}

\subsection{Encoding movies into successive $S(k)$}

\begin{figure}
    \centering
    \includegraphics[width=0.7\textwidth]{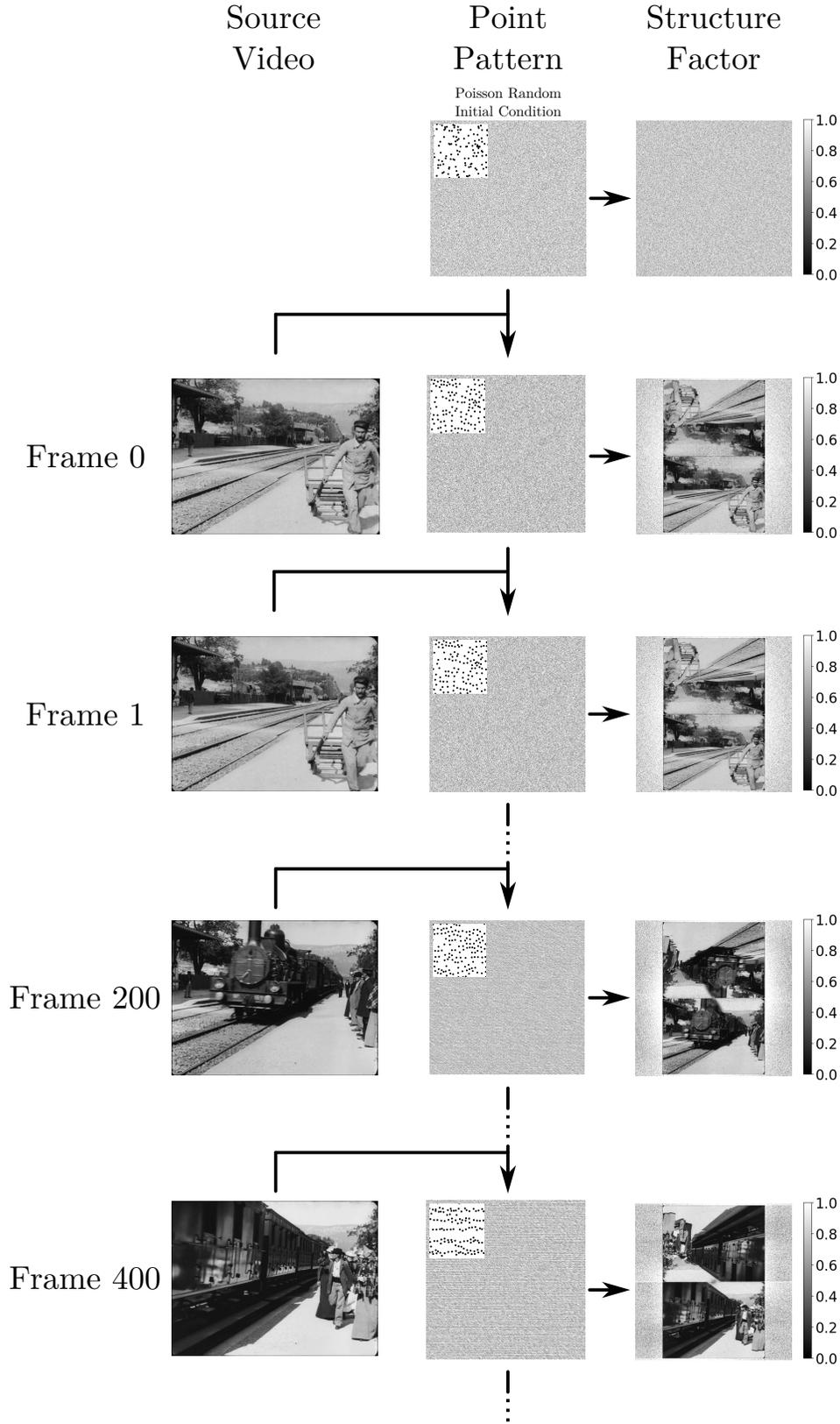}\hspace{20mm}
    \caption{\textbf{Construction and example frames of spectral movie.} Left: Frames from source video \textit{L'Arrivée d'un Train en Gare de La Ciotat}~\cite{LaCiotat}.
    Middle: Minimized point patterns ($N=300,000$) from imposing the source video frame as a constraint in $S(k)$, with an inset showing a zoomed-in region of the point pattern. Right: The calculated structure factor of the resulting minimized point pattern.
    The video is also available at \url{https://youtu.be/2oVJO197Wmc}.
    }
    \label{fig:Movie}
\end{figure}

As we have demonstrated the ability to encode images into the structure factor of point patterns, here we encode a movie by an iterative minimization process (Fig. \ref{fig:Movie}).
We impose each frame extracted from the movie \textit{L'Arrivée d'un train en gare de La Ciotat},``The Arrival of a Train at La Ciotat Station"~\cite{LaCiotat}, ($962\times720$ px) as constraints $S_0(\bm{k})$ on systems of $N=300,000$ points in 2d.
The full 810 frame video depicting the simultaneous evolution of the point pattern and its measured structure factor is available as Supplementary Video 1.
The point pattern based on frame 0 was minimized from a Poisson random initial condition, while all subsequent points patterns are minimized using the previous point pattern as an initial condition.
By performing this successive minimization, we can encode the movie into discrete point trajectories.
Due to the similarity between successive frames in a single-take sequence shot, successive minimizations are faster than minimizations from random configurations, and point trajectories themselves are seemingly close to continuous.
This opens up the possibility of smoothly evolving adaptive point patterns with spectrally-shaped disorder.

\newpage
\subsection{Saturating peaks and echoes in NUwNU constraints}

Here we show that constraining a set of $N_k$ peaks to their maximum real value, $S(\bm{k}_j)=N$, $j=1,\ldots,N_k$, also implicitly constrains peaks at all integer linear combinations of the $\bm{k}_j$ at which the original constraints are imposed.
This can be shown analytically: starting from the expression for the structure factor of a point pattern with positions $\bm{r}_n$, $n=1,\ldots,N$ at the specified coordinates $\bm{k}_j$,
\begin{align}
    S(\bm{k}_j) &\equiv \frac{|\widehat{\rho}\,(\bm{k}_j)|^2}{N} = N,
\end{align}
where
\begin{align}
    \widehat{\rho}(\bm{k}_j) &\equiv \sum_{n=1}^N\exp(i\bm{k}_j\cdot\bm{r}_n).
\end{align}
   
Viewing the definition of $\widehat{\rho}(\bm{k}_j)$ as a sum of $N$ unit length vectors in the complex plane, it is straightforward to conclude that in order to get $|\widehat{\rho}\,(\bm{k}_j)| = N$, all such unit-length vectors must point in the same direction.
This means that, given one of the $N_k$ vectors $\bm{k}_j$ at which the constraint is imposed, all $N$ dot products $\bm{k}_j\cdot\bm{r}_n = \zeta + 2 \pi m$ with $\zeta \in [-\pi; \pi)$ a constant and $m \in \mathbb{Z}$, meaning that the dot products all represent the same phase angle (accounting for shifts of integer multiples of $2\pi$).
In real space, this means that the projections of the points positions onto the direction of $\bm{k}$ are a subset of sites on a periodic $1d$ lattice with period $\lambda = 2\pi / k$.
However, this does not guarantee generically that all $\bm{k}_j\cdot\bm{r}$ are the same, i.e. it is possible that $\bm{k}_j\cdot\bm{r}_n\neq\bm{k}_{l\neq j}\cdot\bm{r}_n$ (they can project to different sites of the $1d$ lattice).

Given the above constraint, the structure factor at a k-space coordinate $\bm{s} = \sum_j C_j \bm{k}_j,$ with $C_j\in \mathbb{Z}$ an arbitrary linear combination of the constrained $\bm{k}_j$, reads
\begin{align}
    \widehat{\rho}(\bm{s}) &= \sum_{n=1}^N\exp\left(i\sum_jC_j\bm{k}_j\cdot\bm{r}_n\right).
\end{align}

For any two generic points $\bm{r}_n$ and $\bm{r}_m$ and one specific $\bm{k}_j$, we showed that $\bm{k}_j\cdot\bm{r}_n = \bm{k}_j\cdot\bm{r}_m +2\pi c$ for some $c\in \mathbb{Z}$.
We then have
\begin{align}
    \exp\left(iC_j\bm{k}_j\cdot\bm{r}_n\right) &= \exp\left(iC_j(\bm{k}_j\cdot\bm{r}_m + 2\pi c)\right)\\
    &= \exp\left(iC_j\bm{k}_j\cdot\bm{r}_m\right)\exp\left(i2\pi cC_j\right)\\
    &= \exp\left(iC_j\bm{k}_j\cdot\bm{r}_m\right).
\end{align}

We can now write $\widehat{\rho}(\bm{s})$ in terms of only one point coordinate $\bm{r}_1$ without loss of generality and conclude:
\begin{align}
    \widehat{\rho}(\bm{s}) &= N\prod_j\exp\left(iC_j\bm{k}_j\cdot\bm{r}_1\right)\\
     &= N\exp\left(i\bm{s}\cdot\bm{r}_1\right)
\end{align}
so that
\begin{align}
    S(\bm{s}) &= N.
\end{align}

Due to the limitations of dimensionality, one cannot enforce arbitrarily many peaks of intensity $N$ in $S$.
In fact, the greatest number of peaks one can arbitrarily impose that will reach an exact value of $S(\bm{k})=N$ is $d$, the number of dimensions, resulting in only the primitive Bravais lattices (generically monoclinic in 2d and triclinic in 3d).
With our algorithm, overconstrained cases such as crystals with motifs of more than one atom or quasicrystals seek to maximize the value of all peaks.
Thus, we still observe quasicrystalline structures emerge from our NUwNU protocol.

\begin{figure*}[h]
    \centering
    \includegraphics[width=0.96\textwidth]{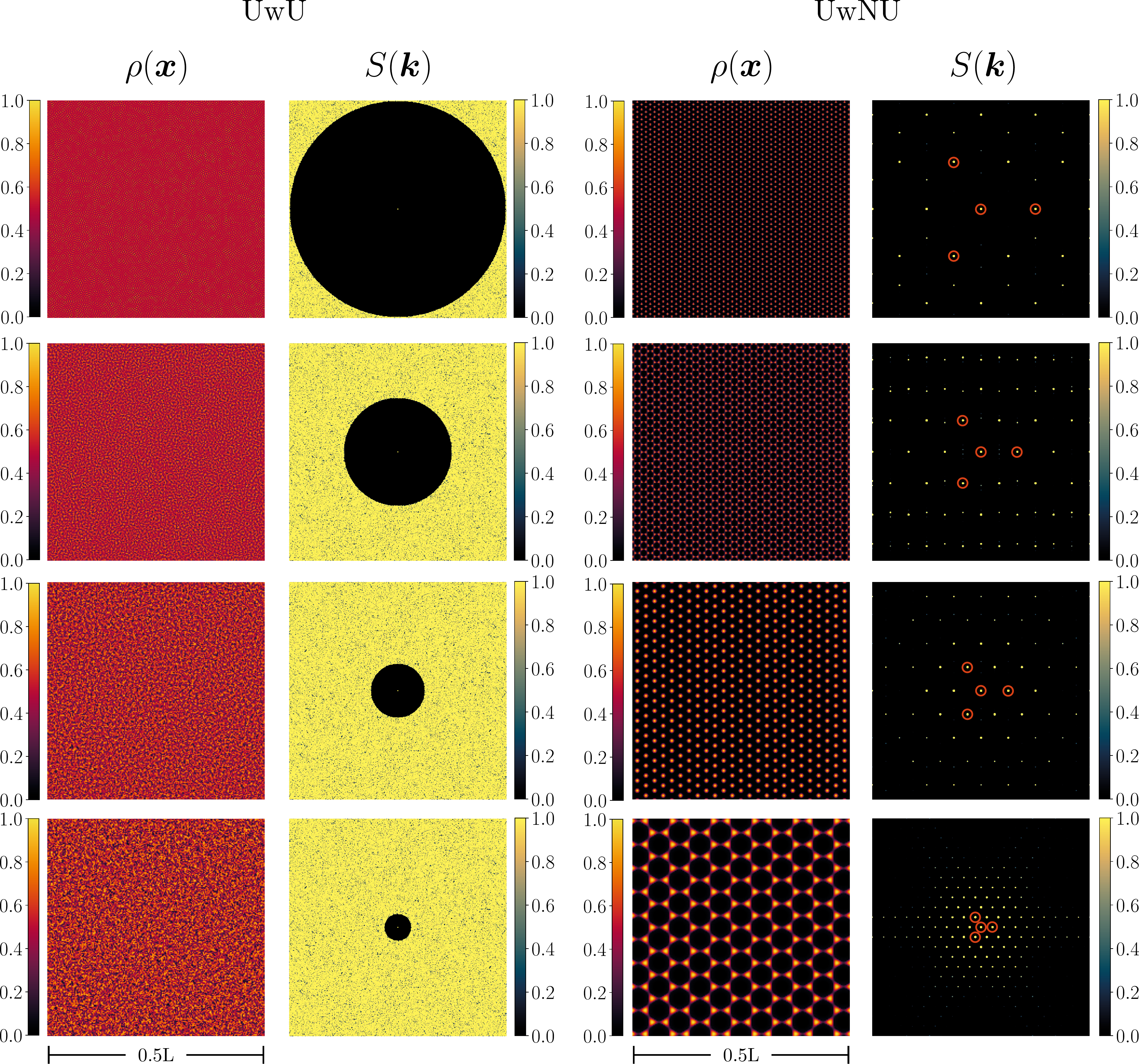}
    \caption{\textbf{Example systems generated using UwU (top) and UwNU  (bottom).}
    Density fields of $403\times 403$ are generated using UwU (left) and UwNU (right) protocols.
    UwU (left) systems are generated to exhibit varying degrees of stealthiness, as demonstrated in the varying radius of $S(k)=0$ in the structure factor.
    The larger the radius, the more uniform the system appears.
    UwNU (right) systems are generated to exhibit 6-fold symmetry, imposing only the peaks of the innermost hexagon (marked with red circles).
    As the radius of the innermost hexagon or peaks changes, we sometimes get a Kagome-like tiling instead of a triangular lattice.
    }
    \label{fig:UwU_UwNU}
\end{figure*}

\subsection{Uniform fields with k-space constraints (UwU and UwNU)}

While our work focuses on non-uniform real space systems (i.e. point patterns), our methodology is easily extendable to uniform systems (e.g. discretized density fields) using the same tools. 
Using the UwU and UwNU protocols (see Methods), we can generate density fields with k-space properties analogous to those of point patterns we have shown elsewhere in this paper (Fig. \ref{fig:UwU_UwNU}).
It is important to note that, in the UwU case, the k-space is exactly as large as the real space.
If one were to fully constrain k-space (i.e constrain both the complex phase and magnitude of $\widehat{\rho}(\bm{k})$ at every grid coordinate), then a single inverse Fourier transform could be used to find a corresponding real space structure rather than a minimization protocol.

However, UwU as defined in this paper only constrains the structure factor (i.e., the modulus $\left|\widehat{\rho}\right|(\bm{k})$) over a set of wavevectors that does not have to cover the whole system.
It may therefore be used to find uniform real-space systems with complete freedom over the phase degree of freedom, and underconstrained power spectra, such as those depicted in Figure~\ref{fig:UwU_UwNU}.
In particular, this could be used to generate first guesses to initialize phase retrieval algorithms in the context of image reconstruction, see e.g. Refs.~\onlinecite{Fienup1982,Bertolotti2012}, to generate textures with suitable properties~\cite{Lagae2010, Dong2020}, or to generate discretized versions of random fields with suitable correlations~\cite{Ma2017a}.

Furthermore, UwNU can be used to impose peaked features at continuous values onto uniform systems with free boundary conditions: this could for instance be useful to generate non-repeating random textures.
In Fig.\ref{fig:UwU_UwNU}, we illustrate UwNU by constraining 6 peaks forming a hexagon as well as the central peak, such that they are all $O(N)$ with $N$ the number of pixels of the real field.
As the radius of the hexagon is varied, we see that the field forms various valid lattices of the triangular family that feature a hexagon of like-valued peaks: namely, a triangular lattice (first and third row from the top) and a Kagome-like structure composed of triangular-shaped peaks forming a honeycomb lattice (second and fourth row).
These fields are all valid solutions for our constraint here as, unlike in NUwNU, the number and the spatial extents of real-space peaks are not constrained.

\bibliography{PostDoc-StefanoMartiniani}